\documentclass[
aps,
reprint,
superscriptaddress,
longbibliography,
nofootinbib
]{revtex4-2}

\usepackage[T1]{fontenc}
\usepackage[utf8]{inputenc}
\usepackage{amsmath,amssymb,amsfonts,amsthm,bm}
\usepackage{physics}
\usepackage{bbm}
\usepackage{slashed}
\usepackage{graphicx}
\usepackage[dvipsnames]{xcolor}
\usepackage{tikz}
\usepackage{tabularx}
\usepackage{multirow}
\usepackage{array}
\usepackage{diagbox}
\usepackage{multibib}
\usepackage{lineno}
\usepackage{booktabs}

\usepackage[colorlinks=true,citecolor=blue,linkcolor=blue,urlcolor=blue]{hyperref}

\begin{document}
	
	\linenumbers
	
	\title{Quantum simulation of traversable-wormhole-inspired quantum teleportation in a chaotic binary sparse SYK model}
	
	\author{Moongul Byun}
	\email{moongulbyun@gm.gist.ac.kr}
	\affiliation{Department of Physics and Photon Science, Gwangju Institute of Science and Technology,\\123 Cheomdan-gwagiro, Gwangju 61005, Republic of Korea}
	\author{Hyeonsoo Lee}
	\email{leeh.soo@gm.gist.ac.kr}
	\affiliation{Department of Physics and Photon Science, Gwangju Institute of Science and Technology,\\123 Cheomdan-gwagiro, Gwangju 61005, Republic of Korea}
	\author{Keun-Young Kim}
	\email{fortoe@gist.ac.kr}
	\affiliation{Department of Physics and Photon Science, Gwangju Institute of Science and Technology,\\123 Cheomdan-gwagiro, Gwangju 61005, Republic of Korea}
	
	\begin{abstract}
		We report the experimental signature of holographically motivated quantum teleportation on a quantum processor, driven by the highly entangled, chaotic dynamics of a many-body system. Specifically, we implement the traversable-wormhole (TW) protocol utilizing a \textit{chaotic} binary sparse $N = 8$ Sachdev--Ye--Kitaev (SYK) model. This optimized approach dramatically reduces circuit depth for noisy intermediate-scale quantum (NISQ) hardware while rigorously preserving the spectral chaos required for gravitational duality. Diagnosing the teleportation signal via mutual information, we find that while inherent noise in NISQ hardware precludes perfect quantitative agreement with exact numerical simulations, our experimental results provide evidence for the essential qualitative signature: a sign-dependent asymmetry. This work establishes a practical framework for holographic quantum simulations, offering a novel empirical testbed for exploring holographic quantum gravity.
	\end{abstract}
	
	\maketitle
	
	\textit{Introduction.---}
	The unification of general relativity and quantum mechanics into a consistent theory of quantum gravity remains a central challenge in modern physics.
	One major obstacle is the Planck scale, whose enormous energy makes direct experimental access essentially impossible and leaves much of the subject without direct experimental evidence.
	To bypass this barrier, a new direction, often called ``Quantum Gravity in the Lab,'' has emerged~\cite{PRXQuantum.4.010320,PRXQuantum.4.010321}.
	Motivated by the holographic principle and gauge/gravity duality~\cite{tHooft:1993dmi,10.1063/1.531249,Maldacena1999,witten1998antisitterspaceholography}, this program pursues controllable quantum many-body platforms as proxy systems for probing aspects of gravitational physics.
	Quantum processors and ultracold atomic systems, for example, have been used to probe holographically motivated phenomena such as black hole information and traversable-wormhole dynamics~\cite{10.1093/ptep/ptx108,Jafferis2022,CHOWDHURY2025117112}.
	
\begin{figure}[tb]
		\centering
		\includegraphics[width=0.75\linewidth]{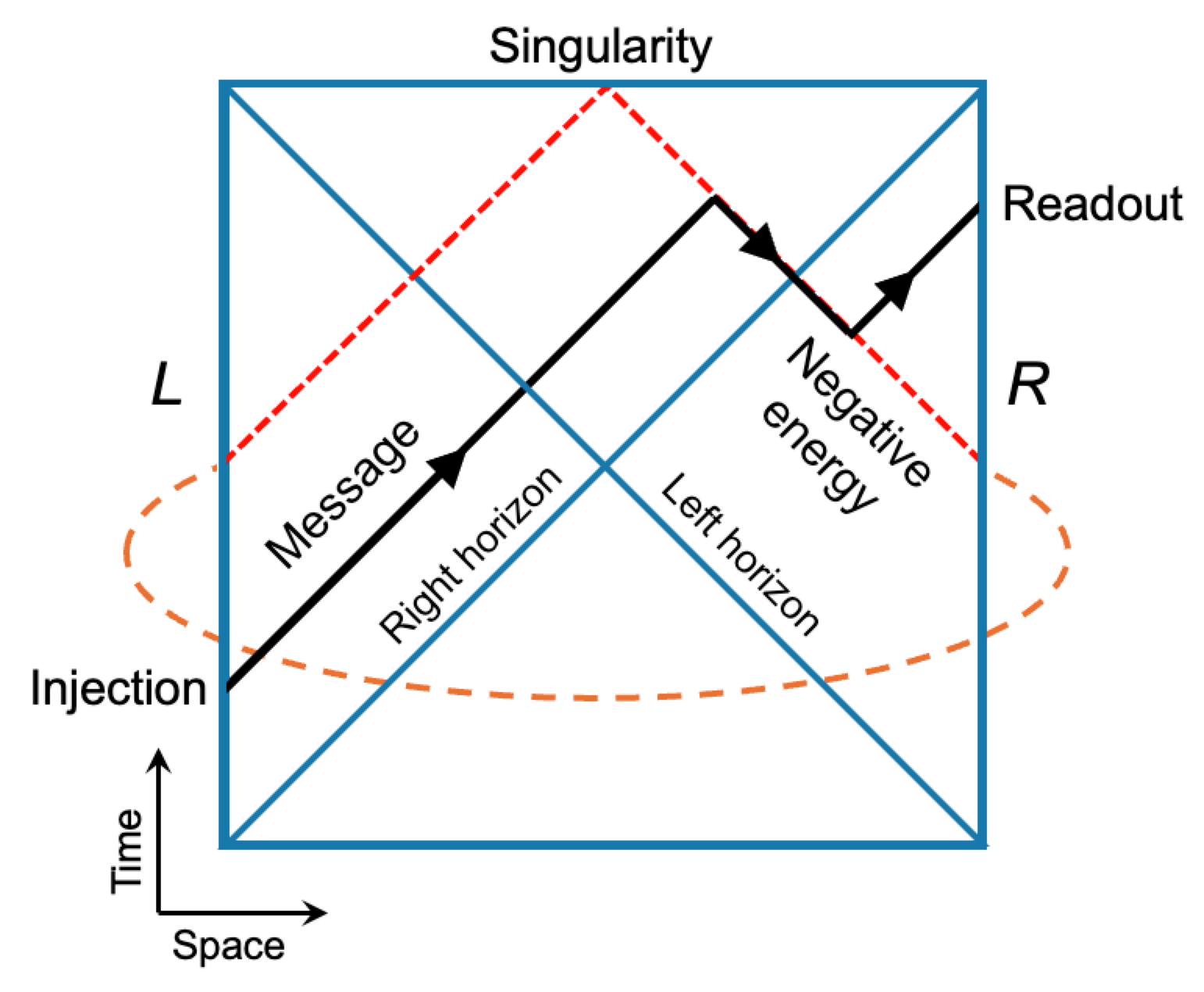}
		\caption{\label{figure1}
			Schematic Penrose diagram of the traversable wormhole.
			A signal injected from the left boundary ($L$) encounters a negative-energy shockwave induced by a double-trace deformation with the appropriate sign (orange) and emerges from the right boundary ($R$) instead of falling into the singularity.
		}
	\end{figure}

	A central object in this program is the eternal Einstein--Rosen bridge~\cite{Juan2003}, a wormhole geometry that is classically non-traversable because the null energy condition forbids causal signal transmission between its two asymptotic boundaries. 
	However, this restriction can be circumvented: the introduction of a negative-energy shockwave violates the averaged null energy condition (ANEC), thereby opening a causal channel through the wormhole~\cite{Gao2017}.
	As illustrated schematically in Fig.~\ref{figure1}, a signal injected from the left boundary ($L$), which would otherwise fall into the singularity, can then emerge from the right boundary ($R$).
	
	In the semiclassical large-$N$ limit, this bulk process admits a holographic dual interpretation as a quantum teleportation protocol between two entangled many-body systems residing at the left and right boundaries~\cite{Maldacena2017,Gao2017,Gao2021}. In this framework, the requisite bulk shockwave is generated by introducing a weak double-trace deformation that explicitly couples these two boundary systems. This interaction facilitates the teleportation of quantum information across the highly entangled background. The Sachdev--Ye--Kitaev (SYK) model~\cite{PhysRevLett.70.3339,Kitaev2015} serves as the quintessential boundary theory for this setup. Characterized by maximal quantum chaos~\cite{Maldacena2016, PhysRevD.94.106002, PhysRevLett.126.030602}, random-matrix spectral statistics~\cite{PhysRevD.94.126010}, and a low-energy effective description governed by the Schwarzian action of nearly AdS$_2$ gravity~\cite{10.1093/ptep/ptw124,PhysRevLett.117.111601,PhysRevD.94.106002}, the SYK model provides the robust properties required to establish a natural paradigm\footnote{Several recent experimental efforts have realized SYK-like dynamics across diverse quantum hardware platforms~\cite{PhysRevD.109.105002, Kundu_2025, CHOWDHURY2025108526, Granet2026}.} for exploring traversable-wormhole physics~\cite{Maldacena2017,maldacena2018eternaltraversablewormhole,Gao2021}.
	
	To investigate these holographically motivated teleportations within an experimentally accessible framework, we employ the traversable-wormhole (TW) protocol applied to a finite-$N$ SYK model~\cite{Maldacena2017,Gao2021,PRXQuantum.4.010320,PRXQuantum.4.010321,PhysRevX.12.031013}.
	A key operational signature of traversability is a sign-dependent asymmetry in the mutual information between the reference and target qubits: the teleportation signal is enhanced when the coupling sign corresponds to a negative-energy shockwave in the bulk~\cite{Gao2021,Jafferis2022}.
	
	A central obstacle to executing this protocol on near-term hardware is the dense $q$-local SYK Hamiltonian, which contains $\binom{N}{q}$ all-to-all interaction terms. This high connectivity necessitates prohibitively deep quantum circuits, rendering the simulation highly susceptible to decoherence and noise accumulation on noisy intermediate-scale quantum (NISQ) devices. To circumvent this limitation, sparse SYK constructions are strongly motivated. In the large-$N$ limit, random sparsification is known to preserve essential holographic and chaotic features provided $O(N)$ terms are retained~\cite{Xu:2020shn,PhysRevD.103.106002,Caceres2021}. Pioneering this approach, the inaugural experimental demonstration of the TW protocol utilized a machine-learned, highly sparsified $N=7$ Hamiltonian with only five nonzero terms. Implemented as a nine-qubit circuit on the Google Sycamore processor, this effectively exhibited the requisite mutual-information asymmetry~\cite{Jafferis2022}.
	
	However, the preservation of quantum chaos under  sparsification at small $N$ is not guaranteed \textit{a priori}. 
	Indeed, recent theoretical scrutiny has questioned whether extremely sparse SYK constructions---including the machine-learned Hamiltonian of Ref.~\cite{Jafferis2022}---faithfully retain the requisite chaotic features for a genuine traversable-wormhole dual~\cite{Kobrin2025,Jafferis2025,Orman2025}. This ongoing debate motivates the central question of our work: can one systematically identify a sparse SYK Hamiltonian that is both sufficiently shallow for near-term quantum devices and robustly chaotic, thereby potentially supporting the traversable-wormhole picture?
	
	In this work, we address this question using the \textit{binary} sparse SYK model~\cite{PhysRevB.107.L081103}, which preserves spectral signatures of quantum chaos---such as the spectral form factor (SFF) and level-spacing statistics---with efficiency under aggressive sparsification. From this framework, we identify a representative, hardware-efficient binary sparse $N=8$ SYK Hamiltonian that maintains consistency with random-matrix-theory (RMT) predictions.
	We also identify this Hamiltonian from other quantum-chaos criteria, such as thermal two- and four-point correlators~\cite{Kobrin2025}.
	
	We subsequently implement the corresponding TW circuit on a superconducting IBM quantum processor.
	The measured mutual-information dynamics exhibit a sign-dependent asymmetry near the teleportation time.
	To the best of our knowledge, this constitutes the first quantum-hardware realization of the TW protocol driven by an explicitly chaotic Hamiltonian.
	
	The key advance of our work is to identify and apply a hardware-efficient yet chaotic regime of the binary sparse SYK model and confirm that the teleportation signal consistently emerges when SYK-like quantum-chaos criteria are simultaneously satisfied.
	This regime is first identified from the satisfaction of these criteria rather than from the teleportation signal itself; within this chaotic regime, the expected teleportation signature is observed experimentally.
	This is an important difference from the foundational experiment in Ref.~\cite{Jafferis2022}.
	The qualitative teleportation-signal behavior observed at $N = 8$ persists up to $N = 20$, indicating that the binary sparse $N = 8$ SYK model provides a hardware-feasible realization that captures the TW behavior at larger $N$.
	Our work therefore provides a systematic route for identifying and validating hardware-efficient many-body models that preserve the physics required for many-body and holography-inspired quantum simulations.
	
	\textit{Traversable wormhole protocol.---}
	The TW protocol starts from the initial state $\ket{\mathrm{Bell}}_{PQ}\otimes \ket{\mathrm{TFD}}_{LR}\otimes \ket{0}_{T}$, where $Q$ is the message qubit injected into the wormhole, and $P$ and $T$ are reference and target qubits, respectively.
	At inverse temperature $\beta$, the TFD state is defined as
	\begin{equation}
		\ket{\mathrm{TFD}}_{LR}=\frac{1}{\sqrt{Z}}\sum_{n}e^{-\beta E_{n}/2}\ket{n}_{L}\otimes\ket{n}_{R},
	\end{equation}
	where $Z$ is the partition function.
	Here the two sides are governed by the SYK Hamiltonians
	\begin{equation}
		\label{eq2}
		H_{L, R} = i^{q/2}\sum_{1 \leq j_{1} < \cdots < j_{q} \leq N}J_{j_{1}\cdots j_{q}}\psi_{L, R}^{j_{1}}\cdots\psi_{L, R}^{j_{q}}
	\end{equation}
	with zero mean and variance $\langle J_{j_{1}\cdots j_{q}}^{2}\rangle = \tfrac{J^{2}(q - 1)!}{N^{q - 1}}$.
	Here $\psi$ are Majorana fermions in a Jordan--Wigner (JW) transformation satisfying $\{\psi_{a}^{i}, \psi_{b}^{j}\} = \delta_{ab}\delta^{ij}$ with $a,b=L,R$ and $i,j=1,\ldots,N$.
	Throughout this work, we consider $J = \sqrt{2}$, $q=4$, and $\beta = 3$.
	
	\begin{figure}[tb]
		\centering
		\includegraphics[width=0.6\linewidth]{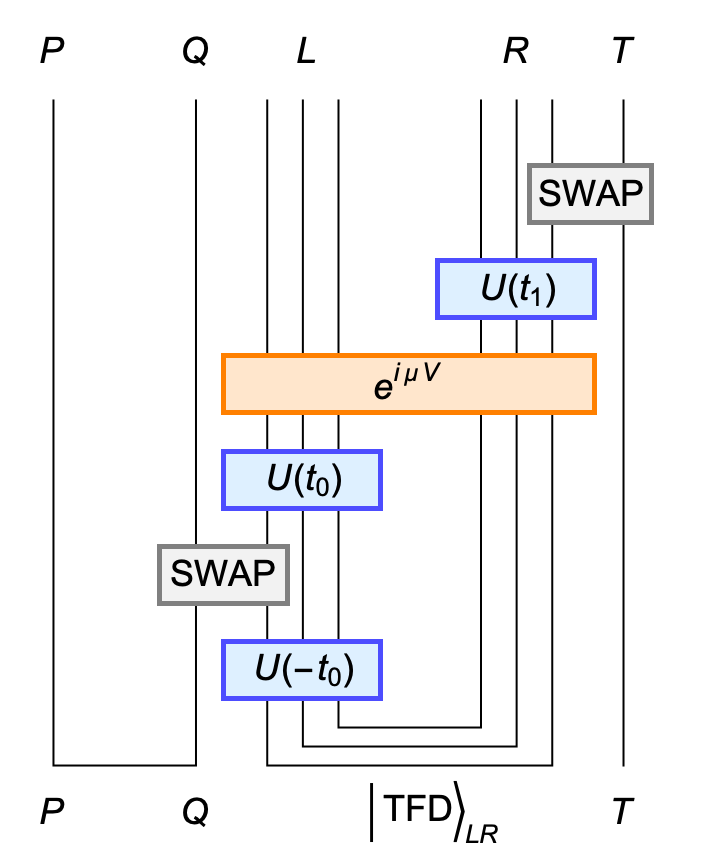}
		\caption{\label{figure2}
			Quantum circuit for the traversable wormhole protocol.
			The message qubit $Q$ is injected into the left system at time $t = -t_{0}$, the two-sided interaction $e^{i\mu V}$ is applied at $t = 0$, and the transmitted signal is read out from the right system onto the target qubit $T$ at time $t = t_{1}$.
		}
	\end{figure}
	
	The protocol injects a qubit into the wormhole at $t = -t_{0}$ by applying a SWAP operation constructed from the Dirac operator $\tfrac{1}{\sqrt{2}}(\psi_{L}^{1} \pm i\psi_{L}^{2})$~\cite{Gao2021,lykken_2024_0b9qh-jt654}.
	Throughout the protocol, the time-evolution operator is $U(t)=e^{-iH_{\text{tot}}t}$ where $H_{\mathrm{tot}}=H_L+H_R$.
	At $t = 0$ we apply the instantaneous fermionic bilinear coupling 
	\begin{equation}
		e^{i\mu V},
		\qquad
		V = \dfrac{1}{qN}H_{\text{int}},
	\end{equation}
	where $\mu$ is the coupling strength and $H_{\text{int}} = i\sum_{j = 1}^{N}\psi_{L}^{j}\psi_{R}^{j}$.
	In the holographic interpretation, with the sign convention adopted here, $\mu<0$ corresponds to ANEC violation and hence to traversability~\cite{Gao2017, Gao2021}.
	In the main text, we select $\abs{\mu} = 12$.
	The qubit is then read out at $t=t_1$ on qubit $T$ by applying $\mathrm{SWAP}$.
	The quantum circuit is summarized in Fig.~\ref{figure2}.
	Further details of the TW protocol are given in Sec.~S1 of the Supplementary Information (SI).
	
	Traversability is diagnosed by the mutual information
	\begin{equation}
		\label{eq4}
		I_{PT}(t_0,t_1)=S_P(t_0,t_1)+S_T(t_0,t_1)-S_{PT}(t_0,t_1),
	\end{equation}
	where $S_{PT}$, $S_P$, and $S_T$ denote the base-2 von Neumann entropies of the reduced density matrices $\rho_{PT}$, $\rho_P$, and $\rho_T$, respectively.
	Following Ref.~\cite{Jafferis2022}, one may probe its time dependence in two representative ways: the fixed-injection-time protocol, in which $t_{0}$ is held fixed while $t_{1}$ is varied, and the symmetric protocol, in which one sets $t_{0}=t_{1}$.
	In both cases, the mutual information in the dense SYK model is generally larger for $\mu < 0$ near the teleportation time, leading to a sign asymmetry quantified by
	\begin{equation}
		\label{eq5}
		\Delta I_{PT} \equiv I_{PT}^{\mu<0} - I_{PT}^{\mu>0},
	\end{equation}
	which is positive in the same time window.
	This asymmetry is one direct information transfer diagnostic~\cite{Gao2021, Jafferis2022}.
	
	\textit{Binary sparse SYK.---}
	For hardware implementation of the TW protocol, the dense SYK Hamiltonian is costly because its all-to-all interactions generate many nonlocal terms, resulting in deep time-evolution circuits.
	This naturally motivates sparsification.
	For the TW protocol, however, sparsification is meaningful only if the resulting Hamiltonian remains in a chaotic regime comparable to that of the dense model.
	
	We first consider the Gaussian sparse SYK Hamiltonian
	\begin{align}
		\label{eq6}
		H_{L, R} = i^{q/2}\sum_{1 \leq j_{1} < \cdots < j_{q} \leq N}\mathcal{J}_{j_{1}\cdots j_{q}}\psi_{L, R}^{j_{1}}\cdots\psi_{L, R}^{j_{q}},\\
		\label{eq7}
		\mathcal{J}_{j_{1}\cdots j_{q}} = J_{j_{1}\cdots j_{q}}x_{j_{1}\cdots j_{q}},
	\end{align}
	where $x_{j_{1}\cdots j_{q}}\in\{0,1\}$ selects the retained interaction terms~\cite{Xu:2020shn, Caceres2021, PhysRevD.103.106002}.
	Its couplings satisfy $\langle J_{j_{1}\cdots j_{q}}^{2}\rangle = \tfrac{J^{2}(q - 1)!}{pN^{q - 1}}$, where in the original sparse-SYK construction $p$ denotes the retention probability for each interaction term. 
	In our finite-$N$ numerics, however, we use $p$ as the retained fraction relative to the dense model and implement sparsification at fixed nonzero term number $K=p\binom{N}{q}$.
	To retain, as much as possible, the uniform connectivity structure of the dense all-to-all SYK model, one can further impose a regular condition that each Majorana fermion appears in the same number of interaction terms and that each interaction term contains the same number of fermions.\footnote{This structure is obtained from $(r, s)$-regular hypergraph, where $r$ denotes the number of interaction terms in which each Majorana fermion appears, and $s$ denotes the number of fermions contained in each interaction term~\cite{Xu:2020shn, Caceres2021}.}
	
	To identify a Hamiltonian within the chaotic regime under such sparsification, we study two spectral diagnostics for the one-sided energy spectrum $\{E_i\}$.
	The first is the ensemble-averaged Gaussian-filtered SFF~\cite{Gharibyan2018, Orman2025},
	\begin{equation}
		\label{eq8}
		h(\alpha, t) = \expval{\dfrac{|Y(\alpha, t)|^{2}}{|Y(\alpha, 0)|^{2}}},
	\end{equation}
	where $Y(\alpha, t) = \sum_{i}e^{-\alpha E_{i}^{2}}e^{-i E_{i}t}$, and the second is the nearest-neighbor gap ratio~\cite{PhysRevB.75.155111, PhysRevLett.110.084101},
	\begin{equation}
		\label{eq9}
		\expval{r} = \expval{\min\left(\dfrac{s_{i}}{s_{i + 1}}, \dfrac{s_{i + 1}}{s_{i}}\right)}_{i},
	\end{equation}
	with $s_{i} = E_{i + 1} - E_{i}$.
	For $N = 8$, the corresponding RMT prediction is GOE~\cite{PhysRevD.94.126010, PhysRevB.95.115150}, so one expects $\expval{r}\approx0.530$ and the characteristic dip--ramp--plateau structure in SFF within the spectrally chaotic regime~\cite{PhysRevD.98.086026}.
	We restrict the level-spacing statistics to the unfolded bulk spectrum of the even-parity sector~\cite{PhysRevB.95.115150}.
	We refer to the transition-point analysis of Ref.~\cite{Orman2025} as a qualitative guide to identify the value of $p$ at which Eqs.~\eqref{eq8} and \eqref{eq9} substantially deviate from the RMT prediction under sparsification.
	The detailed analysis is provided in Sec.~S2 of the SI.
	
	\begin{figure}[t]
		\centering
		\includegraphics[width=\linewidth]{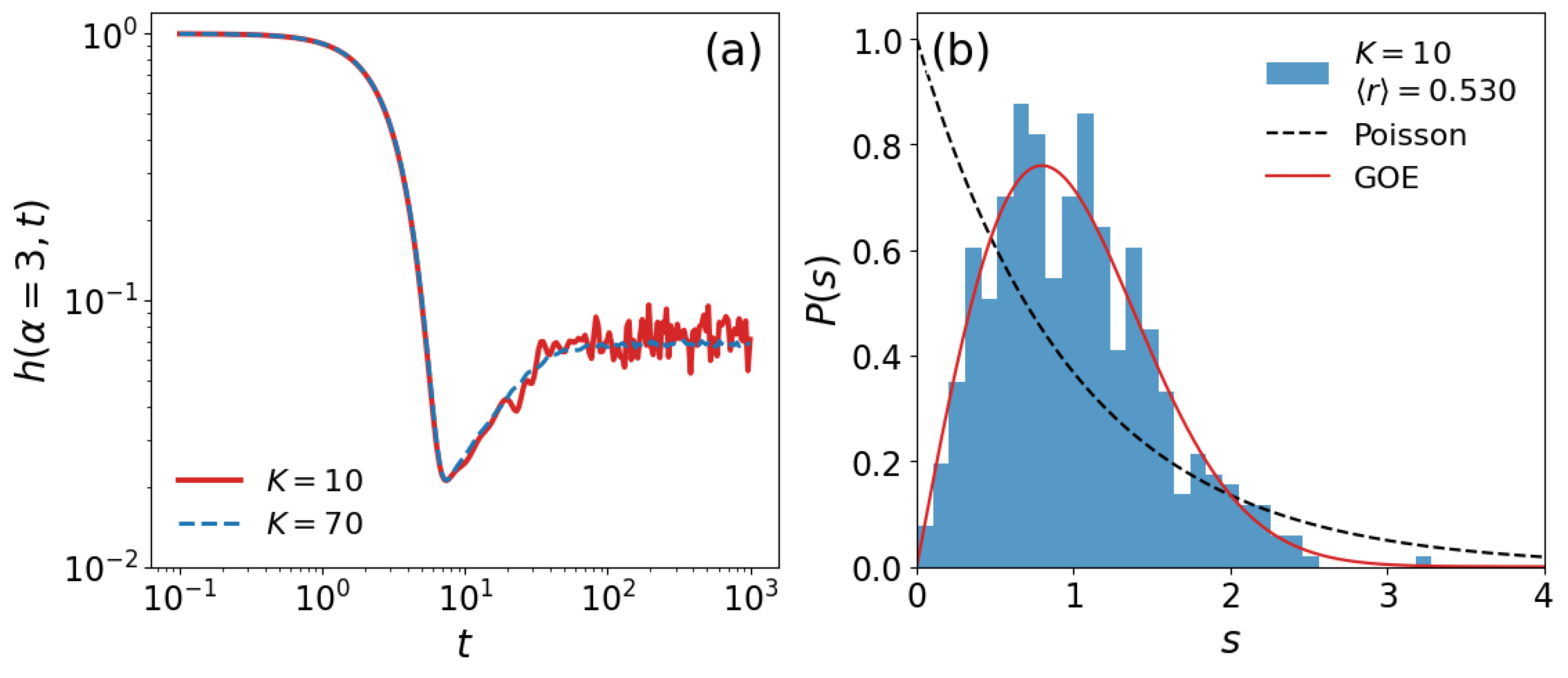}
		\caption{\label{figure3}
			Spectral properties in the binary sparse $N = 8$ SYK model with $K = 10$ ($p\approx0.14$) averaged over 100 samples. 
			(a) Disorder-averaged Gaussian-filtered spectral form factor (SFF) with $\alpha=3$ for the $K = 10$ binary ensemble, compared with that of the $K = 70$ ($p=1$) binary ensemble, averaged over 5,000 samples. 
			(b) Level-spacing distribution for the binary sparse model at $K = 10$, together with the GOE prediction.
		}
	\end{figure}
	
	We find that the Gaussian sparse $N = 8$ SYK model requires at least $K \approx 23$ ($p \approx 0.33$) to retain spectral statistics.
	However, this value of $K$ is still too large for current hardware implementation.
	
	We therefore turn to the \textit{binary} sparse SYK model~\cite{PhysRevB.107.L081103} in search of a more strongly sparsified model that can still remain in the chaotic regime. 
	This model allows a substantially smaller term count while keeping the spectral diagnostics compatible with chaos, and we therefore adopt it in what follows.
	The binary sparse SYK model takes the coefficient in Eq.~\eqref{eq7} in the form
	\begin{equation}
		\mathcal{J}_{j_{1}\cdots j_{q}} = x_{j_{1}\cdots j_{q}}\,\eta_{j_{1}\cdots j_{q}}\,\dfrac{J}{\sqrt{K}},
		\quad
		\eta_{j_{1}\cdots j_{q}}\in\{+1,-1\},
	\end{equation}
	with the two signs chosen with equal probability, while the retained interaction terms are chosen by random pruning.
	For retained terms, all nonzero coefficients have the same magnitude $\mathcal{J} \equiv J/\sqrt{K}$.
	
	For the binary sparse $N = 8$ SYK model, we find that the spectral diagnostics remain close to those of the dense model and to the GOE prediction down to $K \approx 10$.
	As shown in Fig.~\ref{figure3}, both the Gaussian-filtered SFF and the level-spacing distribution at $K = 10$ ($p \approx 0.14$) still exhibit the characteristic signatures of spectral chaos.
	We therefore choose the binary sparse $N = 8$ SYK model at $K = 10$ as the representative hardware-relevant working point for the traversable wormhole protocol.
	
	\textit{Chosen instance.---}
	Within this spectrally validated regime, we select the following $N = 8$ SYK Hamiltonian for hardware implementation:
	\begin{equation}
		\begin{aligned}
			H_{a} =\;&
			-\mathcal{J}\,\psi_{a}^{1}\psi_{a}^{2}\psi_{a}^{4}\psi_{a}^{7}
			+\mathcal{J}\,\psi_{a}^{1}\psi_{a}^{2}\psi_{a}^{6}\psi_{a}^{8}
			-\mathcal{J}\,\psi_{a}^{1}\psi_{a}^{3}\psi_{a}^{4}\psi_{a}^{5}\\
			&+\mathcal{J}\,\psi_{a}^{1}\psi_{a}^{3}\psi_{a}^{7}\psi_{a}^{8}
			-\mathcal{J}\,\psi_{a}^{1}\psi_{a}^{4}\psi_{a}^{5}\psi_{a}^{6}
			+\mathcal{J}\,\psi_{a}^{1}\psi_{a}^{5}\psi_{a}^{6}\psi_{a}^{7}\\
			&-\mathcal{J}\,\psi_{a}^{2}\psi_{a}^{3}\psi_{a}^{4}\psi_{a}^{8}
			+\mathcal{J}\,\psi_{a}^{2}\psi_{a}^{3}\psi_{a}^{5}\psi_{a}^{7}
			-\mathcal{J}\,\psi_{a}^{2}\psi_{a}^{4}\psi_{a}^{6}\psi_{a}^{7}\\
			&+\mathcal{J}\,\psi_{a}^{3}\psi_{a}^{5}\psi_{a}^{6}\psi_{a}^{8},
		\end{aligned}
		\label{eq11}
	\end{equation}
	with $\mathcal{J} = 1/\sqrt{5}$ and $a = L, R$.
	This instance is favorable for hardware compilation because it contains many commuting pairs (34 commuting versus 11 anticommuting pairs), which helps term grouping in Trotterized evolution.
	
\begin{figure}[t]
		\centering
		\includegraphics[width=1.0\linewidth]{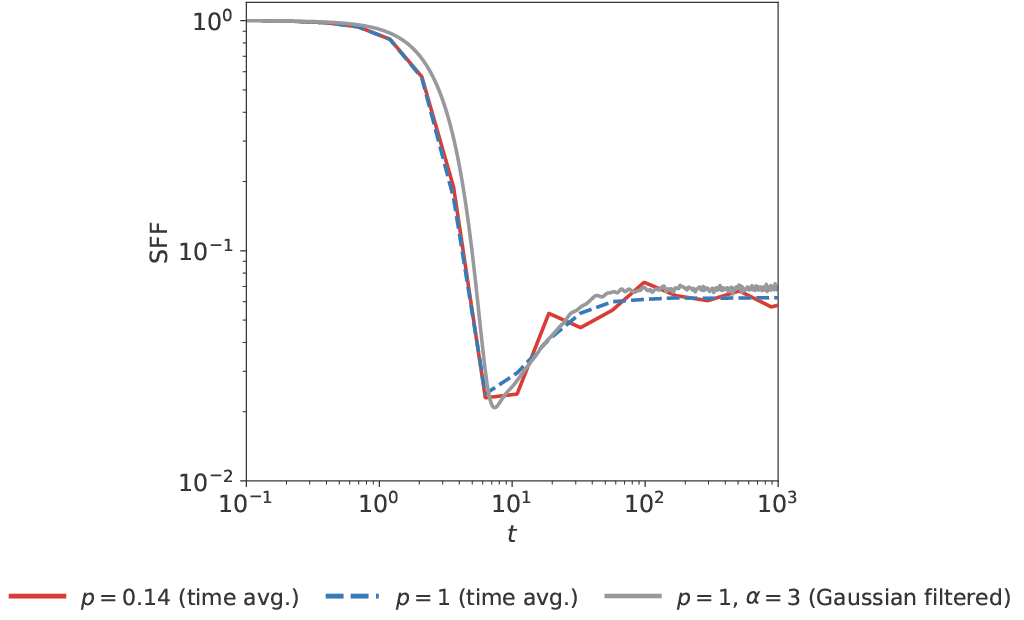}
		\caption{\label{figure4}
			Time-averaged spectral form factor (SFF) for the chosen sparse Hamiltonian, obtained using 18 time windows (red). 
			The dense $K=70$ ($p=1$) reference curves show the raw time-averaged SFF using the same time windows (blue) and the Gaussian-filtered SFF with $\alpha=3$ (gray), each averaged over 5,000 binary-coupling disorder realizations.
		}
	\end{figure}

	While generic disorder realizations at $K = 10$ remain in the chaotic regime, the single Hamiltonian in Eq.~\eqref{eq11} also shows single-instance spectral behavior compatible with the chaotic regime.
	As shown in Fig.~\ref{figure4}, the time-averaged SFF~\cite{Cotler2017,Orman2025} displays a dip--ramp--plateau structure similar to that of the dense model.
	The corresponding gap-ratio values are $\expval{r}_{\mathrm{odd}}\approx0.537$ and $\expval{r}_{\mathrm{even}}\approx0.479$ in each parity sector, which remain closer to the GOE value $\expval{r}\approx0.530$ than to the Poisson value $\expval{r}\approx0.386$~\cite{PhysRevB.75.155111, PhysRevLett.110.084101}.
	
	\begin{figure*}[tb]
		\centering
		\includegraphics[width=1.0\linewidth]{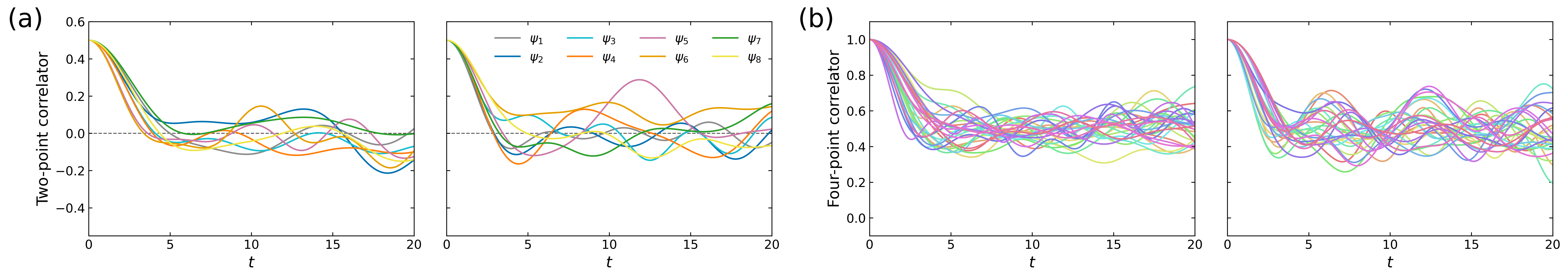}
		\caption{\label{figure5}
			Thermal two- and four-point correlators at $\beta = 3$ for $N = 8$ SYK models.
			(a) Individual thermal two-point correlators for $\psi^{i}$ with $i=1,\ldots,8$ for the (left) Gaussian dense $N = 8$ SYK Hamiltonian and (right) the chosen binary sparse $N = 8$ SYK Hamiltonian.
			(b) Thermal out-of-time-order correlators (OTOCs) for all distinct Majorana pairs $1 \leq i < j \leq 8$ for the (left) Gaussian dense $N = 8$ SYK Hamiltonian and (right) the chosen binary sparse $N = 8$ SYK Hamiltonian.
		}
	\end{figure*}
	
	The chosen Hamiltonian in Eq.~\eqref{eq11} exhibits thermal two-point correlator behavior comparable to that of the Gaussian dense $N = 8$ SYK Hamiltonian, while its parent $K = 10$ ensemble shows thermalization behavior consistent with the dense model.
	This is seen in the individual thermal two-point functions~\cite{Kobrin2025}
	\begin{equation}
		G_{\beta}^{i}(t) =
		\mathrm{Re}\,[\langle\psi^{i}(0)\psi^{i}(t)\rangle_{\beta}].
	\end{equation}
	In Fig.~\ref{figure5}(a), we illustrate $G_{\beta}^{i}(t)$ with $i = 1,\ldots,8$ for the chosen Hamiltonian and the Gaussian dense $N = 8$ SYK Hamiltonian.
	This figure shows that the chosen Hamiltonian exhibits the behavior of thermal two-point correlators comparable to that of the Gaussian dense $N = 8$ SYK Hamiltonian.
	Furthermore, for the binary sparse $N = 8$ SYK model with $K = 10$, the disorder-averaged eigenstate two-point correlators and microcanonical two-point correlators agree with the corresponding canonical thermal correlators~\cite{Sonner2017,wwfl-tkwm} and remain close to those of the Gaussian dense $N = 8$ SYK model.
	Detailed analysis is provided in Sec.~S3 in the SI.
	These results support the conclusion that the chosen Hamiltonian and its ensemble class reproduce the thermalization behavior of the SYK model relevant to the black-hole-like dynamics of the TW protocol~\cite{PhysRevD.62.024027,Kobrin2025}.
	
	We also consider the thermal out-of-time-order correlators (OTOCs)~\cite{Kobrin2025}
		\begin{equation}
			F_{\beta}^{(i, j)}(t) = -\Re[\expval{[\psi^{i}(0), \psi^{j}(t)]^{2}}_{\beta}].
		\end{equation}
		For $1 \leq i < j \leq 8$, the OTOCs for the chosen Hamiltonian exhibit scrambling dynamics comparable to those of the Gaussian dense $N=8$ SYK Hamiltonian, as shown in Fig.~\ref{figure5}(b).
		This consistency also holds when we consider disorder-averaged four-point functions.
		See Sec.~S3 in the SI.
		In particular, the OTOCs show consistent early-time decay and begin to saturate around $t\simeq 2$--$3$, close to the peak time of $\Delta I_{PT}$ in the symmetric protocol.
		In addition, as shown in Sec.~S1 of the SI, the peaks of $\Delta I_{PT}$ for both the symmetric and fixed-injection-time protocols appear consistently within this timescale for different injection pairs $(\psi_{L}^{i},\psi_{L}^{j})$.
		This indicates that the teleportation signal appears on the characteristic scrambling timescale of the system.
		
		Taken together, these spectral quantum chaos, thermalization, and scrambling properties make the chosen Hamiltonian a suitable representative hardware instance that reproduces TW-like dynamics comparable to those of the original Gaussian SYK model.
		In Sec.~S4 of the SI, we show that the teleportation signal degrades when these quantum-chaos criteria are no longer satisfied under strong sparsification.
		In Sec.~S5 of the SI, we also show that the qualitative behavior of $I_{PT}$ and $\Delta I_{PT}$ observed for the binary sparse $N = 8$ SYK model persists up to $N = 20$, in agreement with the Gaussian sparse counterparts within the chaotic regime.
		In this sense, the binary sparse SYK model provides a framework for probing mutual-information dynamics at larger $N$ while retaining tolerable quantum-circuit resource requirements in the NISQ era.
	
	\textit{Hardware implementation.---}
	For hardware implementation, we prepare the approximate TFD state variationally, implement each real-time evolution using a Trotter decomposition, and reconstruct the final two-qubit state by tomography.
	Further implementation details, including the TFD preparation, causal time-ordering analysis, Trotterization error analysis, tomography, and device-calibrated noise analysis are provided in Secs.~S6--S10 of the SI.
	
	\begin{figure*}[t]
		\centering
		\includegraphics[width=1.0\textwidth]{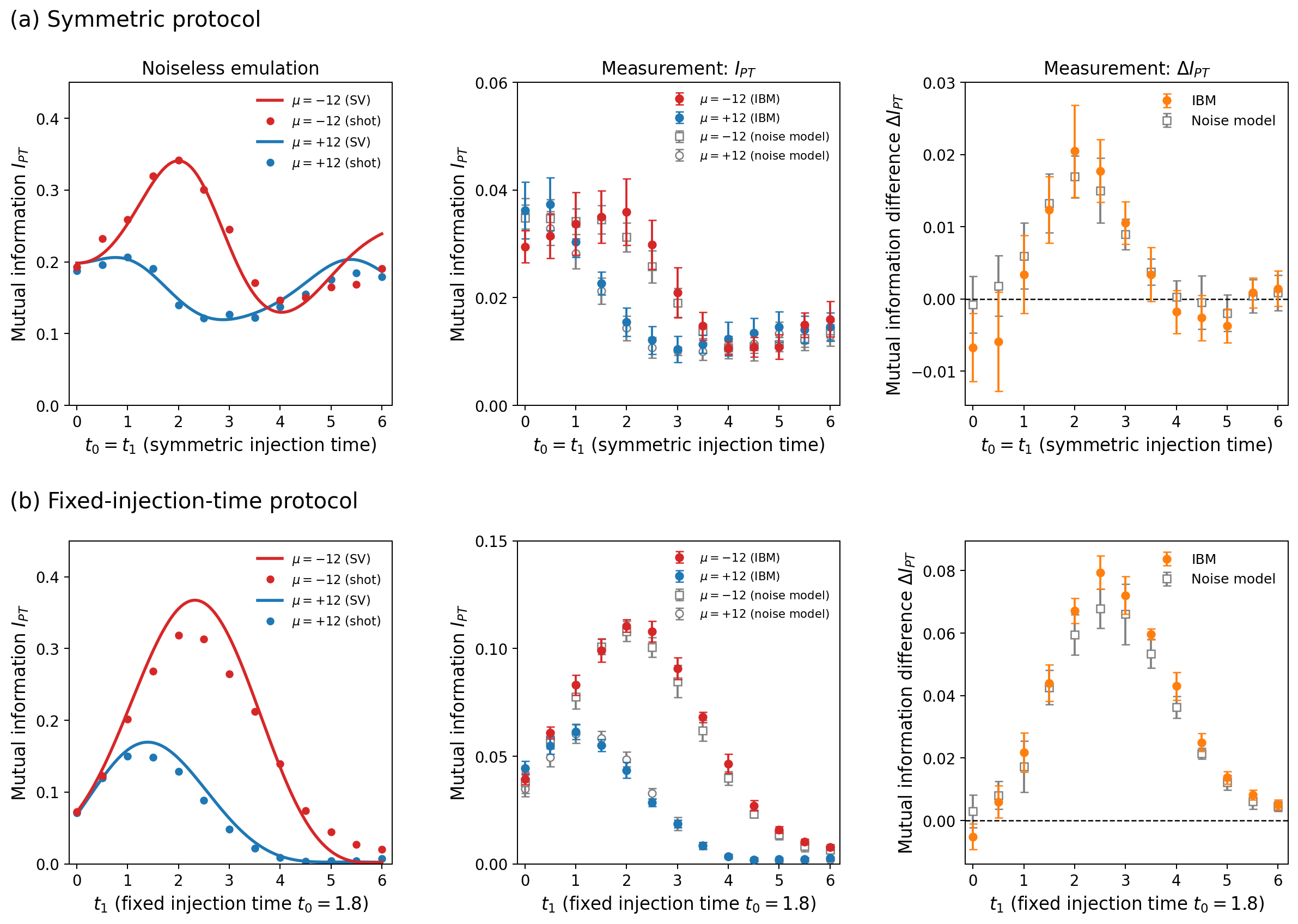}
		\caption{\label{figure6}
			Mutual information $I_{PT}$ for the chosen binary sparse $N = 8$ SYK Hamiltonian with $\beta=3$ and $\mu=\pm12$ in the (a) symmetric protocol with $t_0=t_1$, using single-step second-order Suzuki--Trotterization, and (b) fixed-injection-time protocol at $t_{0}=1.8$, using single-step first-order Lie--Trotterization, both initialized with $\ket{\psi_{\text{VQA}}}$.
			Noiseless emulation: the noiseless statevector (SV) result obtained from exact time evolution (solid) is compared with the noiseless single-step Trotter emulation with 10,000-shot sampling (dots).
			Measurement $I_{PT}$: measured mutual information $I_{PT}$ for $\mu = -12$ (red) and $\mu = +12$ (blue) from the IBM superconducting quantum processor \texttt{ibm\_phoenix}, using 10,000 shots for each tomography circuit, together with the corresponding device-calibrated noise-model results (gray).
			Error bars denote the standard deviation over 10 independent runs.
			Measurement $\Delta I_{PT}$: mutual-information difference $\Delta I_{PT}$ extracted from the IBM data (orange) and the device-calibrated noise model (gray).
		}
	\end{figure*}
	
	The TFD state is approximated by the state, denoted by $\ket{\psi_{\text{VQA}}}$, using a variational quantum algorithm (VQA)~\cite{PhysRevLett.123.220502} with a fixed ansatz of single-qubit $SU(2)$ rotations and entangling gates, similar to that of Ref.~\cite{PhysRevA.104.012427}.
	This variational preparation is motivated by the equivalent form $\ket{\mathrm{TFD}} = (2^{N}/Z)^{1/2}e^{-\beta H_{\mathrm{tot}}/4}\ket{I}$, which realizes the TFD state by imaginary-time evolution from a maximally entangled state $\ket{I}$~\cite{maldacena2018eternaltraversablewormhole}.
	The optimized circuit reaches about $97.3\%$ and $96.8\%$ fidelity $F = |\langle\psi_{\text{VQA}}|\text{TFD}\rangle|^{2}$ for the fixed-injection-time and symmetric protocols, respectively, using 36 two-qubit gates, which reproduces the mutual-information dynamics with good accuracy.
	
	For real-time evolution, we use a single-step first-order Lie--Trotter decomposition for the fixed-injection-time protocol, while we use a single-step second-order Suzuki--Trotter decomposition for the symmetric protocol~\cite{82edc856-4d85-3b98-9b0d-ad55bb9315f6,Seth1996}.
	Since the chosen Hamiltonian in Eq.~\eqref{eq11} still contains 11 anticommuting term pairs, the first-order Trotterization is not expected to reproduce the exact dynamics perfectly.
	Higher-order product formulas improve the approximation but also increase circuit depth and hence hardware noise.
	Also, applying the single-step first-order Lie--Trotterization to the symmetric protocol does not accurately reproduce the exact dynamics over the full range of probe times considered here.
	For this reason, we adopt Suzuki--Trotterization for the symmetric protocol, while the single-step Lie--Trotterization remains sufficient for the fixed-injection-time protocol, as discussed below.
	We compare the exact and Trotterized time evolution in Fig.~\ref{figure6}.
	
	Finally, we reconstruct the density matrix $\rho_{PT}$ by conventional two-qubit tomography~\cite{PhysRevA.64.052312, PhysRevA.93.062320}.
	While the protocol is described using an explicit target qubit $T$, in the hardware implementation $\rho_{PT}$ is reconstructed by measuring the right register directly rather than introducing a separate readout qubit, so the full circuit uses 10 qubits in total.
	
	To identify a hardware working point for the fixed-injection-time protocol, we compare the exact mutual-information dynamics with the corresponding noiseless single-step first-order Lie--Trotter emulation.
	Figure~\ref{figure6}(b) summarizes the comparison between the exact and Trotterized dynamics, both initialized with $\ket{\psi_{\text{VQA}}}$.
	The solid lines show the mutual information for the chosen Hamiltonian obtained from noiseless statevector evaluation of the full protocol, using exact time evolution.
	The dots show the emulation results based on the first-order Lie--Trotter evolution with 10,000 shots.
	For the fixed-injection-time protocol at $t_{0}=1.8$, the first-order Trotterization provides a good approximation to the exact mutual-information curve obtained from $\ket{\psi_{\text{VQA}}}$ over the signal region. Thus, although the chosen Hamiltonian is not fully commuting, $t_{0}=1.8$ defines a working point for which the first-order Trotterization remains reliable in the fixed-injection-time protocol.
	Smaller values of $t_{0}$ lead to more accurate circuit emulation due to reduced Trotter error, but they also yield weaker mutual-information asymmetry.
	The choice $t_{0}=1.8$ therefore provides a practical compromise between Trotter accuracy and signal visibility.
	For the symmetric protocol, the corresponding comparison between exact time evolution and single-step second-order Suzuki--Trotter emulation is shown in Fig.~\ref{figure6}(a).
	
	We examine causal time ordering, that is, whether the ordering of signal injection is preserved in the ordering of signal re-emergence~\cite{Gao2021,Jafferis2022,lykken_2024_0b9qh-jt654}.
	For this purpose, we consider $\arg\max_{t_1} I_{PT}(t_0,t_1)$, and verify whether the choice $t_{0}=1.8$ lies in the regime where wormhole-like time ordering appears.
	In the chosen Hamiltonian, we find a finite time-ordering window for suitable Trotterized interaction kicks.
	Indeed, our chosen point $t_{0}=1.8$ lies inside this window, indicating that the system preserves the causal ordering of injected signals.

	\textit{Hardware result.---}
	We run the quantum circuits for both protocols on \texttt{ibm\_phoenix}, an IBM quantum device.
	For each probe time and each sign of $\mu$, the full quantum circuit for the fixed-injection-time protocol consists of 150 two-qubit gates, 669 single-qubit gates, and has depth 317 for the $ZZ$ measurement.
	For the symmetric protocol, the same measurement circuit consists of 282 two-qubit gates, 992 single-qubit gates, and has depth 580.
	Given these circuit depths, the raw mutual information is substantially suppressed by device noise.
	We therefore adopt two error mitigations provided by the IBM cloud service: dynamical decoupling with the $X_{p}X_{m}$ sequence~\cite{PhysRevApplied.20.064027} and measurement twirling~\cite{PhysRevA.105.032620}.
	We use 10,000 shots for each measurement and repeat the measurement 10 times.

	As shown in Fig.~\ref{figure6}, the measured mutual information retains a visible peak structure and asymmetry between the two signs of $\mu$, even within the error bars.
	The TW protocol is highly sensitive to scrambling--unscrambling dynamics, so the later-time signal is more vulnerable to circuit noise, leading to a suppression of the measured mutual information for the symmetric protocol, consistent with Ref.~\cite{Jafferis2022}.
	Nevertheless, apart from the overall suppression by hardware noise, the individual $I_{PT}$ and $\Delta I_{PT}$ qualitatively follow the noiseless numerical results in terms of the asymmetry, peak structure, and dynamical tendency.
	
	For comparison, we also simulate the same compiled circuits using a device-calibrated noise model based on the calibration data of \texttt{ibm\_phoenix}, as detailed in Sec.~S10 of the SI.
	As shown in Fig.~\ref{figure6}, the noise model closely reproduces the suppression of $I_{PT}$ and $\Delta I_{PT}$ observed in the experimental data.
	This supports the interpretation that the deviations of the observed $I_{PT}$ and $\Delta I_{PT}$ from the exact results are predominantly due to hardware noise.
	Overall, we conclude that, within a binary sparse SYK regime that remains chaotic, the characteristic TW-protocol signal survives execution on present-day quantum hardware
	
\begin{figure}[t]
		\centering
		\includegraphics[width=1.0\linewidth]{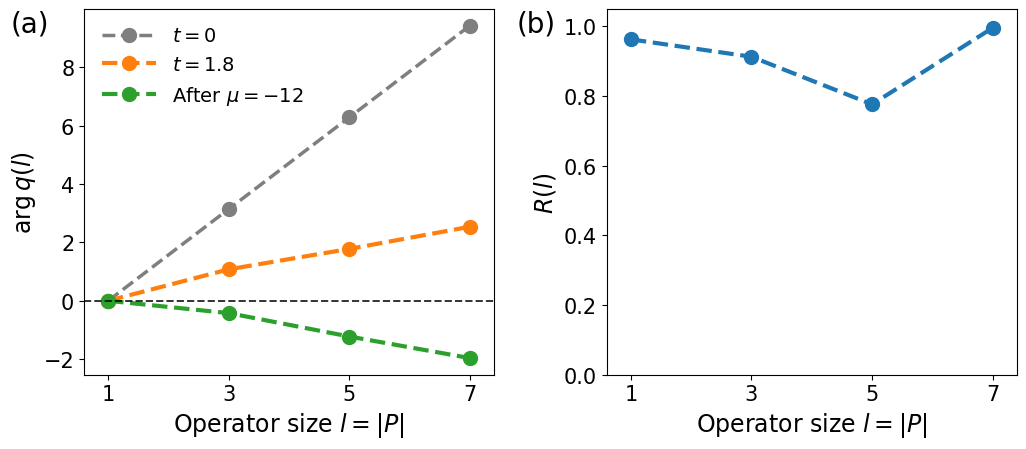}
		\caption{\label{figure7}
			Size-winding diagnostics for the chosen Hamiltonian near the mutual-information peak.
			(a) Phase of the winding size distribution $\arg q(l)$ as a function of operator size $l = \abs{P}$, before (orange) and after (green) the interaction with $\mu = -12$.
			(b) Ratio of the winding size distribution $R(l)$ at $t_{0}=1.8$.
		}
	\end{figure}

	\textit{Size winding.---}
	We examine the size-winding structure of the thermal operator dynamics generated by the Hamiltonian in Eq.~\eqref{eq11} near the mutual-information peak, which serves as a diagnostic of gravitational-like teleportation~\cite{PRXQuantum.4.010320,PRXQuantum.4.010321,Jafferis2022,PhysRevX.12.031013}.
	We consider a thermalized operator $\rho_{\beta}^{1/2}\psi_{L}^{1}(t)=\sum_{P}c_{P}(t)\psi_{L}^{P}$, where $\rho_{\beta} = e^{-\beta H}/\Tr e^{-\beta H}$ and $P$ denotes a multi-index specifying a Majorana string~\cite{Qi2019}.
	We then define the winding size distribution~\cite{PRXQuantum.4.010320,PRXQuantum.4.010321}
	\begin{equation}
		\label{eq14}
		q(l)=\sum_{P\,:\,\abs{P}=l}c_{P}^{2}.
	\end{equation}
	As shown in Fig.~\ref{figure7}(a), near the peak region, the phase $\arg q(l)$ is approximately linear in the operator size $l \equiv \abs{P}$ (orange), and the interaction reverses its slope (green), consistent with the expected size-winding picture of the protocol. 
	The corresponding ratio of winding size distribution to the conventional size distribution,
	\begin{equation}
		\label{eq15}
		R(l) = \frac{\abs{\sum_{P\,:\,\abs{P}=l}c_{P}^{2}}}{\sum_{P\,:\,\abs{P}=l}\abs{c_{P}}^{2}},
	\end{equation}
	also remains close to unity over the relevant size sectors.
	This behavior is shown in Fig.~\ref{figure7}(b), indicating that the chosen Hamiltonian retains the near-perfect size winding near the protocol-relevant time window.
	Although $q(l)$ and $R(l)$ for $\psi_{L}^{1},\ldots,\psi_{L}^{8}$ exhibit time dependence, they all experience near-perfect size winding within the time window near the teleportation time.
	Details of the analysis are provided in Sec.~S11 of the SI.
	
\begin{figure*}[tb]
		\centering
		\includegraphics[width=\linewidth]{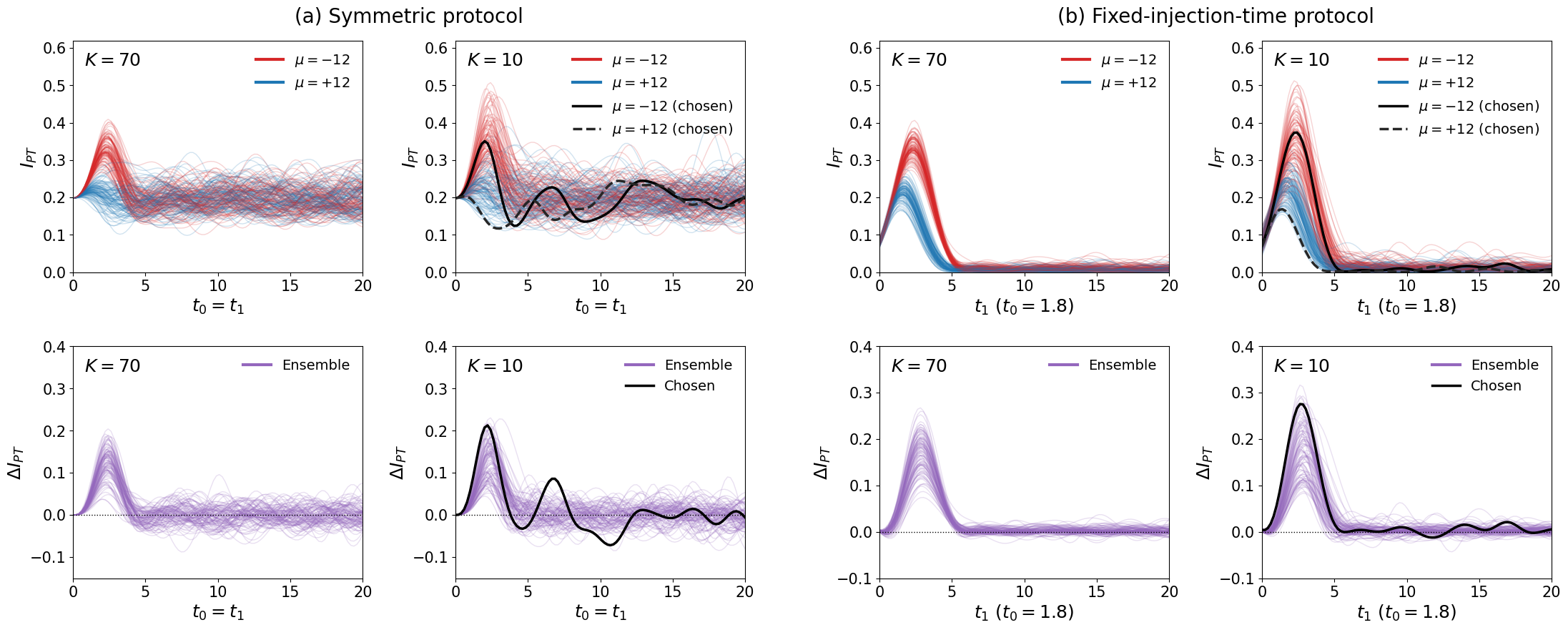}
		\caption{\label{figure8}
			Disorder realizations of mutual information $I_{PT}$ and its difference $\Delta I_{PT}$ for 100 samples of the binary $N = 8$ SYK model at $\beta=3$ and $\mu=\pm12$ in the (a) symmetric protocol with $t_{0}=t_{1}$ and (b) fixed-injection-time protocol with $t_{0}=1.8$. For each protocol, the left and right columns show the dense case $K = 70$ ($p=1$) and $K = 10$ ($p\approx0.14$), respectively. In the $K = 10$ cases, the chosen Hamiltonian is also indicated.
		}
	\end{figure*}

	\textit{Ensemble robustness.---}
	The chosen Hamiltonian in Eq.~\eqref{eq11} is not expected to be an isolated exceptional instance within the chaotic binary sparse ensemble at $K = 10$.
	Figure~\ref{figure8} shows disorder realizations of the mutual information $I_{PT}$ and its difference $\Delta I_{PT}$ for 100 samples of the binary $N=8$ SYK model in the (a) symmetric and (b) fixed-injection-time protocols at $\beta=3$ and $\mu=\pm12$, comparing the dense case $K=70$ with the sparse case $K=10$.
	As shown in this figure, other disorder realizations in the same ensemble also exhibit qualitatively similar mutual-information dynamics, and in particular retain the sign-dependent asymmetry near the teleportation time.
	This indicates that the mutual information behavior is a generic feature of the ensemble rather than a peculiarity of the chosen Hamiltonian.
	At the same time, the chosen Hamiltonian was selected not only for hardware efficiency but also because its asymmetry is among the largest within the ensemble, making it especially favorable for resolving the mutual-information signal on noisy quantum hardware.
	
	In this sense, the chosen instance should be regarded as a representative yet practically optimized member of the chaotic binary sparse ensemble: it captures the same qualitative mutual information dynamics expected more broadly in the ensemble, while providing an enhanced asymmetry signal for experimental observation.
	
	Other Hamiltonians satisfying the same criteria and yielding similar circuit depths also exist, so we emphasize that the chosen Hamiltonian should be viewed as one representative hardware-friendly choice rather than a unique instance.
	
	\textit{Discussion.---}
	In this work, we implemented the traversable-wormhole protocol on a quantum processor utilizing a binary sparse $N = 8$ SYK Hamiltonian drawn from a chaotic ensemble. Leveraging the resilience of the binary model's chaotic features under sparsification, we identified a hardware-efficient Hamiltonian at $K = 10$ that drastically reduces circuit complexity. This specific Hamiltonian accurately reproduces the theoretical mutual-information dynamics in exact numerical emulations and successfully demonstrates the hallmark sign-dependent asymmetry on superconducting IBM hardware.  To the best of our knowledge, this represents the first quantum-hardware realization of the TW protocol utilizing an explicitly chaotic Hamiltonian.
	
	Looking forward, the rigorous preservation of many-body quantum chaos will remain a central criterion for future hardware realizations of holographic dynamics and quantum teleportation. Consequently, a compelling theoretical extension of our approach is the investigation of commuting~\cite{gao2024commutingsykpseudoholographicmodel} and $d$-commuting SYK models~\cite{1k6k-xhmk}, which may offer further algorithmic simplifications while retaining the requisite spectral properties.
	
	Experimentally, scaling these protocols to larger system sizes necessitates both advanced error-mitigation techniques and systematic searches for maximally hardware-efficient, chaos-preserving Hamiltonians. 
	Transitioning to alternative, highly connected quantum architectures, such as trapped-ion processors~\cite{Granet2026}, represents another crucial frontier.
	
	Finally, the robust binary sparsification framework established here can be generalized to other quantum-informational probes, including Krylov complexity. Furthermore, it offers a highly viable path toward executing complex phenomena, most notably the Hayden--Preskill protocol~\cite{Hayden2007,Chandrasekaran2022_1,Chandrasekaran2022_2,10.1093/ptep/ptad147}, thereby providing an empirical foundation for experimentally probing the black hole information paradox.
	
	\textit{Acknowledgements.---}
	We would like to thank Jeongho Bang and Kyoungho Cho for valuable discussions, helpful correspondence, and for providing access to the quantum processor. We acknowledge the Yonsei University Quantum Computing Project Group for providing support and access to the Quantum System One (Eagle Processor), which is operated at Yonsei University.
	We also acknowledge the Yonsei University Quantum Computing Project Group for providing access to the IBM cloud service for the Nighthawk processor.
	This work was supported by the Basic Science Research Program through the National Research Foundation of Korea(NRF) funded by the Ministry of Science, ICT \& Future Planning(NRF-2021R1A2C1006791), the Korea government(MSIT)(RS-2025-02311201), (RS-2024-00445164) and the framework of international cooperation program managed by the NRF of Korea(RS-2025-02307394), the Creation of the Quantum Information Science R\&D Ecosystem(Grant No. RS-2023-NR068116) through the National Research Foundation of Korea(NRF) funded by the Korean government(Ministry of Science and ICT). 
	This research was also supported by GIST research fund (Future leading Specialized Research Project, 2026, and the Regional Innovation System \& Education(RISE) program through the(Gwangju RISE Center), funded by the Ministry of Education(MOE) and the(Gwangju Metropolitan City), Republic of Korea(2025-RISE-05-001). 
	
	\textit{Competing interests.---}
	The authors declare no competing interests.
	
	\textit{Data availability.---}
	The data supporting the findings of this study are available from the corresponding author upon reasonable request.
	
	\textit{Ethics declaration.---}
	Not applicable.
	
	\textit{Author contributions.---}
	M. B. and H. L. performed the analytical calculations and numerical simulations, including simulations on quantum computers. 
	M. B. took the lead in writing the manuscript, with contributions from H. L. and K.-Y. K. 
	K.-Y. K. conceived and supervised the project. 
	All authors contributed to the analysis, interpretation, and discussion of the results.

	\bibliographystyle{jhep}
	\bibliography{ref}

\begin{thebibliography}{99}
		
		\bibitem{Gao2021_2}
		P.~Gao and D.~L.~Jafferis,
		\textit{A traversable wormhole teleportation protocol in the SYK model},
		\href{https://doi.org/10.1007/JHEP07(2021)097}
		{\textit{JHEP} \textbf{07} (2021) 097}.
		
		\bibitem{https://doi.org/10.7907/0b9qh-jt654_2}
		J.~Lykken, A.~Zlokapa, S.~I.~Davis, D.~K.~Kolchmeyer, H.~Neven,
		M.~Spiropulu and QCCFP Collaboration,
		\textit{Long-range wormhole teleportation},
		California Institute of Technology, Accepted version, May 2024,
		\href{https://doi.org/10.7907/0b9qh-jt654}
		{doi:10.7907/0b9qh-jt654}.
		
		\bibitem{PhysRevD.94.126010_2}
		A.~M.~Garc\'{\i}a-Garc\'{\i}a and J.~J.~M.~Verbaarschot,
		\textit{Spectral and thermodynamic properties of the Sachdev--Ye--Kitaev model},
		\href{https://doi.org/10.1103/PhysRevD.94.126010}
		{\textit{Phys. Rev. D} \textbf{94} (2016) 126010}.
		
		\bibitem{PhysRevD.103.106002_2}
		A.~M.~Garc\'{\i}a-Garc\'{\i}a, Y.~Jia, D.~Rosa, and J.~J.~M.~Verbaarschot,
		\textit{Sparse Sachdev-Ye-Kitaev model, quantum chaos, and gravity duals},
		\href{https://doi.org/10.1103/PhysRevD.103.106002}
		{\textit{Phys. Rev. D} \textbf{103} (2021) 106002}.
		
		\bibitem{Orman2025_2}
		P.~Orman, H.~Gharibyan and J.~Preskill,
		\textit{Quantum chaos in the sparse SYK model},
		\href{https://doi.org/10.1007/JHEP02(2025)173}
		{\textit{JHEP} \textbf{02} (2025) 173}.
		
		\bibitem{PhysRevB.95.115150_2}
		Y.-Z.~You, A.~W.~W.~Ludwig and C.~Xu,
		\textit{Sachdev--Ye--Kitaev model and thermalization on the boundary of many-body localized fermionic symmetry-protected topological states},
		\href{https://doi.org/10.1103/PhysRevB.95.115150}
		{\textit{Phys. Rev. B} \textbf{95} (2017) 115150}.
		
		\bibitem{PhysRevB.75.155111_2}
		V.~Oganesyan and D.~A.~Huse,
		\textit{Localization of interacting fermions at high temperature},
		\href{https://doi.org/10.1103/PhysRevB.75.155111}
		{\textit{Phys. Rev. B} \textbf{75} (2007) 155111}.
		
		\bibitem{PhysRevLett.110.084101_2}
		Y.~Y.~Atas, E.~Bogomolny, O.~Giraud, and G.~Roux,
		\textit{Distribution of the ratio of consecutive level spacings in random matrix ensembles},
		\href{https://doi.org/10.1103/PhysRevLett.110.084101}
		{\textit{Phys. Rev. Lett.} \textbf{110} (2013) 084101}.
		
        \bibitem{PhysRevD.62.024027_2}
		G.~T.~Horowitz and V.~E.~Hubeny,
		\textit{Quasinormal modes of AdS black holes and the approach to thermal equilibrium},
		\href{https://doi.org/10.1103/PhysRevD.62.024027}
		{\textit{Phys. Rev. D} \textbf{62} (2000) 024027}.
		
		\bibitem{Kobrin2025_2}
		B.~Kobrin, T.~Schuster, and N.~Y.~Yao,
		\textit{Experiments implementing small commuting models lack gravitational features},
		\href{https://doi.org/10.1038/s41586-025-08939-7}
		{\textit{Nature} \textbf{643} (2025) E17--E19}.
		
		\bibitem{Kobrin2021}
		B.~Kobrin, Z.~Yang, G.~D.~Kahanamoku-Meyer, C.~T.~Olund, J.~E.~Moore,
		D.~Stanford, and N.~Y.~Yao,
		\textit{Many-body chaos in the Sachdev--Ye--Kitaev model},
		\href{https://doi.org/10.1103/PhysRevLett.126.030602}
		{\textit{Phys. Rev. Lett.} \textbf{126} (2021) 030602}.
		
		\bibitem{Sonner2017_2}
		J.~Sonner and M.~Vielma,
		\textit{Eigenstate thermalization in the Sachdev-Ye-Kitaev model},
		\href{https://doi.org/10.1007/JHEP11(2017)149}
		{\textit{JHEP} \textbf{11} (2017) 149}.
		
		\bibitem{PhysRevB.107.L081103_2}
		M.~Tezuka, O.~Oktay, E.~Rinaldi, M.~Hanada and F.~Nori,
		\textit{Binary-coupling sparse Sachdev--Ye--Kitaev model: An improved model of quantum chaos and holography},
		\href{https://doi.org/10.1103/PhysRevB.107.L081103}
		{\textit{Phys. Rev. B} \textbf{107} (2023) L081103}.
		
		\bibitem{maldacena2018eternaltraversablewormhole_2}
		J.~Maldacena and X.-L.~Qi,
		\textit{Eternal traversable wormhole},
		\href{https://arxiv.org/abs/1804.00491}{arXiv:1804.00491}.
		
		\bibitem{PhysRevLett.123.220502_2}
		J.~Wu and T.~H.~Hsieh,
		\textit{Variational thermal quantum simulation via thermofield double states},
		\href{https://doi.org/10.1103/PhysRevLett.123.220502}
		{\textit{Phys. Rev. Lett.} \textbf{123} (2019) 220502}.
		
		\bibitem{PhysRevA.104.012427_2}
		V.~P.~Su,
		\textit{Variational preparation of the thermofield double state of the Sachdev--Ye--Kitaev model},
		\href{https://doi.org/10.1103/PhysRevA.104.012427}
		{\textit{Phys. Rev. A} \textbf{104} (2021) 012427}.
		
		\bibitem{Jafferis2022_2}
		D.~Jafferis, A.~Zlokapa, J.~D.~Lykken, D.~K.~Kolchmeyer, S.~I.~Davis,
		N.~Lauk et al.,
		\textit{Traversable wormhole dynamics on a quantum processor},
		\href{https://doi.org/10.1038/s41586-022-05424-3}
		{\textit{Nature} \textbf{612} (2022) 51}.
		
		\bibitem{PhysRevA.64.052312_2}
		D.~F.~V.~James, P.~G.~Kwiat, W.~J.~Munro and A.~G.~White,
		\textit{Measurement of qubits},
		\href{https://doi.org/10.1103/PhysRevA.64.052312}
		{\textit{Phys. Rev. A} \textbf{64} (2001) 052312}.
		
		\bibitem{PhysRevA.93.062320_2}
		O.~Gamel,
		\textit{Entangled Bloch spheres: Bloch matrix and two-qubit state space},
		\href{https://doi.org/10.1103/PhysRevA.93.062320}
		{\textit{Phys. Rev. A} \textbf{93} (2016) 062320}.
		
		\bibitem{McKay:2023nxa}
		D.~C.~McKay, I.~Hincks, E.~J.~Pritchett, M.~Carroll, L.~C.~G.~Govia and S.~T.~Merkel,
		\textit{Benchmarking quantum processor performance at scale},
		\href{https://arxiv.org/abs/2311.05933}{arXiv:2311.05933}.
		
		\bibitem{Maciejewski2020mitigationofreadout}
		F.~B.~Maciejewski, Z.~Zimbor{\'a}s and M.~Oszmaniec,
		\textit{Mitigation of readout noise in near-term quantum devices by classical post-processing based on detector tomography},
		\href{https://doi.org/10.22331/q-2020-04-24-257}
		{\textit{Quantum} \textbf{4} (2020) 257}.
		
		\bibitem{susskind2018thingsfall_2}
		L.~Susskind,
		\textit{Why do Things Fall?},
		\href{https://arxiv.org/abs/1802.01198}{arXiv:1802.01198}.
		
		\bibitem{PhysRevD.98.126016_2}
		A.~R.~Brown, H.~Gharibyan, A.~Streicher, L.~Susskind, L.~Thorlacius and Y.~Zhao,
		\textit{Falling toward charged black holes},
		\href{https://doi.org/10.1103/PhysRevD.98.126016}
		{\textit{Phys. Rev. D} \textbf{98} (2018) 126016}.
		
		\bibitem{PRXQuantum.4.010320_2}
		A.~R.~Brown, H.~Gharibyan, S.~Leichenauer, H.~W.~Lin, S.~Nezami,
		G.~Salton et al.,
		\textit{Quantum gravity in the lab. i. teleportation by size and traversable wormholes},
		\href{https://doi.org/10.1103/PRXQuantum.4.010320}
		{\textit{PRX Quantum} \textbf{4} (2023) 010320}.
		
		\bibitem{PRXQuantum.4.010321_2}
		S.~Nezami, H.~W.~Lin, A.~R.~Brown, H.~Gharibyan, S.~Leichenauer,
		G.~Salton et al.,
		\textit{Quantum gravity in the lab. ii. teleportation by size and traversable wormholes},
		\href{https://doi.org/10.1103/PRXQuantum.4.010321}
		{\textit{PRX Quantum} \textbf{4} (2023) 010321}.
		
		\bibitem{Qi2019_2}
		X.-L.~Qi and A.~Streicher,
		\textit{Quantum epidemiology: operator growth, thermal effects, and SYK},
		\href{https://doi.org/10.1007/JHEP08(2019)012}
		{\textit{JHEP} \textbf{08} (2019) 012}.
		
		\bibitem{PhysRevX.12.031013_2}
		T.~Schuster, B.~Kobrin, P.~Gao, I.~Cong, E.~T.~Khabiboulline,
		N.~M.~Linke et al.,
		\textit{Many-body quantum teleportation via operator spreading in the traversable wormhole protocol},
		\href{https://doi.org/10.1103/PhysRevX.12.031013}
		{\textit{Phys. Rev. X} \textbf{12} (2022) 031013}.
		
		\bibitem{Roberts2018_2}
		D.~A.~Roberts, D.~Stanford and A.~Streicher,
		\textit{Operator growth in the SYK model},
		\href{https://doi.org/10.1007/JHEP06(2018)122}
		{\textit{JHEP} \textbf{06} (2018) 122}.
		
%
		
	\end{thebibliography}

	\onecolumngrid  
	\clearpage

	\linenumbers
	
	\onecolumngrid
	\makeatletter
	\def\set@footnotewidth@two{\hsize\textwidth\linewidth\hsize}
	\let\set@footnotewidth\set@footnotewidth@two
	\let\compose@footnotes@two\compose@footnotes@one
	\let\compose@footnotes\compose@footnotes@one
	\makeatother
	
	\setcounter{section}{0}
	\setcounter{subsection}{0}
	\setcounter{equation}{0}
	\setcounter{figure}{0}
	\setcounter{table}{0}
	\setcounter{footnote}{0}
	
	\renewcommand{\thesection}{S\arabic{section}}
	\renewcommand{\thesubsection}{S\arabic{section}.\arabic{subsection}}
	\renewcommand{\theequation}{S\arabic{equation}}
	\renewcommand{\thefigure}{S\arabic{figure}}
	\renewcommand{\thetable}{S\arabic{table}}
	
	\setcounter{page}{1}
	
	\begin{center}
		
		{\large \textbf{Supplementary Information for ``Quantum simulation of traversable-wormhole-inspired quantum teleportation in a chaotic binary sparse SYK model''}}\\[1em]
		
	\end{center}
	
	\begin{center}
		\textbf{Contents}
	\end{center}
	
	\noindent
	S1. Details of the traversable wormhole protocol \dotfill \pageref{S_section1}\\[3pt]
	\hspace*{1.5em}S1.1 Jordan--Wigner transformation \dotfill \pageref{S_section1.1}\\[3pt]
	\hspace*{1.5em}S1.2 SWAP gates \dotfill \pageref{S_section1.2}\\[3pt]
	\hspace*{1.5em}S1.3 Teleportation signal for different injection pairs \dotfill \pageref{S_section1.3}\\[3pt]
	S2. Quantum chaos in sparse SYK models \dotfill \pageref{S_section2}\\[3pt]
	\hspace*{1.5em}S2.1 Spectral form factor \dotfill \pageref{S_section2.1}\\[3pt]
	\hspace*{1.5em}S2.2 Level-spacing distribution \dotfill \pageref{S_section2.2}\\[3pt]
	S3. Thermalization and scrambling dynamics \dotfill \pageref{S_section3}\\[3pt]
	\hspace*{1.5em}S3.1 Thermalization \dotfill \pageref{S_section3.1}\\[3pt]
	\hspace*{1.5em}S3.2 Scrambling dynamics \dotfill \pageref{S_section3.2}\\[3pt]
	S4. Teleportation signal under sparsification \dotfill \pageref{S_section4}\\[3pt]
	S5. System-size scaling \dotfill \pageref{S_section5}\\[3pt]
	S6. Thermofield double state preparation \dotfill \pageref{S_section6}\\[3pt]
	S7. Causal time ordering \dotfill \pageref{S_section7}\\[3pt]
	S8. Trotterization error analysis \dotfill \pageref{S_section8}\\[3pt]
	S9. Measurements and tomography \dotfill \pageref{S_section9}\\[3pt]
	S10. Device-calibrated noise analysis \dotfill \pageref{S_section10}\\[3pt]
	\hspace*{1.5em}S10.1 Noise model \dotfill \pageref{S_section10.1}\\[3pt]
	\hspace*{1.5em}S10.2 Statistical significance \dotfill \pageref{S_section10.2}\\[3pt]
	S11. Size winding \dotfill \pageref{S_section11}\\[3pt]
	\hspace*{1.5em}S11.1 Size-winding ansatz \dotfill \pageref{S_section11.1}\\[3pt]
	\hspace*{1.5em}S11.2 Winding size distribution \dotfill \pageref{S_section11.2}\\[3pt]
	\hspace*{1.5em}S11.3 Size winding before and after interaction \dotfill \pageref{S_section11.3}\\
	
	\section{\label{S_section1}Details of the traversable wormhole protocol}
	Here we provide additional details on the traversable-wormhole (TW) protocol introduced in the main text, including the Jordan--Wigner (JW) transformation and the injection and readout SWAP gates.
	We also discuss the mutual information under different injection operations.
	
	\subsection{\label{S_section1.1}Jordan--Wigner transformation}
	The JW transformation used in the TW protocol maps Majorana fermions to Pauli strings acting on the qubit Hilbert space used in the circuit implementation.
	In our convention, the Majorana operators are normalized as 
	\begin{equation}
		\label{S_eq1}
		\{\psi_a^i,\psi_b^j\}=\delta_{ab}\delta^{ij}, \qquad i, j = 1, \ldots, N, \quad a, b = L, R.
	\end{equation}
	Accordingly, each Majorana operator carries an overall factor of $1/\sqrt{2}$. 
	We choose the JW transformation with the aim of reducing the gate count for each protocol.
	The explicit transformation for the fixed-injection-time protocol is given by
	\begin{equation}
		\label{S_eq2}
		\begin{aligned}
			\psi_{L}^{1} &= \frac{ZX}{\sqrt{2}},&
			\psi_{L}^{2} &= \frac{ZY}{\sqrt{2}},&
			\psi_{L}^{3} &= \frac{Z^{5}X}{\sqrt{2}},&
			\psi_{L}^{4} &= \frac{Z^{6}X}{\sqrt{2}},&
			\psi_{L}^{5} &= \frac{Z^{4}X}{\sqrt{2}},&
			\psi_{L}^{6} &= \frac{Z^{2}X}{\sqrt{2}},&
			\psi_{L}^{7} &= \frac{Z^{7}X}{\sqrt{2}},&
			\psi_{L}^{8} &= \frac{Z^{3}X}{\sqrt{2}},\\
			\psi_{R}^{1} &= \frac{X}{\sqrt{2}},&
			\psi_{R}^{2} &= \frac{Y}{\sqrt{2}},&
			\psi_{R}^{3} &= \frac{Z^{5}Y}{\sqrt{2}},&
			\psi_{R}^{4} &= \frac{Z^{6}Y}{\sqrt{2}},&
			\psi_{R}^{5} &= \frac{Z^{4}Y}{\sqrt{2}},&
			\psi_{R}^{6} &= \frac{Z^{2}Y}{\sqrt{2}},&
			\psi_{R}^{7} &= \frac{Z^{7}Y}{\sqrt{2}},&
			\psi_{R}^{8} &= \frac{Z^{3}Y}{\sqrt{2}},
		\end{aligned}
	\end{equation}
	while the transformation for the symmetric protocol is given by
	\begin{equation}
		\label{S_eq3}
		\begin{aligned}
			\psi_{L}^{1} &= \frac{ZX}{\sqrt{2}},&
			\psi_{L}^{2} &= \frac{ZY}{\sqrt{2}},&
			\psi_{L}^{3} &= \frac{Z^{6}X}{\sqrt{2}},&
			\psi_{L}^{4} &= \frac{Z^{7}X}{\sqrt{2}},&
			\psi_{L}^{5} &= \frac{Z^{3}X}{\sqrt{2}},&
			\psi_{L}^{6} &= \frac{Z^{4}X}{\sqrt{2}},&
			\psi_{L}^{7} &= \frac{Z^{2}X}{\sqrt{2}},&
			\psi_{L}^{8} &= \frac{Z^{5}X}{\sqrt{2}},\\
			\psi_{R}^{1} &= \frac{X}{\sqrt{2}},&
			\psi_{R}^{2} &= \frac{Y}{\sqrt{2}},&
			\psi_{R}^{3} &= \frac{Z^{6}Y}{\sqrt{2}},&
			\psi_{R}^{4} &= \frac{Z^{7}Y}{\sqrt{2}},&
			\psi_{R}^{5} &= \frac{Z^{3}Y}{\sqrt{2}},&
			\psi_{R}^{6} &= \frac{Z^{4}Y}{\sqrt{2}},&
			\psi_{R}^{7} &= \frac{Z^{2}Y}{\sqrt{2}},&
			\psi_{R}^{8} &= \frac{Z^{5}Y}{\sqrt{2}}.
		\end{aligned}
	\end{equation}
	Here $X$, $Y$, and $Z$ denote Pauli operators. 
	The notation in Eqs.~\eqref{S_eq2} and \eqref{S_eq3} is shorthand for tensor products of Pauli matrices and identity operators. 
	For example, $Z^{p}X$ denotes the Pauli string $Z\otimes \cdots \otimes Z \otimes X \otimes I \otimes \cdots \otimes I$. 
	This representation is chosen so that the first two Majoranas form complex Dirac fermions, $(\psi_{L,R}^{1}\pm i\psi_{L,R}^{2})/\sqrt{2}$, which are used to construct the SWAP operations~\cite{Gao2021_2, https://doi.org/10.7907/0b9qh-jt654_2}.
	
	For the analysis of the one-sided Hamiltonian, the two-sided representation in Eqs.~\eqref{S_eq2} and \eqref{S_eq3} contains redundant degeneracies.
	We therefore use the conventional one-sided JW transformation of $N$ Majorana fermions on a $2^{N/2}$-dimensional Hilbert space, such that
	\begin{equation}
		\label{S_eq4}
		\chi^{2m - 1} = \dfrac{Z^{m - 1}X}{\sqrt{2}},
		\qquad
		\chi^{2m} = \dfrac{Z^{m - 1}Y}{\sqrt{2}},
		\qquad
		m=1,\ldots,\frac{N}{2}.
	\end{equation}
	This representation also satisfies $\{\chi^i,\chi^j\}=\delta^{ij}$.
	Unless otherwise stated, all calculations involving the two-sided system in this Supplementary Information are performed in the full $2^{N}$-dimensional JW Hilbert space defined by Eqs.~\eqref{S_eq2} and \eqref{S_eq3}, including traces of operators that are supported only on a single side.
	The $2^{N/2}$-dimensional representation in Eq.~\eqref{S_eq4} is used only for analysis of an isolated single-side SYK Hamiltonian.
	
	\subsection{\label{S_section1.2}SWAP gates}
	Based on the JW transformation in Eqs.~\eqref{S_eq2} and \eqref{S_eq3}, we construct the SWAP gates acting on the left and right systems in the TW protocol.
	
	We first consider the injection of the message qubit through $\mathrm{SWAP}_{QL}$, which acts on the Hilbert space $\mathcal{H}_{Q}\otimes\mathcal{H}_{L}$. 
	To express this operation in the fermionic language, we combine two Majorana operators into a complex Dirac fermion, such that
	\begin{equation}
		\label{S_eq5}
		\psi_{L}^{-} = \dfrac{1}{\sqrt{2}}\left(\psi_{L}^{1} + i\psi_{L}^{2}\right),
		\qquad
		\psi_{L}^{+} = \dfrac{1}{\sqrt{2}}\left(\psi_{L}^{1} - i\psi_{L}^{2}\right).
	\end{equation}
	With the normalization in Eq.~\eqref{S_eq1}, these operators satisfy $\{\psi_{L}^{-}, \psi_{L}^{+}\}=1$. 
	Under the JW transformation used in Eqs.~\eqref{S_eq2} and \eqref{S_eq3}, $\psi_{L}^{-}$ and $\psi_{L}^{+}$ annihilate and create the occupation state of the corresponding left-system qubit, respectively.
	Then, in the computational basis, the first SWAP gate in the TW protocol is given by~\cite{Gao2021_2, https://doi.org/10.7907/0b9qh-jt654_2}
	\begin{equation}
		\mathrm{SWAP}_{QL} =
		\begin{pmatrix}
			1 & 0\\
			0 & 0
		\end{pmatrix}_{Q}\otimes\psi_{L}^{-}\psi_{L}^{+} +
		\begin{pmatrix}
			0 & 1\\
			0 & 0
		\end{pmatrix}_{Q}\otimes\psi_{L}^{+} +
		\begin{pmatrix}
			0 & 0\\
			1 & 0
		\end{pmatrix}_{Q}\otimes\psi_{L}^{-} +
		\begin{pmatrix}
			0 & 0\\
			0 & 1
		\end{pmatrix}_{Q}\otimes\psi_{L}^{+}\psi_{L}^{-}.
	\end{equation}
	With the qubit ordering of Eqs.~\eqref{S_eq2} and \eqref{S_eq3}, this gate exchanges the message qubit $Q$ with the second qubit of the $LR$ register, which represents the left injection mode.
	
	Likewise, the readout is implemented by $\mathrm{SWAP}_{RT}$, which acts on $\mathcal{H}_{R}\otimes\mathcal{H}_{T}$.
	In this case, we define
	\begin{equation}
		\psi_{R}^{-} = \dfrac{1}{\sqrt{2}}\left(\psi_{R}^{1} + i\psi_{R}^{2}\right),
		\qquad
		\psi_{R}^{+} = \dfrac{1}{\sqrt{2}}\left(\psi_{R}^{1} - i\psi_{R}^{2}\right).
	\end{equation}
	These operators annihilate and create the occupation state of the corresponding right-system qubit, respectively. 
	Then the last SWAP gate is defined by
	\begin{equation}
		\mathrm{SWAP}_{RT} = \psi_{R}^{-}\psi_{R}^{+}\otimes
		\begin{pmatrix}
			1 & 0\\
			0 & 0
		\end{pmatrix}_{T} + \psi_{R}^{+}\otimes
		\begin{pmatrix}
			0 & 1\\
			0 & 0
		\end{pmatrix}_{T} + \psi_{R}^{-}\otimes
		\begin{pmatrix}
			0 & 0\\
			1 & 0
		\end{pmatrix}_{T} + \psi_{R}^{+}\psi_{R}^{-}\otimes
		\begin{pmatrix}
			0 & 0\\
			0 & 1
		\end{pmatrix}_{T}.
	\end{equation}
	With the same qubit ordering, this gate exchanges the first qubit of the $LR$ register with the target qubit $T$.
	In the protocol setup, however, this SWAP operation can be replaced by a direct measurement of the corresponding right-system qubit after the evolution $U(t_1)$. 
	Since this gives the same reduced density matrix $\rho_{PT}$, we use this direct readout scheme in the hardware implementation described in the main text to reduce the circuit depth.
	
	\subsection{\label{S_section1.3}Teleportation signal for different injection pairs}
\begin{figure}[t]
		\includegraphics[width=1.0\linewidth]{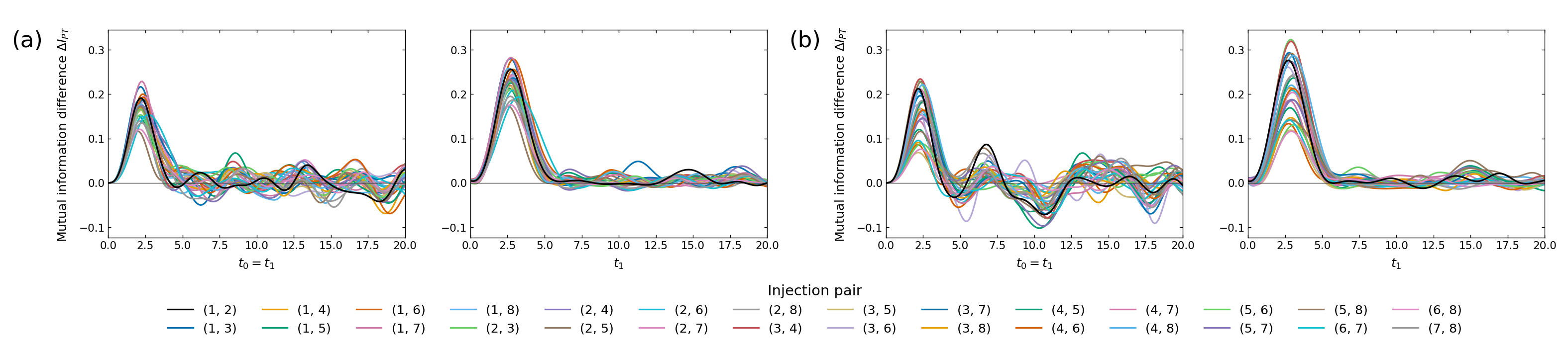}
		\caption{\label{S_figure1}
			Mutual-information difference $\Delta I_{PT}$ for each injection pair $(\psi_{L}^{i}, \psi_{L}^{j})$, denoted $(i, j)$, for (a) the Gaussian dense and (b) the chosen binary sparse $N = 8$ SYK Hamiltonians at $\beta = 3$ with $\abs{\mu} = 12$.
			Symmetric protocol with $t_{0} = t_{1}$ and fixed-injection-time protocol are shown in the left and right figures, respectively, for each SYK Hamiltonian.
			In the fixed-injection-time protocol, for each pair, $t_{0}$ is chosen to maximize the asymmetry near the teleportation time.
			The black curve is the injection pair $(1, 2)$ used in the main text.
		}
	\end{figure}

	We confirm that, for the chosen binary sparse $N = 8$ SYK Hamiltonian in the main text, the teleportation signal remains clear when we choose different injection pairs $(\psi_{L}^{i},\psi_{L}^{j})$, denoted $(i, j)$.
	This is quantified by the mutual-information asymmetry
	\begin{equation}
		\Delta I_{PT}^{(i, j)}(t_{0}, t_{1}) = I_{PT}^{(i, j)}(t_{0}, t_{1}; -\lvert\mu\rvert) - I_{PT}^{(i, j)}(t_{0}, t_{1}; +\lvert\mu\rvert),
	\end{equation}
	similar to Eq.~(5) of the main text, with $\abs{\mu} = 12$, where the Dirac mode in Eq.~\eqref{S_eq5} is replaced with the corresponding pair of Majorana operators.
	We consider both the symmetric ($t_{0} = t_{1}$) and fixed-injection-time protocols.
	For the fixed-injection-time protocol, the values of $t_{0}$ are chosen to maximize $\Delta I_{PT}$ near the teleportation time.
	As shown in Fig.~\ref{S_figure1}, the asymmetry produces a relatively dominant peak across all injection pairs at a common time comparable to that of the Gaussian dense case, for both protocols.
	The mutual-information asymmetry is therefore not restricted to the single injection pair used in the main text, but persists across different injection choices for the chosen binary sparse $N=8$ Hamiltonian.
	
	\section{\label{S_section2}Quantum chaos in sparse SYK models}
	We further examine how the spectral signatures of quantum chaos change under sparsification and how their breakdown is related to the teleportation signal in the TW protocol.
	Universal spectral chaos consistent with random matrix theory (RMT) is regarded as a necessary condition for retaining the gravitational dual of the SYK model~\cite{PhysRevD.94.126010_2,PhysRevD.103.106002_2,Orman2025_2}.
	Accordingly, the loss of spectral rigidity in a highly sparse SYK model signals a breakdown of its holographic interpretation.
	
	For the sparse $N=8$ SYK models, we parameterize the sparsity by the retained fraction $p$, defined from $K = p\binom{N}{4}$, where $K$ is the number of nonzero interaction terms in the sparse SYK Hamiltonian.
	Following the analysis of Ref.~\cite{Orman2025_2}, we consider two characteristic transition points, $p_{1}$ and $p_{2}$, identified from the spectral form factor (SFF) and the nearest-neighbor gap ratio, respectively.
	The transition point $p_{1}$ marks the onset of a significant deviation of the ramp in the sparse model from its dense counterpart.
	This transition point indicates the onset of the weakening of long-range spectral rigidity present in the dense SYK model.
	Another transition point $p_{2}$ is determined from the nearest-neighbor gap ratio, which probes nearest-neighbor eigenvalue repulsion.
	Accordingly, we identify the spectrally chaotic regime as the range of $p$ for which both $p \geq p_{1}$ and $p \geq p_{2}$ are simultaneously satisfied.
	
	For $N=8$, the RMT prediction for the dense SYK model within a fixed parity sector is the Gaussian orthogonal ensemble (GOE)~\cite{PhysRevD.94.126010_2, PhysRevB.95.115150_2}.
	Therefore, retaining comparable GOE-like spectral statistics is a necessary requirement for a sparse realization intended to reproduce the physics of the chaotic SYK model relevant to the holographically motivated TW protocol.
	Here and in the main text, we use $p_{1}$ and $p_{2}$ as the criteria for identifying spectral chaos for the binary sparse SYK model at finite $N$.\footnote{We note that these transition points can fluctuate depending on the choice of disorder ensemble at finite $N$.}
	
	\subsection{\label{S_section2.1}Spectral form factor}
	We first examine the SFF, which probes correlations in the energy spectrum and provides a diagnostic of spectral rigidity.
	For the one-sided Hamiltonian with eigenvalues $\{E_{i}\}$, we consider the Gaussian-filtered SFF
	\begin{equation}
		h(\alpha,t) = \expval{\frac{|Y(\alpha,t)|^{2}}{|Y(\alpha,0)|^{2}}},
		\qquad
		Y(\alpha,t) = \sum_{i} e^{-\alpha E_{i}^{2}}e^{-iE_{i}t},
	\end{equation}
	as defined in Eq.~(8) of the main text.
	The value of $\alpha$ is chosen such that the filtering sufficiently washes out the oscillations along the ramp before too much filtering moves the dip upward.
	As in the main text, we use $\alpha = 3$.
	In a spectrally chaotic regime, the SFF of the SYK models exhibits the characteristic dip--ramp--plateau structure, with the ramp reflecting long-range spectral rigidity.
	
	To characterize the breakdown of the SFF under strong sparsification, we consider the relative error $\epsilon(p)$, defined as the ratio of the minimum of the disorder-averaged, unnormalized Gaussian-filtered SFF for the sparse model to that of the corresponding dense model~\cite{Orman2025_2},
	\begin{equation}
		\epsilon(p) = \dfrac{\min_{t}\langle\abs{Y_{p}(\alpha, t)}^{2}\rangle}{\min_{t}\langle\abs{Y_{p = 1}(\alpha, t)}^{2}\rangle}.
	\end{equation}
	Following Ref.~\cite{Orman2025_2}, the transition point $p_{1}$ is identified by $\epsilon(p_{1}) = 1.2$.
	For $p < p_{1}$, the ramp deviates significantly from that of the corresponding dense model, indicating a modification of the long-range spectral correlations under strong sparsification relative to the dense SYK model.
	
\begin{figure}[t]
		\centering
		\includegraphics[width=\linewidth]{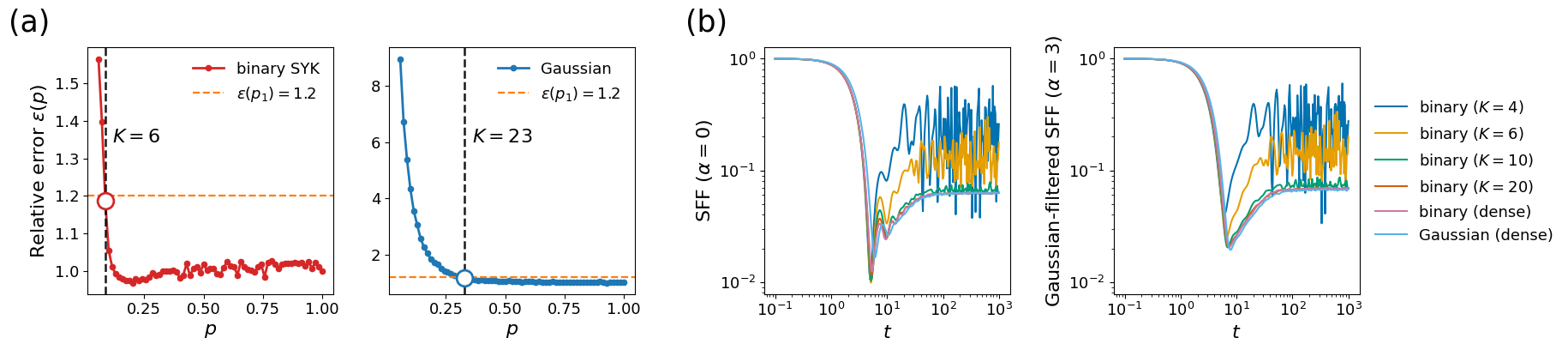}
		\caption{\label{S_figure2}
			Spectral form factors (SFFs) and their relative errors for the sparse $N=8$ SYK models, averaged over $10{,}000$ disorder realizations.
			(a) Relative error $\epsilon(p)$ for the (left) binary sparse and (right) Gaussian sparse $N = 8$ SYK models.
			The horizontal dashed line denotes $\epsilon(p_{1})=1.2$, and the vertical dashed lines indicate the resulting transition points at $K=6$ ($p_{1}\approx0.086$) for the binary sparse SYK model and $K=23$ ($p_{1}\approx0.33$) for the Gaussian sparse SYK model.
			(b) Disorder-averaged SFFs of the binary sparse SYK model for $K=4$, $6$, $10$, $20$, and $70$ (the dense case), together with the Gaussian dense SYK model, shown for the (left) raw SFF with $\alpha=0$ and the (right) Gaussian-filtered SFF with $\alpha=3$.
		}
	\end{figure}

	In Fig.~\ref{S_figure2}, we show the SFFs and their relative errors for the binary and Gaussian sparse $N=8$ SYK models.
	All SFFs are disorder-averaged over $10{,}000$ realizations for each value of $p$.
	As shown in Fig.~\ref{S_figure2}(a), for the binary sparse $N=8$ SYK model, we find $p_{1}\approx0.086$, corresponding to $K\approx6$, while for the Gaussian sparse $N=8$ SYK model, we find $p_{1}\approx0.33$, corresponding to $K\approx23$.
	Thus, the binary sparse $N = 8$ SYK model with $K=10$ chosen in the main text lies above the transition point $p_{1}$.
	Figure~\ref{S_figure2}(b) further shows how the SFF changes under sparsification.
	For $K=10$ and $K=20$, the binary sparse $N = 8$ SYK model retains a dip--ramp--plateau structure close to that of the dense SYK models, whereas for $K=6$ and especially $K=4$, the ramp becomes increasingly distorted and large late-time oscillations develop.
	These results show that long-range spectral rigidity is progressively modified under strong sparsification, indicating that this case can no longer reproduce the properties of the SYK model required in the setting of the holographically motivated protocol.
	
	\subsection{\label{S_section2.2}Level-spacing distribution}
	We next examine the correlations between neighboring energy levels.
	As in the main text, the level-spacing statistics are evaluated using the unfolded bulk spectrum in the even-parity sector.
	For consecutive eigenvalues, we define the spacings $s_{i} = E_{i + 1} - E_{i}$, and compare their distribution with the GOE prediction appropriate for the Gaussian dense $N=8$ SYK model.
	We specifically compute the nearest-neighbor gap ratio
	\begin{equation}
		\expval{r} = \expval{\min\left(\dfrac{s_{i}}{s_{i + 1}}, \dfrac{s_{i + 1}}{s_{i}}\right)}_{i},
	\end{equation}
	as defined in Eq.~(9) of the main text.
	The GOE prediction is $\expval{r}_{\rm GOE} \approx 0.530$, whereas the Poisson distribution gives $\expval{r}_{\rm Poisson} \approx 0.386$~\cite{PhysRevB.75.155111_2,PhysRevLett.110.084101_2}.
	
	We identify a second transition point $p_{2}$ from $\expval{r}$.
	In Ref.~\cite{Orman2025_2}, $p_{2}$ is defined as the value of $p$ at which $\expval{r}$ decreases to $99\%$ of its value for the dense SYK model.\footnote{While Ref.~\cite{Orman2025_2} defines $p_{2}$ without fixing the parity sector, we apply this definition to the even-parity sector. We also restrict the disorder realizations in the sparse SYK models to those whose degeneracies are identical to those of the corresponding dense models following Ref.~\cite{Orman2025_2}.}
	For the binary sparse $N=8$ SYK model, it turns out that $\expval{r}$ increases upon further sparsification.
	We therefore identify $p_{2}$ as the point at which $\expval{r}$ deviates by $1\%$ from $\expval{r}_{\rm GOE}$ and starts to increase monotonically.
	
	\begin{figure}[t]
		\includegraphics[width=\linewidth]{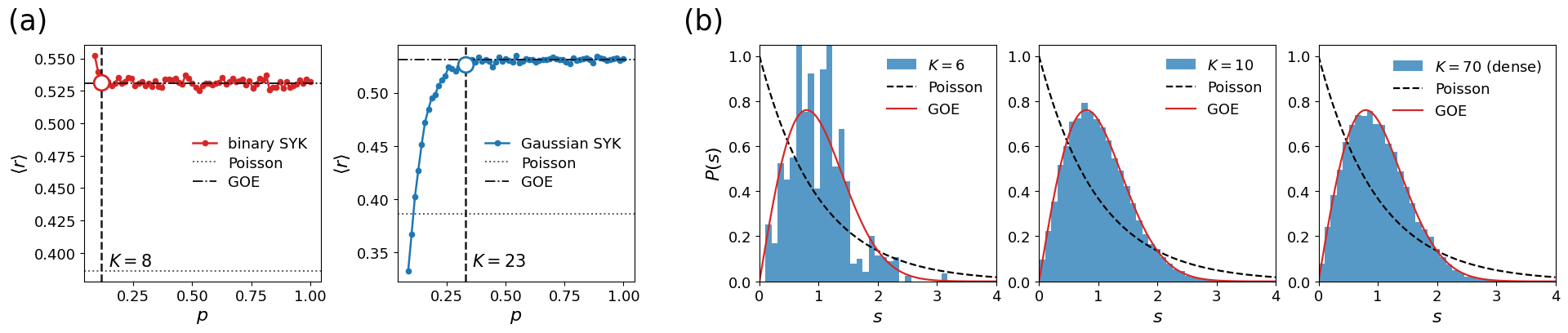}
		\caption{\label{S_figure3}
			Level-spacing statistics of the sparse $N=8$ SYK models.
			(a) Nearest-neighbor gap ratio $\expval{r}$ as a function of the retained fraction $p$ for the (left) binary sparse and (right) Gaussian sparse SYK models, averaged over $10{,}000$ disorder realizations.
			The horizontal dotted and dash-dotted lines indicate the Poisson and GOE values, respectively, and the vertical dashed lines indicate the corresponding transition points at $K=8$ for the binary sparse model and $K=23$ for the Gaussian sparse model.
			(b) Level-spacing distributions $P(s)$ of the binary sparse SYK model for $K=6$, $K=10$, and $70$ (the dense case), compared with the Poisson (black dashed) and GOE (red solid) predictions.
		}
	\end{figure}
	
	In Fig.~\ref{S_figure3}, we show $\expval{r}$ and the level-spacing distributions for the binary and Gaussian sparse $N=8$ SYK models under sparsification with $10{,}000$ and $5{,}000$ samples, respectively.
	From Fig.~\ref{S_figure3}(a), for the binary sparse $N=8$ SYK model, we identify $p_{2}\approx0.11$, corresponding to $K\approx8$.
	Therefore, $K=10$ chosen in the main text lies above this transition point and remains in the spectrally chaotic regime.
	On the other hand, for the Gaussian sparse $N=8$ SYK model, we find $p_{2}\approx0.33$, corresponding to $K\approx23$.
	Furthermore, Fig.~\ref{S_figure3}(b) shows that the binary sparse $N=8$ SYK model with $K = 10$ retains a GOE-like $P(s)$, whereas the case with $K = 6$ clearly deviates from the GOE prediction.
	These results show that the binary sparse $N=8$ SYK model with $K=10$ retains spectral signatures of quantum chaos consistent with the GOE prediction and comparable to those of the Gaussian dense $N = 8$ SYK model.
	
	\section{\label{S_section3}Thermalization and scrambling dynamics}
	\subsection{\label{S_section3.1}Thermalization}
	The chosen binary sparse $N=8$ Hamiltonian used in the main text exhibits thermalization comparable to that of the dense SYK model.
	This analysis is motivated in part by the observation that fully commuting models can restrict the thermalization required for the holography-like interpretation~\cite{PhysRevD.62.024027_2,Kobrin2025_2}. 
	In contrast, the chosen Hamiltonian studied here contains 11 anticommuting term pairs and exhibits GOE-like spectral statistics, as discussed in the main text and Sec.~\ref{S_section2} in this Supplementary Information. 
	The correlators below test whether the chosen binary sparse $N = 8$ SYK Hamiltonian and its parent ensemble also reproduce the dense-SYK-like thermalization behavior.
	
\begin{figure}[t]
		\centering
		\includegraphics[width=\linewidth]{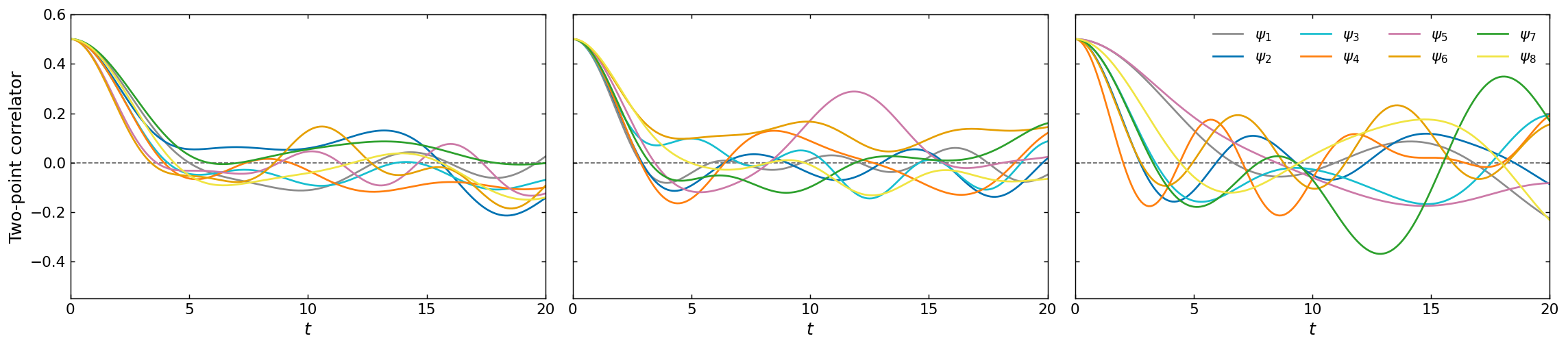}
		\caption{\label{S_figure4}Thermal two-point correlators at $\beta=3$ of the (left) Gaussian dense $N = 8$ SYK Hamiltonian, (middle) chosen binary sparse $N=8$ SYK Hamiltonian with $K = 10$, and (right) binary sparse $N=8$ SYK Hamiltonian with $K = 6$.}
	\end{figure}

	For each Majorana operator, we consider the thermal two-point correlator~\cite{Kobrin2025_2}
	\begin{equation}
		G_{\beta}^{i}(t) =
		\mathrm{Re}\,[\langle\psi^{i}(0)\psi^{i}(t)\rangle_{\beta}],
	\end{equation}
	at $\beta=3$ where $\expval{O}_{\beta} = \Tr(\rho_{\beta}O)$.
	As shown in Fig.~\ref{S_figure4}, the individual correlators of the chosen binary sparse $N = 8$ Hamiltonian relax similarly to those of the Gaussian dense $N=8$ SYK model, up to finite-$N$ revivals and fluctuations.
	Thus, the chosen Hamiltonian follows the same qualitative relaxation dynamics as the dense SYK model at the level of a single disorder realization and individual operators.
	However, when the sparsification becomes stronger, individual Hamiltonians often show less consistent early-time decay and strong late-time revivals, as shown in Fig.~\ref{S_figure4} for a binary sparse $N = 8$ SYK Hamiltonian with $K = 6$, which lies outside the spectrally chaotic regime as discussed in Sec.~\ref{S_section2}.
	
\begin{figure}[t]
		\centering
		\includegraphics[width=\linewidth]{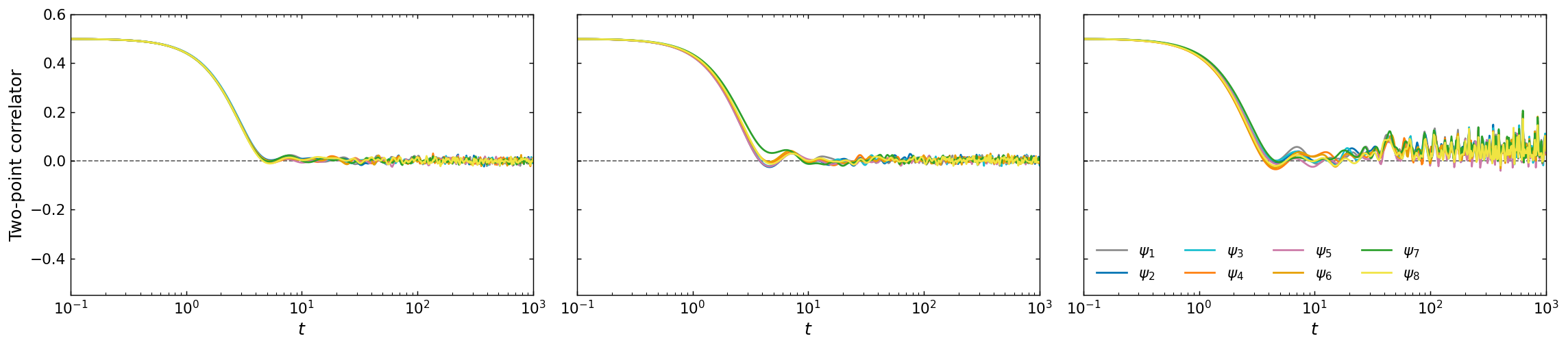}
		\caption{\label{S_figure5}Disorder-averaged thermal two-point correlators at $\beta=3$ with 100 disorder realizations for the (left) Gaussian dense $N=8$ SYK model, (middle) binary sparse $N = 8$ SYK model with $K = 10$, and (right) binary sparse $N = 8$ SYK model with $K = 6$.}
	\end{figure}

	We next examine disorder-averaged correlators. Since the SYK model is naturally defined as an ensemble, disorder averaging provides a more direct comparison of the sparse and dense thermalization dynamics~\cite{Kobrin2021, Sonner2017_2}. 
	We therefore consider the canonical thermal correlator
	\begin{equation}
		\label{S_eq14}
		\overline{G_{\beta}^{i}(t)}^{\,J}
		= \mathrm{Re}\,\overline{\langle\psi^{i}(0)\psi^{i}(t)\rangle}_{\beta}^{\,J}.
	\end{equation}
	As shown in Fig.~\ref{S_figure5}, the binary sparse $N=8$ model with $K=10$ exhibits decay of the individual correlators similar to that of the Gaussian dense $N=8$ model, with comparable late-time saturation values and finite-$N$ fluctuations.
	This shows that the binary sparse $N = 8$ SYK model exhibits relaxation behavior comparable to the Gaussian dense $N = 8$ SYK model for the individual operators considered here.
	On the other hand, as shown in Fig.~\ref{S_figure5} for a binary sparse $N = 8$ SYK model with $K = 6$, the disorder-averaged two-point functions develop late-time oscillations and do not saturate to the values observed in the Gaussian dense $N = 8$ SYK model.
	
\begin{figure}[t]
		\centering
		\includegraphics[width=\linewidth]{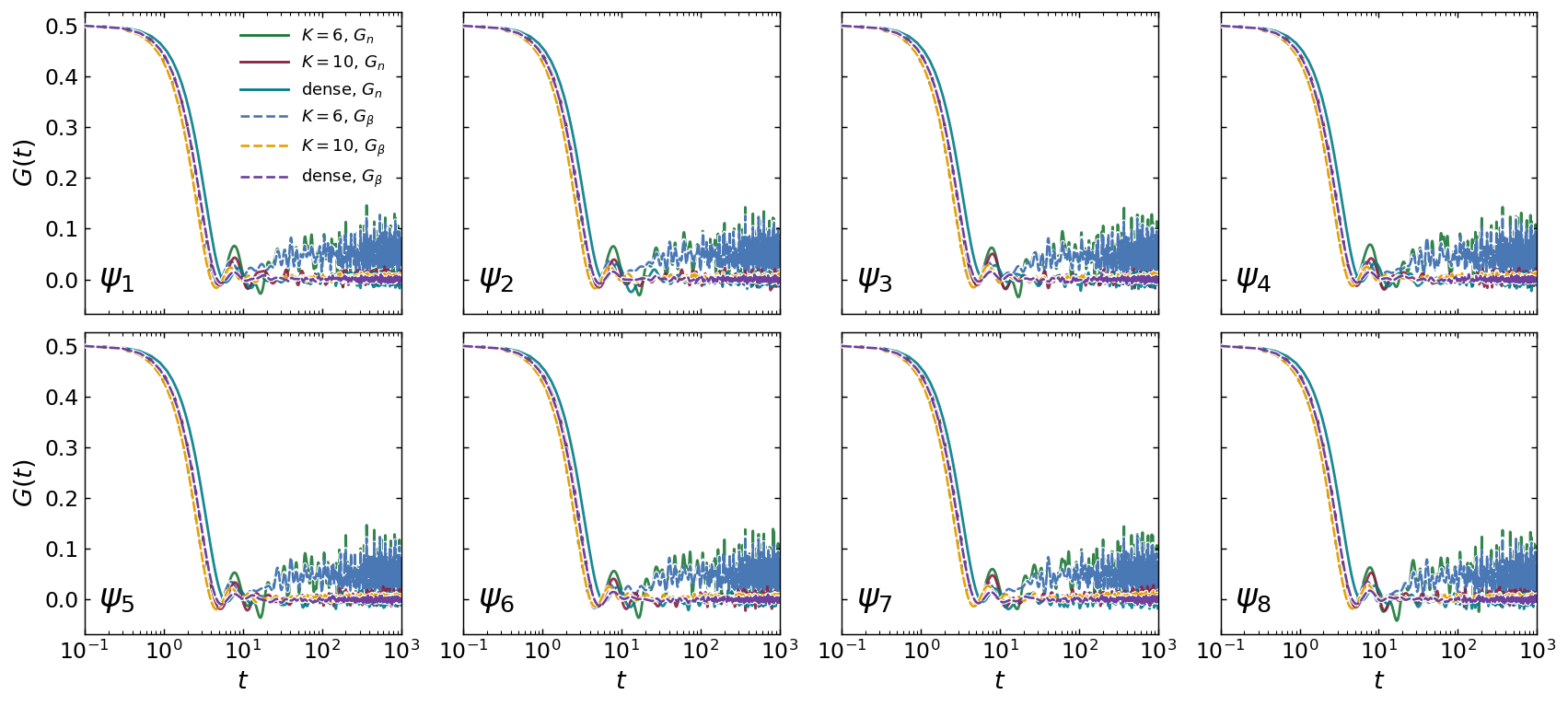}
		\caption{\label{S_figure6}Disorder-averaged eigenstate correlator $G_n$ (solid) and thermal correlator $G_{\beta}$ (dashed) for the binary sparse $N=8$ SYK model with $K = 6$ and 10 and the Gaussian dense $N=8$ SYK model.
		We take 1{,}000 disorder realizations.
		}
	\end{figure}

	We then test eigenstate thermalization by comparing the disorder-averaged thermal correlator in Eq.~\eqref{S_eq14} with the eigenstate correlator
	\begin{equation}
		\label{S_eq15}
		\overline{G_{n}^{i}(t)}^{J}
		= \mathrm{Re}\,\overline{\bra{n}\psi^{i}(0)\psi^{i}(t)\ket{n}}^{J},
	\end{equation}
	where, for each disorder realization, the eigenstate $\ket{n}$ is chosen such that its energy is closest to the thermal energy,
	\begin{equation}
		\label{S_eq16}
		E(\beta) = \frac{1}{Z}\Tr(e^{-\beta H}H),
	\end{equation}
	where $Z = \Tr e^{-\beta H}$.
	We consider 1{,}000 disorder realizations.
	For the thermalization analysis, we restrict to the even-parity sector.
	Under eigenstate thermalization, eigenstate correlators reproduce thermal correlators up to finite-size fluctuations~\cite{Sonner2017_2}. 
	Figure~\ref{S_figure6} compares $G_n$ and $G_{\beta}$ for the sparse SYK model with their dense-model correlators for each operator.
	For the binary sparse $N=8$ SYK model with $K=10$, $G_n$ and $G_{\beta}$ agree with each other and remain close to the dense counterparts within finite-$N$ fluctuations.
	
	\begin{figure}[t]
		\centering
		\includegraphics[width=\linewidth]{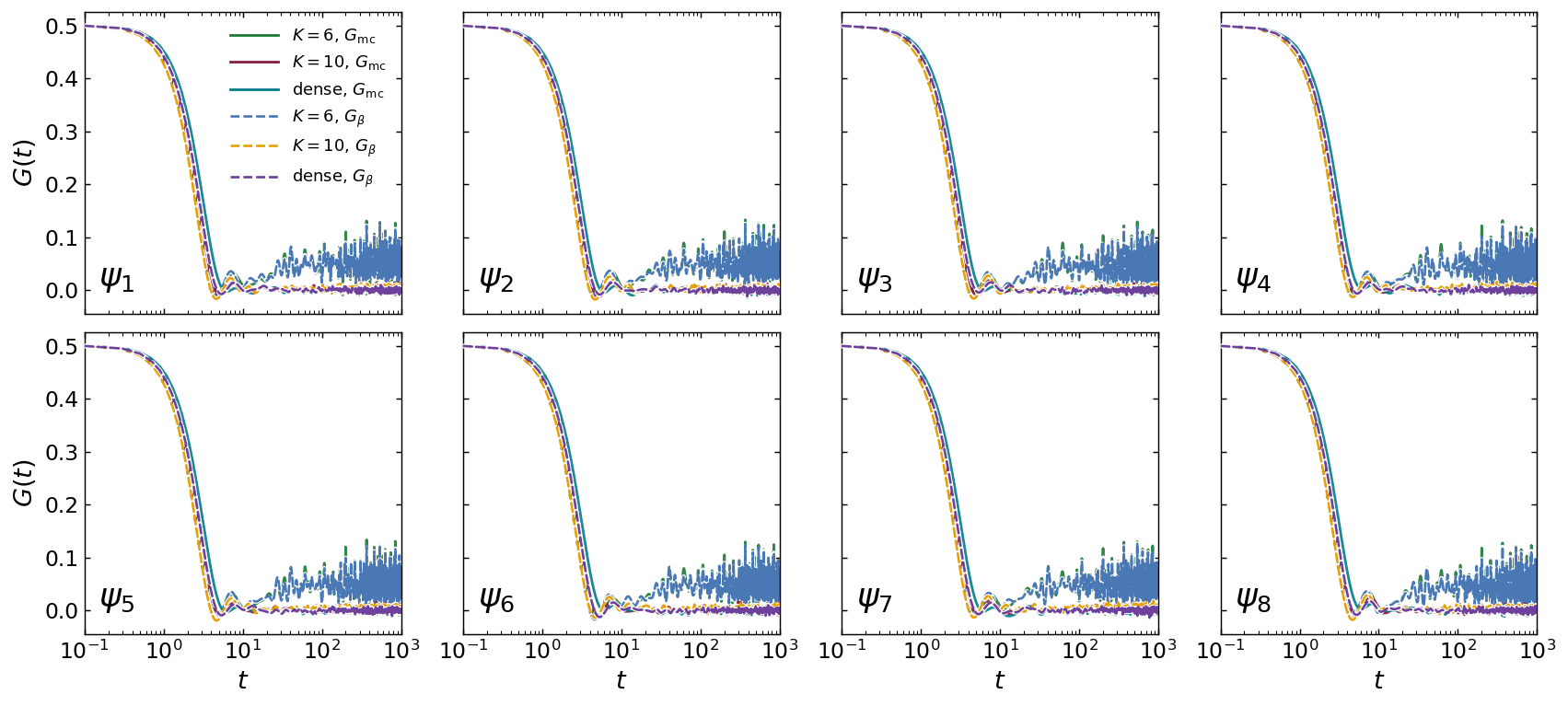}
		\caption{\label{S_figure7}Disorder-averaged microcanonical correlator $G_{\rm mc}$ (solid) and thermal correlator $G_{\beta}$ (dashed) for the binary sparse $N=8$ SYK model with $K = 6$ and 10 and the Gaussian dense $N=8$ SYK model.
			We take 1{,}000 disorder realizations.
		}
	\end{figure}

	Finally, we consider the microcanonical correlator
	\begin{equation}
		\label{S_eq17}
		\overline{G_{\rm mc}^{i}(t)}^{J}
		= \dfrac{1}{N_{E}}\overline{\sum_{n\in\mathcal{W}_{E}}G_{n}^{i}(t)}^{J},
	\end{equation}
	for the given energy window $\mathcal{W}_{E}$ with its size $N_{E}$.
	For the finite-$N$ analysis, we choose $N_{E} = 3$ consecutive energies $E_{n - 1} < E_{n} \approx E(\beta) < E_{n + 1}$.
	In Fig.~\ref{S_figure7}, we show the comparison between the microcanonical correlators and thermal correlators for each Majorana operator for the binary sparse $N = 8$ SYK model with $K = 6$ and 10, and the Gaussian dense $N = 8$ SYK model.
	As this figure shows, the disorder-averaged microcanonical correlators are similar between the binary sparse $N = 8$ SYK model with $K = 10$ and the Gaussian dense $N = 8$ SYK model.
	This implies that the binary sparse $N = 8$ SYK model with $K = 10$ reproduces the thermalization observed in the ordinary Gaussian dense SYK model.
	On the other hand, the microcanonical correlators for the binary sparse $N = 8$ SYK model with $K = 6$ do not saturate to the values of the Gaussian dense model and show strong late-time revivals, indicating that the strongly sparsified model fails to reproduce the thermalization behavior of the target SYK model.
	This thermalization criterion is also consistent with the spectral quantum chaos criteria in Sec.~\ref{S_section2}.
    
	\subsection{\label{S_section3.2}Scrambling dynamics}
	We show that the chosen binary sparse $N=8$ SYK Hamiltonian and its ensemble with $K = 10$ exhibit scrambling dynamics comparable to those of the Gaussian dense $N = 8$ SYK model.
	To see this, we compute the thermal out-of-time-order correlator (OTOC) for each one-sided Majorana operator pair $(\psi^{i}, \psi^{j})$, denoted by $(i, j)$, as defined by
	\begin{equation}
		F_{\beta}^{(i, j)}(t) = -\Re[\expval{[\psi^{i}(0), \psi^{j}(t)]^{2}}_{\beta}],
	\end{equation}
	which directly probes the scrambling dynamics of the SYK model~\cite{Kobrin2025_2}.
	We set $\beta = 3$ as in the main text.
	Due to the Clifford algebra, this OTOC starts from zero for $i = j$ or unity for $i \neq j$; if the Hamiltonian shows sufficient scrambling, it is expected to saturate near one half in both cases on the scrambling-time scale.
	
\begin{figure}[t]
		\centering
		\includegraphics[width=\linewidth]{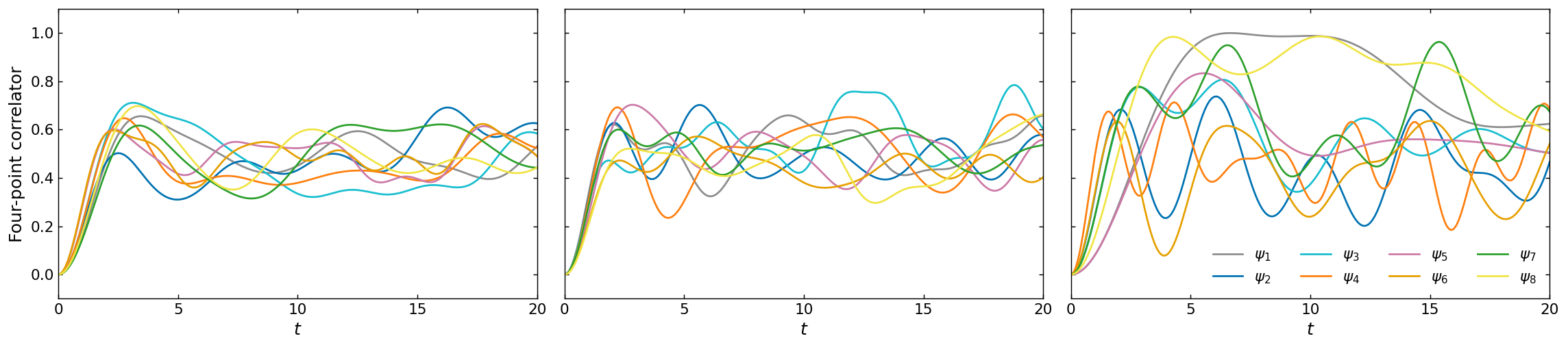}
		\caption{\label{S_figure8}Thermal four-point correlators with $i = j$ at $\beta=3$ of the (left) Gaussian dense $N=8$ SYK Hamiltonian, (middle) chosen binary sparse $N=8$ SYK Hamiltonian with $K=10$, and (right) binary sparse $N=8$ SYK Hamiltonian with $K=6$.}
	\end{figure}

\begin{figure}[t]
		\centering
		\includegraphics[width=\linewidth]{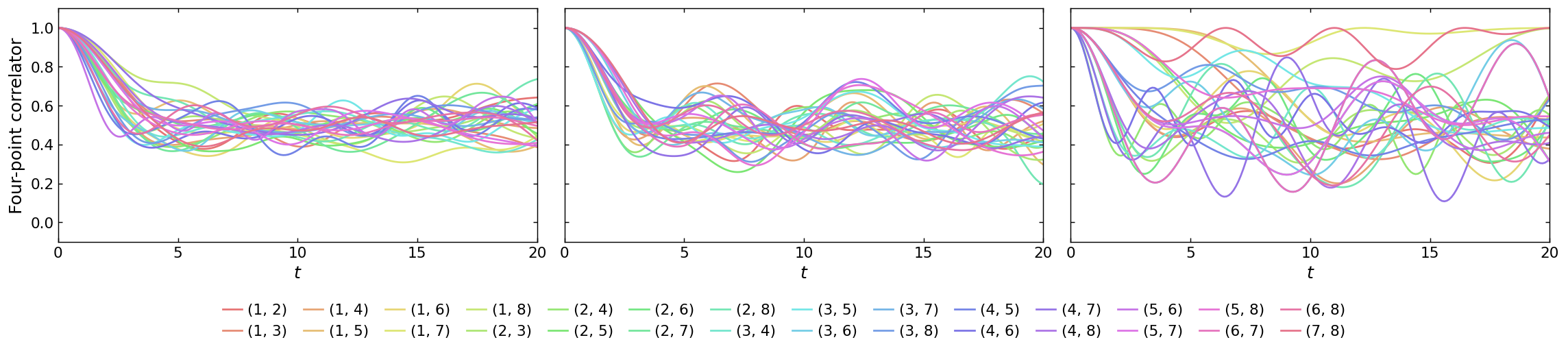}
		\caption{\label{S_figure9}Thermal four-point correlators with $i < j$ at $\beta=3$ of the (left) Gaussian dense $N=8$ SYK Hamiltonian, (middle) chosen binary sparse $N=8$ SYK Hamiltonian with $K=10$, and (right) binary sparse $N=8$ SYK Hamiltonian with $K=6$.}
	\end{figure}

	We first consider the case of $i = j$ and confirm that the chosen Hamiltonian used in the hardware experiment exhibits OTOC behavior comparable to that of the Gaussian dense $N = 8$ SYK Hamiltonian, as shown in Fig.~\ref{S_figure8}, with consistent early-time operator growth and comparable late-time fluctuations.
	However, for the binary sparse $N = 8$ SYK Hamiltonian with $K = 6$, which lies outside of the spectrally chaotic regime, individual OTOCs show less consistent early-time growth and large late-time oscillations.
	Also, the correlators for $1 \leq i < j \leq N$ for the chosen Hamiltonian show the early-time decays and late-time fluctuations comparable to those of the Gaussian dense $N = 8$ SYK Hamiltonian, as shown in Fig.~\ref{S_figure9}, while the case of $K = 6$ does not.
	
\begin{figure}[t]
		\centering
		\includegraphics[width=\linewidth]{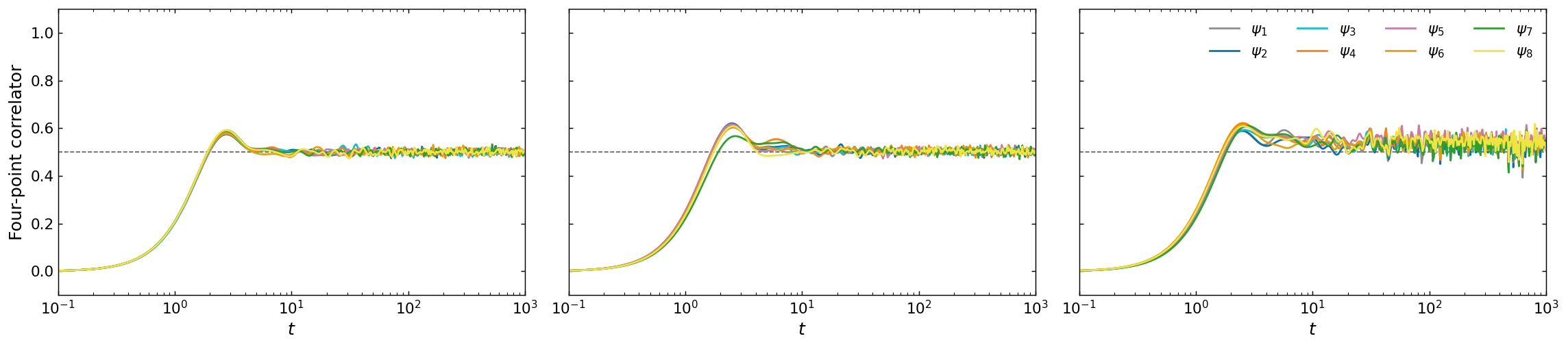}
		\caption{\label{S_figure10}Disorder-averaged thermal four-point correlators with $i = j$ at $\beta=3$ with 100 disorder realizations for the (left) Gaussian dense $N=8$ SYK model, (middle) binary sparse $N=8$ SYK model with $K=10$, and (right) binary sparse $N=8$ SYK model with $K=6$.}
	\end{figure}

\begin{figure}[t]
		\centering
		\includegraphics[width=\linewidth]{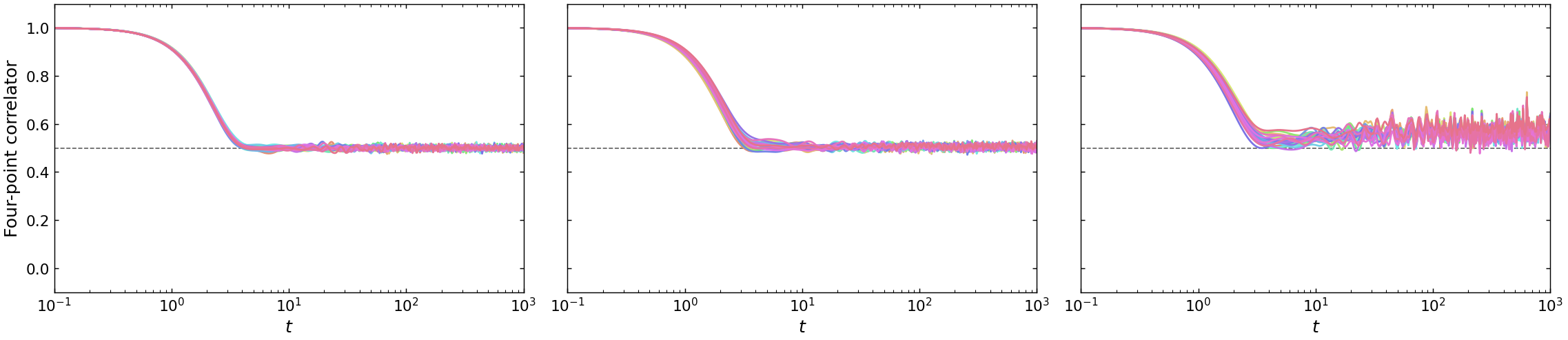}
		\caption{\label{S_figure11}Disorder-averaged thermal four-point correlators with $i < j$ at $\beta=3$ with 100 disorder realizations for the (left) Gaussian dense $N=8$ SYK model, (middle) binary sparse $N=8$ SYK model with $K=10$, and (right) binary sparse $N=8$ SYK model with $K=6$.}
	\end{figure}

	Indeed, the disorder-averaged OTOCs for the binary sparse $N = 8$ SYK model with $K = 10$ show consistent behavior compared to those of the Gaussian dense $N=8$ SYK model, as shown in Figs.~\ref{S_figure10} and \ref{S_figure11}.
	Also, for the binary sparse $N = 8$ SYK model with $K = 6$, the ensemble-averaged OTOCs show relatively large late-time fluctuations and do not saturate to the values of the Gaussian dense $N = 8$ SYK model.
	These results confirm that the binary sparse $N=8$ SYK model with $K=10$ and its chosen instance exhibit clear scrambling dynamics comparable to those of the Gaussian dense SYK model, while the model with $K = 6$ does not.
	
	Notably, the OTOCs undergo rapid initial growth or decay and start to saturate around $t\simeq 2$--$3$, which is close to the mutual-information peak time of the symmetric protocol, as shown in Fig.~\ref{S_figure1}.
	This agreement in early-time dynamics indicates that the strongest teleportation signal occurs on the characteristic scrambling timescale of the system.
	
	\section{\label{S_section4}Teleportation signal under sparsification}
	In Secs.~\ref{S_section2} and \ref{S_section3}, we verify that the case of $K = 10$ in the binary sparse $N = 8$ SYK model exhibits quantum chaos, thermalization, and scrambling dynamics comparable to those of the Gaussian dense $N = 8$ SYK model, while the model with $K = 6$ does not.
	We finally investigate how the breakdown of these properties under sparsification affects the teleportation signal in the TW protocol.
	We also show the advantage of using the binary sparse SYK model for the hardware implementation of the TW protocol compared to the Gaussian sparse SYK model.
	For this purpose, we compare the mutual-information dynamics of $100$ disorder realizations for the binary and Gaussian sparse $N=8$ SYK models at different values of $K$, using $\beta=3$ and $|\mu|=12$.
	We consider both the symmetric protocol with $t_{0}=t_{1}$ and the fixed-injection-time protocol with $t_{0} = 1.8$.
	
	\begin{figure}[t]
		\centering
		\includegraphics[width=\linewidth]{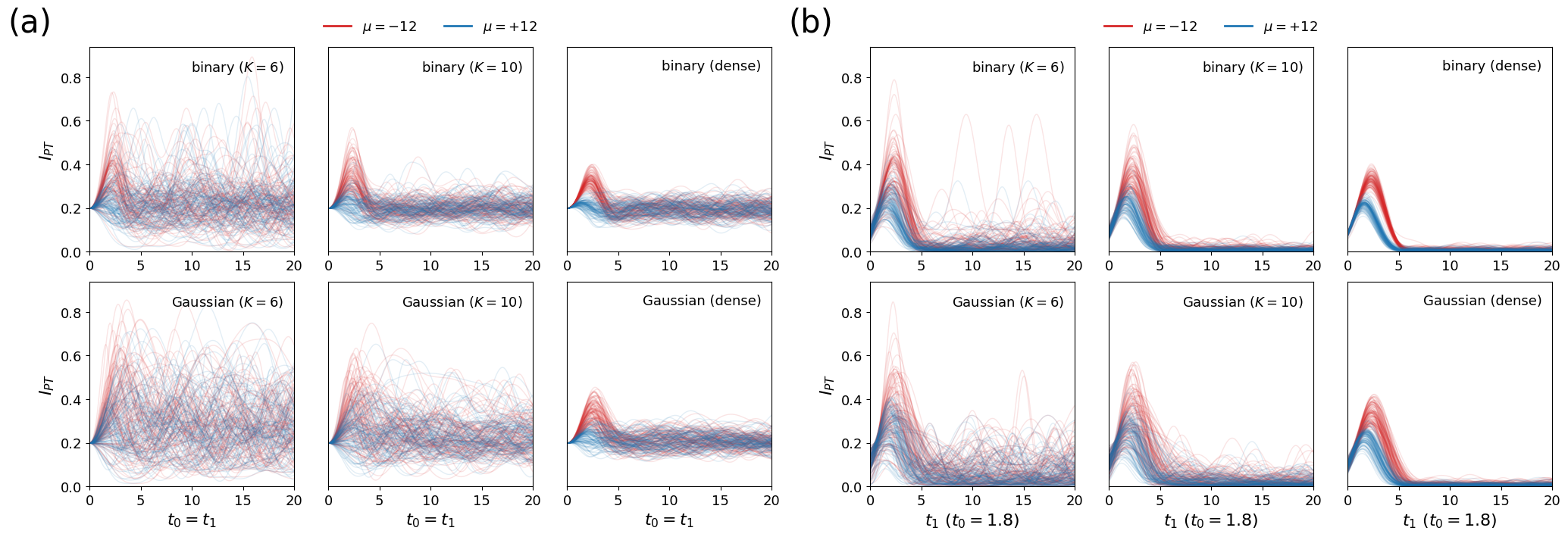}
		\caption{\label{S_figure12}
			Mutual-information dynamics of $100$ disorder realizations for the binary (top) and Gaussian (bottom) $N=8$ SYK models with $K=6$, $K=10$, and $K = 70$ (the dense case) at $\beta=3$ and $\mu=\pm12$.
			(a) Symmetric protocol with $t_{0}=t_{1}$.
			(b) Mutual information as a function of $t_{1}$ with the insertion time fixed at $t_{0}=1.8$.
			For the dense models, the disorder realizations exhibit a consistent early-time peak within a common time window and relatively weak late-time fluctuations.
			As the models are sparsified, realization-dependent late-time revivals and irregular oscillations become increasingly pronounced.
		}
	\end{figure}
	
	Figure~\ref{S_figure12} shows that the dense binary and Gaussian SYK ensembles exhibit a relatively consistent mutual-information peak within a common early-time window across the $100$ disorder realizations.
	The binary sparse $N = 8$ model with $K=10$ retains this characteristic peak structure in both time protocols, with the late-time dynamics remaining comparatively close to the dense behavior.
	In contrast, for $K=6$, the mutual-information dynamics become strongly dependent on the disorder realization, with pronounced late-time revivals and irregular oscillations.
	A similar tendency is observed for the Gaussian sparse model, where the realization dependence and late-time revivals become increasingly pronounced as $K$ is reduced below the chaotic regime.
	Thus, strong sparsification progressively destroys the consistent early-time teleportation peak and late-time saturation shared across the disorder ensemble.
	
	\begin{figure}[t]
		\centering
		\includegraphics[width=\linewidth]{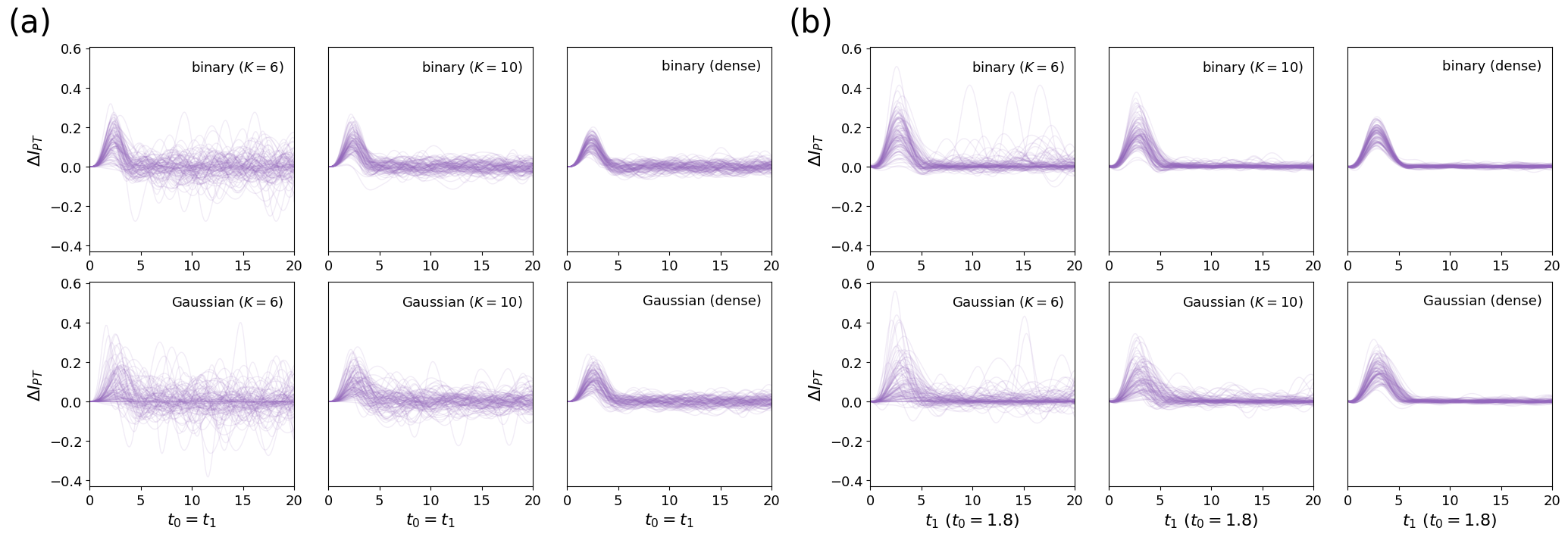}
		\caption{\label{S_figure13}
			Mutual-information asymmetry for the same $100$ disorder realizations shown in Fig.~\ref{S_figure12}.
			(a) Symmetric protocol with $t_{0} = t_{1}$.
			(b) Fixed-injection-time protocol with $t_{0}=1.8$.
			The binary $K=10$ ensemble retains a consistent positive asymmetry peak within a common early-time window and weak late-time fluctuations, similar to the dense case, whereas the strongly sparse ensemble with $K = 6$ exhibits substantially larger realization-dependent fluctuations and late-time revivals.
			The Gaussian sparse $N = 8$ model shows the same progressive loss of a common asymmetry peak structure under stronger sparsification.
		}
	\end{figure}
	
	To more directly characterize the sign dependence of the teleportation signal, we further consider the mutual-information asymmetry $\Delta I_{PT}$ for each disorder realization, as shown in Fig.~\ref{S_figure13}.
	For the dense models, the disorder realizations are concentrated around a consistent positive asymmetry peak at early times, followed by a rapid suppression of the asymmetry at late times.
	Remarkably, the binary sparse $K=10$ ensemble preserves this common asymmetry peak structure despite retaining only a small fraction of the interaction terms.
	By contrast, at $K=6$, the peak heights and positions become substantially more realization dependent, and sizable late-time revivals appear.
	The Gaussian sparse $N = 8$ model exhibits the same qualitative evolution, with increasingly broad realization dependence and late-time fluctuations as $K$ decreases.
	
	Taken together with the quantum-chaos, thermalization, and scrambling analysis above, these results show that the consistent information-transfer dynamics of the TW protocol are retained in the regime where these many-body properties are preserved.
	In the regime where all these properties are retained, the teleportation signal exhibits a consistent peak across disorder realizations within a common time window around the scrambling time, while stronger sparsification is accompanied by increasingly pronounced late-time revivals and irregular oscillations.
	This supports the interpretation that the teleportation dynamics observed for the chosen binary sparse $N = 8$ SYK Hamiltonian with $K = 10$ originate from the chaotic $N = 8$ SYK dynamics relevant to the TW protocol.
	Furthermore, at $K = 10$, the Gaussian sparse $N = 8$ SYK model lies outside the spectrally chaotic regime, and the corresponding teleportation signals also deviate from the full SYK model.
	This also coincides with the observation in Sec.~\ref{S_section2} that the quantum-chaos transition point for the binary sparse SYK model is smaller than that for the Gaussian sparse SYK model, consistent with Ref.~\cite{PhysRevB.107.L081103_2}.
	As we further describe in the following section, this shows that the binary sparse SYK model serves as a hardware-feasible yet chaotic model for implementing the TW protocol, maintaining the essential teleportation signal observed at larger $N$.
	
	\section{\label{S_section5}System-size scaling}
	
	The chosen system size $N = 8$ in the main text serves as a hardware-feasible working point for implementing the TW protocol while capturing the essential qualitative wormhole-like behavior and many-body physics of the SYK model.
	To demonstrate this, we numerically analyze the behavior of the quantum-chaos transition point $K_{c}$ determined by the criteria of Sec.~\ref{S_section2}, mutual information, and the corresponding gate counts by varying $N$.\footnote{We note that the precise value of $K_{c}$ required to retain spectral quantum chaos can vary among individual disorder realizations at a given $N$.}
	
	\begin{table}[t]
		\centering
		\begin{tabular*}{\linewidth}{@{\extracolsep{\fill}}c rrrrr rrrrr@{}}
			\toprule
			& \multicolumn{5}{c}{Binary}
			& \multicolumn{5}{c}{Gaussian} \\
			\cmidrule(lr){2-6} \cmidrule(lr){7-11}
			$N$
			& $K_{c}$ & $K_1$ & $K_2$ & $2q$ & $D$ 
			& $K_{c}$ & $K_1$ & $K_2$ & $2q$ & $D$\\
			\midrule
			8 & 8 &  6 &  8 &   32 ($K=10$) & 78 ($K=10$) & 24 &  23 & 23 &  81 & 210 \\
			10 & 16 & 12 & 15 &  133 & 326 & 58 &  57 & 32 &  277 & 619 \\
			12 & 26 & 26 & 19 &  243 & 522 & 84 &  83 & 41 &  626 & 1{,}414 \\
			14 & 46 & 46 & 22 &  509 & 904 & 140 & 140 & 48 & 1{,}142 & 2{,}292 \\
			16 & 46 & 45 & 20 &  805 & 1{,}479 & 148 & 147 & 34 & 1{,}774 & 3{,}186 \\
			18 & 54 & 54 & 24 & 1{,}162 & 1{,}822 & 174 & 174 & 38 & 2{,}811 & 4{,}857 \\
			20 & 70 & 70 & 26 & 1{,}656 & 2{,}544 & 238 & 238 & 40 & 3{,}734 & 6{,}467 \\
			\bottomrule
		\end{tabular*}
		\caption{\label{S_table1}
			Transition points, two-qubit gate counts, and circuit depth for binary and Gaussian sparse SYK Hamiltonians. 
			$K_1$ and $K_2$ denote the transition points of spectral form factor (SFF) and nearest-neighbor gap ratio, respectively, with $K_{i} = \lceil p_{i}\binom{N}{4}\rceil$. 
			The quantum-chaos transition point $K_{c}$ is defined as $K_{c} = 2\left\lceil \max(K_1,K_2)/2\right\rceil$, i.e., the smallest even integer not smaller than $\max(K_1,K_2)$. 
			The $2q$ and $D$ columns report the two-qubit gate counts and gate depths, respectively, for the single time evolution $e^{-iH_{L}t_{0}}$ at $t_{0} = 1.8$ at $K_{c}$, using one first-order Lie--Trotter step transpiled for \texttt{ibm\_phoenix}.
			For the binary $N = 8$ SYK Hamiltonian, the $2q$ and $D$ entries correspond to the chosen Hamiltonian with $K = 10$ used in the hardware experiment.
		}
	\end{table}
	
	\begin{figure}[t]
		\centering
		\includegraphics[width=\linewidth]{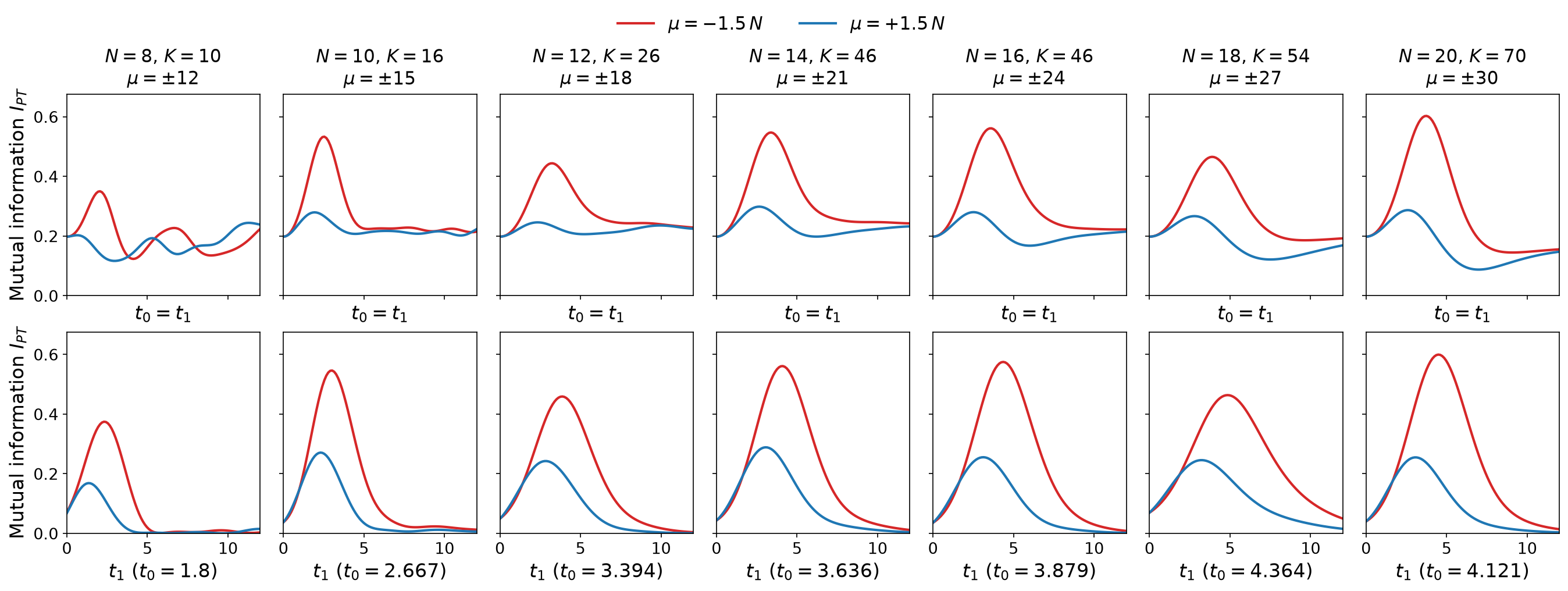}
		\caption{\label{S_figure14}
			Mutual information $I_{PT}$ at $\beta = 3$ for binary sparse SYK Hamiltonians with $N = 8, 10, \ldots, 20$, using the corresponding transition point $K_{c}$ for $N \geq 10$.
			The upper figures show the symmetric protocol, whereas the lower figures show the fixed-injection-time protocol.
			For each $N \geq 10$, the injection time $t_{0}$ for the fixed-injection-time protocol is chosen as the time at which the mutual-information asymmetry is maximal in the corresponding symmetric protocol.
			Also, the interaction strength is chosen such that $\abs{\mu}/N = 1.5$, which is identical to our $N = 8$ setup.
			For $N = 8$, we adopt the chosen binary sparse SYK Hamiltonian in the main text and use $t_{0} = 1.8$ which is used in the hardware experiment.
		}
	\end{figure}
	
	\begin{figure}[t]
		\centering
		\includegraphics[width=\linewidth]{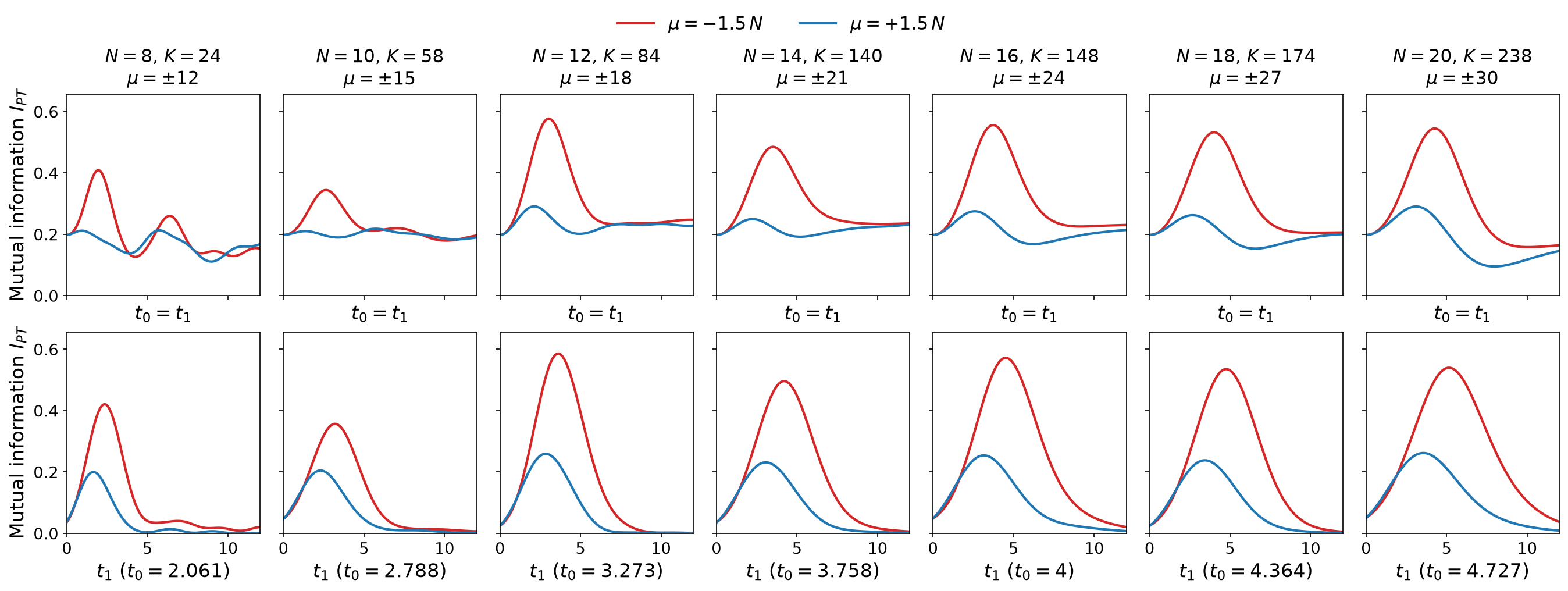}
		\caption{\label{S_figure15}
			Mutual information $I_{PT}$ at $\beta = 3$ for Gaussian sparse SYK Hamiltonians with $N = 8, 10, \ldots, 20$, using the corresponding transition point $K_{c}$ for $N \geq 8$.
			The upper figures show the symmetric protocol, whereas the lower figures show the fixed-injection-time protocol.
			For each $N \geq 8$, the injection time $t_{0}$ for the fixed-injection-time protocol is chosen as the time at which the mutual-information asymmetry is maximal in the corresponding symmetric protocol.
			Also, the interaction strength is chosen such that $\abs{\mu}/N = 1.5$, which is identical to our binary $N = 8$ setup.
		}
	\end{figure}    
	
	In Table~\ref{S_table1}, we show the transition points $K_{i} = \lceil p_{i}\binom{N}{4}\rceil$ with $i = 1, 2$ and the two-qubit gate counts required for transpiling the single time evolution $e^{-iH_{L}t_{0}}$ on \texttt{ibm\_phoenix} for $N = 10, 12, \ldots, 20$.
	We also compare them to those for the chosen binary sparse $N = 8$ SYK Hamiltonian used in the hardware experiment.
	As this table shows, the quantum-chaos transition point $K_{c}$ shows an overall moderate increase with $N$.
	Importantly, across all $N$ considered, the binary sparse SYK model retains the spectral quantum chaos down to substantially smaller $K_{c}$ than the Gaussian sparse SYK model.
	This stronger retention of quantum chaos under sparsification also leads to a persistent resource advantage, with substantially fewer two-qubit gates and lower circuit depth required for the binary sparse model than for the Gaussian sparse model.
	
	For the binary sparse SYK Hamiltonians with $N = 8, 10, \ldots, 20$ used here, we also illustrate $I_{PT}$ for the symmetric and fixed-injection-time protocols in Fig.~\ref{S_figure14}.
	As this figure shows, the mutual-information asymmetry persists across the system sizes, while the characteristic peak time and magnitude of the teleportation signal depend on $N$.
	The mutual-information dynamics for all $N$ considered here are also similar to those for the Gaussian sparse SYK models shown in Fig.~\ref{S_figure15}.
	Despite the relative resource advantage over the Gaussian sparse model, the transpiled two-qubit gate counts for the binary sparse model increase monotonically with $N$.
	For the sparse Hamiltonians with $N \geq 10$ used for this table, the required two-qubit gate counts are expected to yield extremely deep quantum circuits, so the experimental mutual information is also expected to be heavily suppressed.
	Together, these results indicate that the binary sparse SYK model maintains a robust mutual-information behavior for larger $N$ in the TW protocol while retaining the relevant wormhole-like dynamics and many-body physics, with $N = 8$ providing the most hardware-efficient model for the experiment.
	
	\section{\label{S_section6}Thermofield double state preparation}
	In this section, we describe the preparation of the thermofield double (TFD) state used for the hardware implementation in the main text. 
	In general, the exact TFD state at inverse temperature $\beta$ is defined by~\cite{maldacena2018eternaltraversablewormhole_2}
	\begin{equation}
		\label{S_eq19}
		\ket{\mathrm{TFD}} = \frac{2^{N/2}}{\sqrt{Z}} e^{-\beta H_{\mathrm{tot}}/4}\ket{I}, 
		\qquad
		H_{\mathrm{tot}}=H_{L} + H_{R},
	\end{equation}
	where $Z = \Tr e^{-\beta H_{L}} = \Tr e^{-\beta H_{R}}$.
	Here $\ket{I}$ denotes the normalized maximally entangled state between the left and right Hilbert spaces, satisfying $(\psi_{L}^{j}+i\psi_{R}^{j})\ket{I}=0$ $\forall j$.
	Also, this state is a ground state of $H_{\text{int}} = i\sum_{j = 1}^{N}\psi_{L}^{j}\psi_{R}^{j}$.
	For the state preparation for quantum simulation, we first classically construct the TFD state for the chosen binary sparse $N = 8$ Hamiltonian in the main text using Eq.~\eqref{S_eq19}.
	This is used as the target state for the following variational circuit optimization.
	As we described in the main text, we use $\beta = 3$.
	
	In practice, instead of directly implementing imaginary-time evolution on the quantum device, we approximate the TFD state using a variational quantum circuit~\cite{PhysRevLett.123.220502_2, PhysRevA.104.012427_2}. 
	The circuit is initialized from the maximally entangled state $\ket{I}$ in the computational basis.
	We then apply a hardware-efficient ansatz $U_{\rm VQA}$ consisting of single-qubit rotations, $R_x(\theta_x^{ij})R_z(\theta_z^{ij})$, acting on the $i$th qubit in the $j$th layer, followed by linear entangling layers composed of CNOT gates. 
	The ansatz consists of $d = 6$ single-qubit rotation layers and five linear CNOT layers, giving 96 variational parameters and 35 CNOT gates before transpilation, while 3 single-qubit gates ($X$, $H$, $S$) and one CNOT gate are additionally required for constructing $\ket{I}$.
	
	\begin{figure}[t]
		\centering
		\includegraphics[width=0.65\linewidth]{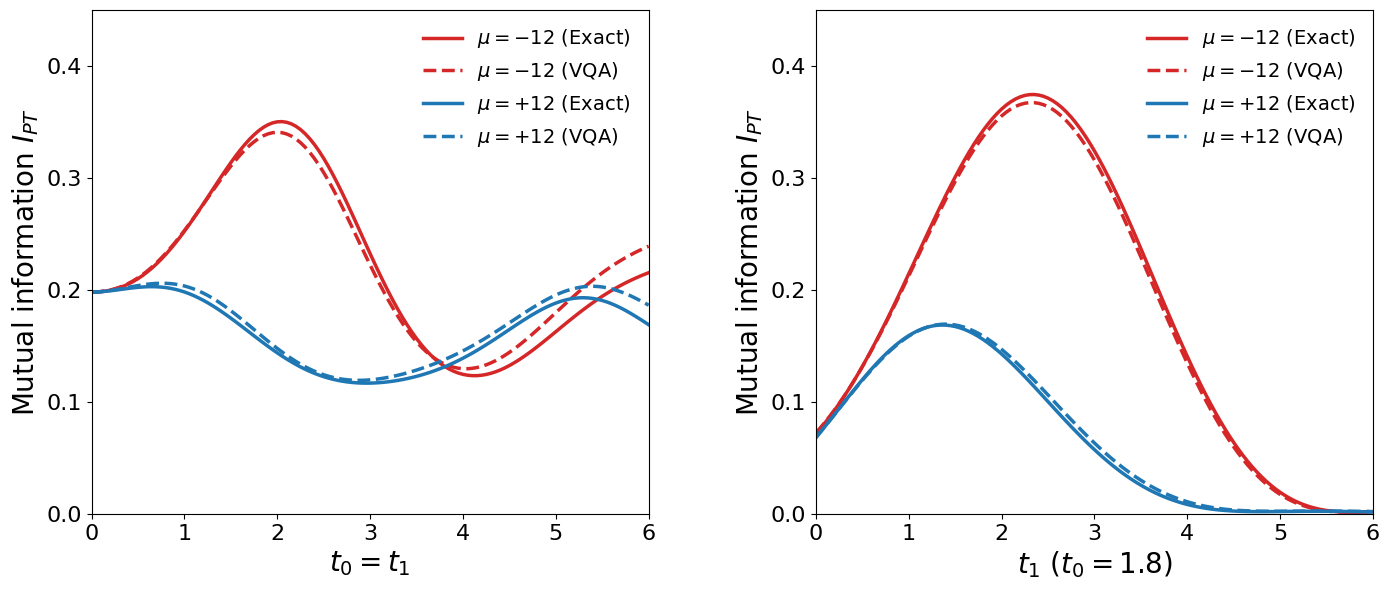}
		\caption{
			\label{S_figure16}
			Mutual information for the symmetric (left) and fixed-injection-time protocols at $t_{0} = 1.8$ (right), computed using the exact TFD statevector (solid) and the variationally prepared TFD state (dashed).
			The red and blue curves correspond to $\mu=-12$ and $\mu=+12$, respectively.
			This shows that the variationally prepared states with 96.8\% and 97.3\% fidelities for each protocol provide a close approximation to the exact mutual-information dynamics.
		}
	\end{figure}
	
	The variationally prepared state is written in terms of the variational parameters, such that
	\begin{equation}
		\ket{\psi_{\mathrm{VQA}}(\bm{\theta})} = U_{\mathrm{VQA}}(\bm{\theta}) \ket{I},
		\qquad
		\bm{\theta} = \left\{ \theta_x^{ij},\theta_z^{ij} \mid i=1,\ldots,N, \; j=1,\ldots,d\right\}.
	\end{equation}
	The parameters are optimized by maximizing the fidelity between the variationally prepared state and the exact TFD statevector from Eq.~\eqref{S_eq19}, which is given by
	\begin{equation}
		F(\bm{\theta}) = \abs{\braket{\psi_{\mathrm{VQA}}(\bm{\theta})}{\mathrm{TFD}}}^{2}.
	\end{equation}
	Equivalently, we minimize the infidelity cost function $\mathcal{L}(\bm{\theta})=1-F(\bm{\theta})$.
	For the circuit used in the hardware implementation in the main text, the resulting optimized state $\ket{\psi_{\mathrm{VQA}}}$ achieves a TFD-state fidelity $F\simeq0.968$ and $F \simeq 0.973$ for the symmetric and fixed-injection-time protocols, respectively.
	After compilation to the \texttt{ibm\_phoenix} topology, the TFD preparation circuit has depth 52 and contains 36 two-qubit gates for both protocols.
	In Fig.~\ref{S_figure16}, we compare the mutual information for $\beta=3$ and $\abs{\mu}=12$, computed from $\ket{\psi_{\text{VQA}}}$ and from the exact TFD statevector, both evolved using exact time evolution.
	As this figure shows, $\ket{\psi_{\text{VQA}}}$ closely approximates the mutual information for both signs of $\mu$ compared to the exact case.
	
	\section{\label{S_section7}Causal time ordering}
	In a semiclassical traversable geometry, signals injected consecutively are expected to emerge in the same temporal order~\cite{Gao2021_2}. 
	In the TW protocols, this causal time ordering is seen as a wormhole-like diagnostic, together with the sign-dependent teleportation signal~\cite{Jafferis2022_2, https://doi.org/10.7907/0b9qh-jt654_2}.

	For each injection time $t_0$, we scan the readout time $t_1$ and define the arrival time, denoted as $t_{1}^{\ast}(t_{0})$, as the location of the mutual-information peak, such that
	\begin{equation}
		\label{S_eq22}
		t_{1}^{\ast}(t_{0}) \equiv \arg\max_{t_1} I_{PT}(t_0,t_1).
	\end{equation}
	Causal time ordering is then seen from the behavior of $t_1^\ast(t_0)$ using two protocols. 
	The first is the instantaneous protocol, which is the main protocol used in the main text. 
	In this case, the interaction $e^{i\mu V}$ is applied as a single kick at $t = 0$.
	The second is a Trotterized protocol, in which the interaction $e^{i\mu'V}$ with the coupling constant $\mu'$ is applied as three kicks at times $-t_{\mu}/2$, $0$, and $t_{\mu}/2$. 
	The corresponding entire protocol is
	\begin{equation}
		\begin{aligned}
			\ket{\text{out}} = \text{SWAP}_{RT}U(t_{1} - t_{\mu}/2)e^{i\mu' V}U(t_{\mu}/2)e^{i\mu' V}U(t_{\mu}/2)e^{i\mu' V}U(t_0 - t_{\mu}/2)\text{SWAP}_{QL}U(-t_0)\ket{\text{in}},
		\end{aligned}
	\end{equation}
	where $\ket{\mathrm{in}}=\ket{\mathrm{Bell}}_{PQ}\otimes \ket{\mathrm{TFD}}_{LR}\otimes \ket{0}_{T}$ and $U(t) = e^{-iH_{\text{tot}}t}$.
	The coupling $\mu'$ and step $t_{\mu}/2$ are chosen such that the Trotterized protocol produces a mutual-information difference, defined by $\max_{t_{1}}\Delta I_{PT}$, comparable to that of the instantaneous protocol. 
	For the Trotterized protocol we restrict the region $t_{0}, t_{1} > t_{\mu}/2$ so that all time-evolution intervals are positive after the insertion at $t = -t_{0}$.
	For this protocol, the causal time-ordering window is identified as the region where $t_{1}^{\ast}(t_{0})$ decreases as $t_{0}$ increases.
	This corresponds to first-in--first-out behavior: a signal injected earlier, namely at a smaller value of $-t_{0}$, is associated with a smaller readout time $t_{1}$.
	In the same time window, the instantaneous protocol shows a comparatively weaker slope in $t_{1}^{\ast}(t_{0})$.
	
\begin{figure}[t]
		\includegraphics[width=0.7\linewidth]{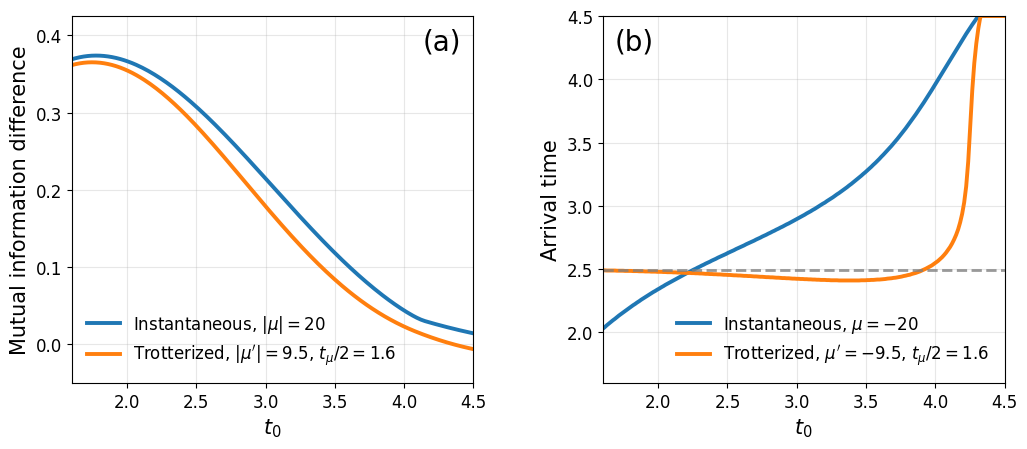}
		\caption{
			\label{S_figure17}
			Causal time-ordering diagnostic for the binary dense $N = 8$ SYK model at $\beta=3$. 
			(a) Mutual-information difference $\max_{t_{1}}\Delta I_{PT}$ for the instantaneous ($\abs{\mu} = 20$) and Trotterized ($\abs{\mu'} = 9.5$, $t_{\mu}/2 = 1.6$) protocols. 
			(b) Arrival time $t_1^\ast(t_0)$ extracted from the mutual-information peak for the negative coupling constants. 
			The Trotterized protocol exhibits a negative slope in the time-ordering window.
			In the same time window, the instantaneous protocol shows a weaker slope inside than outside the window.
		}
	\end{figure}

	To analyze the time-ordering behavior of the protocol, we first examine the binary dense $N = 8$ SYK model at $\beta = 3$.
	For the instantaneous protocol with $\abs{\mu} = 20$, we choose $\abs{\mu'} = 9.5$ and $t_{\mu}/2 = 1.6$ for the Trotterized protocol, which gives a comparable mutual-information difference, as shown in Fig.~\ref{S_figure17}(a). 
	\begin{figure}[t]
		\includegraphics[width=0.7\linewidth]{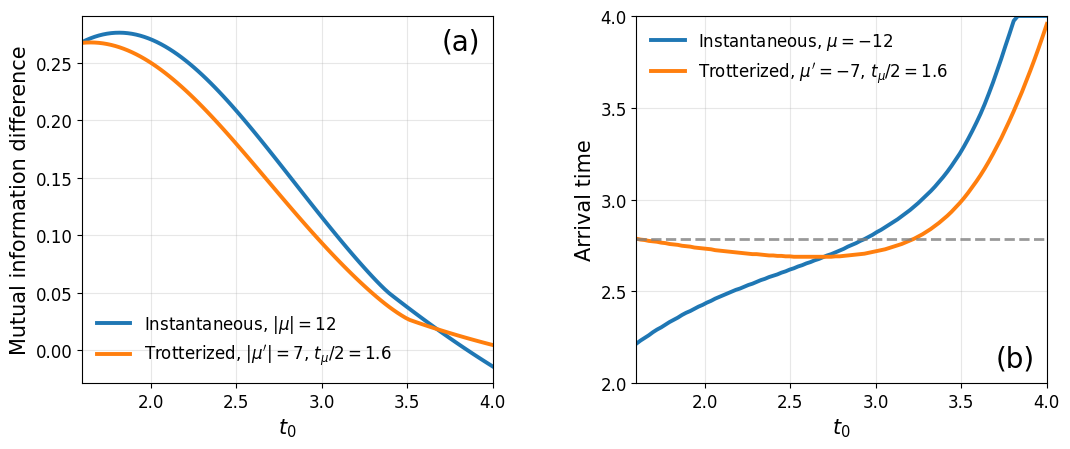}
		\caption{
			\label{S_figure18}
			Causal time-ordering diagnostic for the chosen binary sparse $N = 8$ SYK Hamiltonian used in the hardware implementation.
			(a) Mutual-information difference $\max_{t_{1}}\Delta I_{PT}$ for the instantaneous ($\abs{\mu} = 12$) and Trotterized ($\abs{\mu'} = 7$, $t_{\mu}/2 = 1.6$) protocols. 
			(b) Arrival time $t_1^\ast(t_0)$ extracted from the mutual-information peak for the negative coupling constants. 
			The Trotterized protocol exhibits a negative slope in the time-ordering window.
			In the same time window, the instantaneous protocol shows a weaker slope inside than outside the window.
		}
	\end{figure}
	
	Figure~\ref{S_figure17}(b) shows the arrival time $t_{1}^{\ast}(t_0)$ for each protocol with the negative coupling. 
	The Trotterized protocol exhibits a clear negative slope over a finite window. 
	These features indicate that causal time ordering appears in the binary dense SYK model.
	
	We next apply the same analysis to the chosen binary sparse $N = 8$ SYK Hamiltonian used in the hardware implementation at the same temperature. 
	For the instantaneous protocol with $\abs{\mu} = 12$, we choose $\abs{\mu'} = 7$ and $t_{\mu}/2 = 1.6$ for the Trotterized protocol. 
	This choice produces a mutual-information difference comparable to that of the instantaneous kick, as shown in Fig.~\ref{S_figure18}(a). 
	The corresponding arrival-time curve in Fig.~\ref{S_figure18}(b) shows a finite time-ordering window for the Trotterized protocol. 
	The working point $t_{0} = 1.8$ used in our hardware implementation lies inside this window, confirming that this point is in a regime where the TW protocol exhibits wormhole-like causal dynamics over a finite time interval.
	
	\section{\label{S_section8}Trotterization error analysis}
	We analyze the Trotterization error associated with the single-step first-order Lie--Trotter approximation used in the hardware implementation for the fixed-injection-time protocol. 
	The hardware working point, $t_{0} = 1.8$, is chosen because the teleportation signal remains large, while the Trotterization error remains tolerable, providing a practical upper bound on the injection time for the hardware implementation.

	To quantify this, we compare the exact and single-step first-order Trotterized mutual-information asymmetries at each injection time $t_0$ for the state $\ket{\psi_{\text{VQA}}}$ described in Sec.~\ref{S_section6}.
	We consider the exact and Trotterized mutual-information curves, denoted by $I_{PT}^{\mathrm{ex}}$ and $I_{PT}^{\mathrm{tr}}$, respectively. 
	We then compute the sign asymmetry at each $t_{1}$, denoted by $A_{\mathrm{ex}} \equiv \Delta I_{PT}^{\mathrm{ex}}$ and $A_{\mathrm{tr}} \equiv \Delta I_{PT}^{\mathrm{tr}}$.
	The corresponding peak asymmetries, denoted by $\max A_{\mathrm{ex}}$ and $\max A_{\mathrm{tr}}$, are obtained by maximizing over the readout time $t_1$.
	We quantify the single-step first-order Trotterization error in the mutual-information asymmetry by defining
	\begin{equation}
		\delta(t_{0}) \equiv \max A_{\text{ex}} - \max A_{\text{tr}}.
	\end{equation}
	Also we define the relative error
	\begin{equation}
		\eta(t_0) \equiv \dfrac{\delta(t_{0})}{\max A_{\text{ex}}},
	\end{equation}
	which measures the Trotterization error relative to the exact peak asymmetry.
	
\begin{figure}[t]
		\centering
		\includegraphics[width=0.7\linewidth]{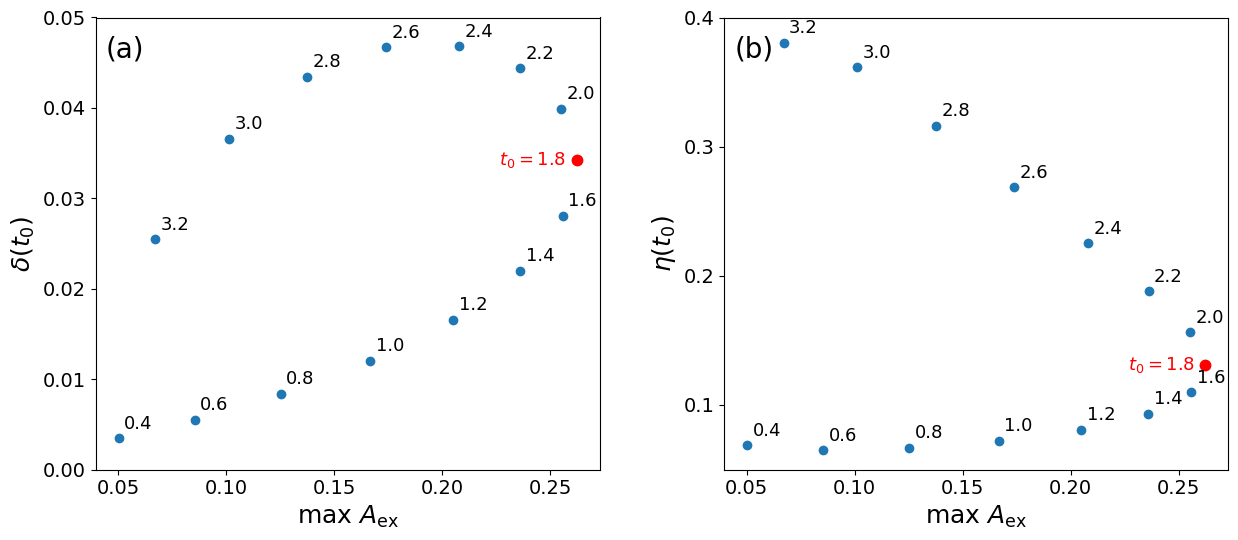}
		\caption{
			\label{S_figure19}
			First-order Trotterization error analysis for the mutual-information asymmetry as a function of the injection time $t_0$.
			(a) Trotterization error $\delta(t_0)$ and (b) the relative error $\eta(t_{0})$ plotted against the exact peak asymmetry $\max A_{\text{ex}}$.
			The working point $t_{0} = 1.8$ (red points) marks a practical boundary beyond which the signal strength decreases while the Trotter error keeps increasing.
		}
	\end{figure}

	For the chosen Hamiltonian, Fig.~\ref{S_figure19}(a) shows the relation between $\max A_{\mathrm{ex}}$ and $\delta(t_0)$ for different injection times.
	We find that $t_{0} = 1.8$ marks a turning point.
	After this turning point, the Trotterization error initially continues to increase, whereas the exact mutual-information asymmetry no longer improves and instead begins to decrease. 
	Figure~\ref{S_figure19}(b) shows the corresponding $\eta(t_{0})$, which supports the same conclusion. 
	Therefore, we choose $t_{0} = 1.8$ as the hardware working point because it is close to the largest accessible signal before the signal starts to decrease while the Trotter error keeps increasing.
	
\begin{figure}[t]
		\centering
		\includegraphics[width=0.40\linewidth]{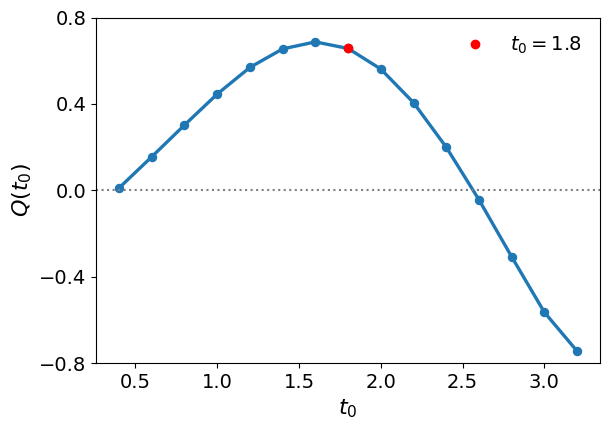}
		\caption{\label{S_figure20}
			Normalized signal-error difference $Q(t_{0})$ for the first-order Trotterized protocol.
			The red marker denotes the hardware working point $t_{0} = 1.8$.
		}
	\end{figure}

	To further identify a useful working regime, we compare the normalized exact asymmetry signal with the normalized relative Trotter error. 
	We define two dimensionless quantities: 
	$S(t_{0})$ as $\max A_{\text{ex}}$ normalized by its maximum value over $t_0$, and $E(t_0)$ as $\eta(t_{0})$ normalized by its maximum value over $t_{0}$.
	We then consider their difference
	\begin{equation}
		Q(t_0)  =  S(t_0) - E(t_0).
	\end{equation}
	This quantity provides a normalized measure of the balance between signal strength and Trotter error.
	A larger $Q(t_0)$ indicates that the normalized exact asymmetry signal is large compared with the normalized relative Trotter error.
	We illustrate $Q(t_{0})$ for the chosen Hamiltonian in Fig.~\ref{S_figure20}.
	As shown in Fig.~\ref{S_figure20}, together with the causal time-ordering analysis in Sec.~\ref{S_section7}, the region around $t_0\simeq1.6$--$1.8$ provides practical working points for hardware implementation, where the signal remains large while the relative Trotter error is still controlled.
	
	\section{\label{S_section9}Measurements and tomography}
	In this section we explain how the reduced density matrix $\rho_{PT}$ is reconstructed from the quantum-computer data used to obtain the mutual information $I_{PT}$ shown in the main text.
	
	For each value of $t_{1}$ and each sign of $\mu$, we run nine measurement circuits in the Pauli bases $XX, XY, \ldots, ZZ$ on $P$ and $T$.
	The $X$ and $Y$ measurements are implemented by applying single-qubit basis rotations before the final computational-basis measurement.
	From the measurement outcomes, we estimate $\expectationvalue{\sigma_{\mu}\otimes\sigma_{\nu}}$ with $\mu,\nu\in\{I,X,Y,Z\}$.
	The one-body expectation values $\expectationvalue{\sigma_{\alpha}\otimes I}$ and $\expectationvalue{I\otimes\sigma_{\beta}}$, with $\alpha,\beta\in\{X,Y,Z\}$, are obtained from the corresponding single-qubit marginals by averaging over the repeated measurements of $\sigma_{\alpha}$ on $P$ and $\sigma_{\beta}$ on $T$, respectively.
	Then the density matrix is reconstructed by two-qubit tomography using the Pauli expansion, such that~\cite{PhysRevA.64.052312_2, PhysRevA.93.062320_2}
	\begin{equation}
		\rho_{PT}
		=
		\dfrac{1}{4}
		\sum_{\mu,\nu\in\{I,X,Y,Z\}}
		\expectationvalue{\sigma_{\mu}\otimes\sigma_{\nu}}
		\sigma_{\mu}\otimes\sigma_{\nu},
	\end{equation}
	with $\expectationvalue{I\otimes I}=1$.
	The reduced density matrices $\rho_{P}$ and $\rho_{T}$ are obtained by taking partial traces of $\rho_{PT}$, such that $\rho_{P} = \Tr_{T}\rho_{PT}$ and $\rho_{T} = \Tr_{P}\rho_{PT}$.
	
\begin{figure}[t]
		\centering
		\includegraphics[width=0.8\linewidth]{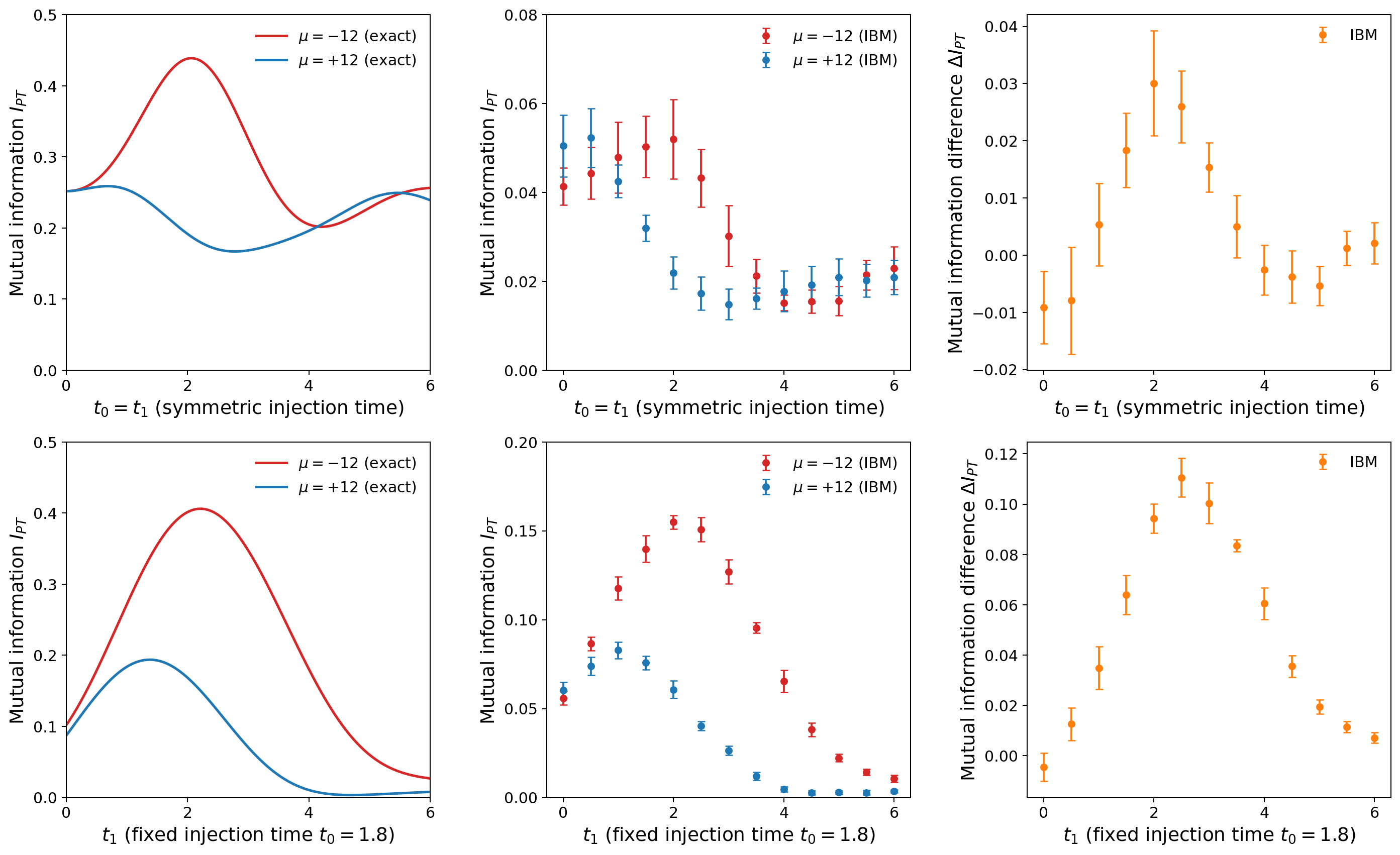}
		\caption{\label{S_figure21}
			Mutual information computed from the second R\'enyi entropy for the symmetric (top) and fixed-injection-time (bottom) protocols.
			Left: Mutual information $I_{PT}$ for $\mu=-12$ (red) and $\mu=+12$ (blue) from the variationally prepared TFD statevector (SV) and Trotterization used in the hardware experiment.
			Middle: Mutual information $I_{PT}$ for $\mu=-12$ (red) and $\mu=+12$ (blue) from the hardware experiment.
			Right: Mutual-information asymmetry $\Delta I_{PT}$ from the hardware experiment.
		}
	\end{figure}

	From the obtained reduced density matrices, we compute the mutual information using the von Neumann entropy, given by $S(\rho) = -\Tr(\rho\log_{2}\rho)$.
	In Fig.~\ref{S_figure21}, we also report the mutual information from the hardware experiment using the second R\'enyi entropy $S^{(2)}(\rho) = -\ln\Tr\rho^{2}$ adopted in Ref.~\cite{Jafferis2022_2}.
	For a direct comparison with the quantum-hardware experiment in Ref.~\cite{Jafferis2022_2}, we also compare the corresponding circuit resources and experimental signal.
	The experiment in Ref.~\cite{Jafferis2022_2} was performed using 164 two-qubit gates, whereas our experiment requires 282 and 150 two-qubit gates after compilation for the symmetric and fixed-injection-time protocols, respectively.
	Despite this comparable or larger circuit cost, the maximum of the mean second-R\'enyi mutual-information asymmetry $\Delta I_{PT}$ measured in Fig.~\ref{S_figure21} is approximately 4 and 34 times as large as that reported in Ref.~\cite{Jafferis2022_2} for the symmetric and fixed-injection-time protocols, respectively.
	
	In Fig.~6 of the main text, the measurements are performed at 13 values of $t_{1}\in[0, 6.0]$ with intervals of $0.5$ for each sign of the interaction.
	Since each value of $(\mu,t_1)$ requires nine Pauli-basis measurement circuits, the total number of distinct circuits per repetition is $9\times 13\times 2=234$.
	Each circuit is measured with 10,000 shots, and the full measurement is repeated 10 times.
	For all measurement points, the reconstructed two-qubit density matrix $\rho_{PT}$ is verified to be positive semidefinite and trace-normalized up to numerical precision.
	Each set of nine measurements with 10,000 shots takes about 3.6~s and 2.2~s of QPU time for the symmetric and fixed-injection-time protocols, respectively.
	The total QPU time is therefore approximately 15.5 and 9.5~min, respectively.

	\section{\label{S_section10}Device-calibrated noise analysis}
	To assess how device noise degrades the teleportation signal, we perform a calibration-based noise-scaling analysis of the mutual information.
	The analysis reuses the quantum circuits executed on the backend \texttt{ibm\_phoenix} for both protocols, together with the device calibration data recorded during the experiment, without fitting any parameters to the measured mutual information.
	The main calibrated error parameters of the ten physical qubits used in the compiled circuits are summarized in Table~\ref{S_table2} for both protocols.
	We introduce a dimensionless noise scale $s \in \left(0,1\right]$ that scales the calibrated gate, readout, and initialization errors together with the relaxation and dephasing rates, so that $s = 1$ reproduces the calibrated device and the $s \to 0$ limit recovers the noiseless quantum circuit.

\begin{table}[t]
	\centering
	\renewcommand{\arraystretch}{1.0}
	\begin{tabular*}{\linewidth}{@{\extracolsep{\fill}} c c c c c c c | c c c}
		\hline\hline
		Qubit & Role & $T_{1}$ ($\mu$s) & $T_{2}$ ($\mu$s)
		& $\sqrt{X}$ ($10^{-4}$) & Readout ($10^{-2}$) & Init. ($10^{-4}$)
		& Edge & CZ ($10^{-3}$) & $R_{ZZ}$ ($10^{-3}$) \\
		\hline
		\multicolumn{10}{@{}l}{Symmetric protocol} \\
		\hline
		$q_{57}$ & $P$ & 252 & 180 & 1.01 & 0.49 (0.07) & 0.36 & $(q_{39},q_{49})$ & 1.39 & 2.28 \\
		$q_{67}$ & $Q$ & 155 & 171 & 1.94 & 0.30 (0.04) & 0.10 & $(q_{47},q_{48})$ & 2.44 & --- \\
		$q_{68}$ & $T$ & 165 & 191 & 1.54 & 0.46 (0.12) & 0.31 & $(q_{47},q_{57})$ & 1.88 & --- \\
		$q_{47}$ & sys   & 213 & 268 & 1.16 & 1.19 (0.04) & --- & $(q_{48},q_{49})$ & 1.31 & --- \\
		$q_{58}$ & sys   & 189 & 239 & 1.68 & 0.36 (0.05) & 0.22 & $(q_{48},q_{58})$ & 1.08 & 1.45 \\
		$q_{69}$ & sys   & 172 & 282 & 1.84 & 0.37 (0.12) & 0.21 & $(q_{49},q_{59})$ & 1.42 & 2.45 \\
		$q_{59}$ & sys   & 285 & 155 & 1.24 & 0.64 (0.06) & 6.20 & $(q_{57},q_{58})$ & 0.98 & 1.15 \\
		$q_{49}$ & sys   & 173 & 227 & 1.57 & 0.54 (0.04) & 0.20 & $(q_{57},q_{67})$ & 1.62 & --- \\
		$q_{48}$ & sys   & 228 & 230 & 2.09 & 0.90 (0.03) & 0.14 & $(q_{58},q_{59})$ & 1.32 & 1.77 \\
		$q_{39}$ & sys   & 222 & 168 & 2.73 & 0.59 (0.04) & 0.33 & $(q_{58},q_{68})$ & 1.08 & --- \\
		& & & & & & & $(q_{59},q_{69})$ & 1.38 & --- \\
		& & & & & & & $(q_{67},q_{68})$ & 1.90 & 1.46 \\
		& & & & & & & $(q_{68},q_{69})$ & 1.04 & 2.22 \\
		\hline
		\multicolumn{10}{@{}l}{Fixed-injection-time protocol} \\
		\hline
		$q_{68}$ & $P$ & 169 & 191 & 2.28 & 0.48 & 0.13 & $(q_{38},q_{48})$ & 0.86 & --- \\
		$q_{69}$ & $Q$ & 167 & 249 & 2.37 & 0.28 & 0.21 & $(q_{46},q_{47})$ & 0.89 & --- \\
		$q_{58}$ & $T$ & 216 & 232 & 1.61 & 0.78 & 0.68 & $(q_{46},q_{56})$ & 1.09 & --- \\
		$q_{59}$ & sys   & 285 & 158 & 1.42 & 0.79 & 6.20 & $(q_{47},q_{48})$ & 1.23 & --- \\
		$q_{57}$ & sys   & 278 & 186 & 1.24 & 0.31 & 0.12 & $(q_{47},q_{57})$ & 1.24 & --- \\
		$q_{56}$ & sys   & 180 & 193 & 1.78 & 0.43 & 1.67 & $(q_{48},q_{58})$ & 1.05 & 1.29 \\
		$q_{46}$ & sys   & 215 & 151 & 1.60 & 0.52 & 0.85 & $(q_{56},q_{57})$ & 0.78 & --- \\
		$q_{47}$ & sys   & 237 & 268 & 1.37 & 1.12 & --- & $(q_{57},q_{58})$ & 0.75 & 1.41 \\
		$q_{48}$ & sys   & 261 & 245 & 2.62 & 0.79 & 0.14 & $(q_{58},q_{59})$ & 0.90 & 1.66 \\
		$q_{38}$ & sys   & 196 & 195 & 1.42 & 0.37 & 0.78 & $(q_{58},q_{68})$ & --- & 1.23 \\
		& & & & & & & $(q_{59},q_{69})$ & 1.18 & --- \\
		& & & & & & & $(q_{68},q_{69})$ & 1.13 & --- \\
		\hline\hline
	\end{tabular*}
	\caption{\label{S_table2}
		Main calibration parameters of \texttt{ibm\_phoenix} for the ten physical qubits used in the compiled circuits of the symmetric and fixed-injection-time protocols.
		The first seven columns show the single-qubit energy-relaxation time $T_{1}$, dephasing time $T_{2}$, $\sqrt{X}$ gate error, readout assignment error, and initialization error.
		The roles $P$, $Q$, and $T$ denote the reference, injection, and target qubits, respectively, and ``sys'' labels the remaining SYK register qubits.
		The last three columns show two-qubit CZ and $R_{ZZ}$ gate errors on the physical couplers used by each protocol; --- marks a gate that is not used on that coupler.
		In the initialization-error column, --- denotes a value that is not reported in the calibration data.
		These are the isolated calibrated values; in the noise model, they are multiplied by the layer-fidelity factor $r$.
		For the symmetric protocol, values are the mean (standard deviation) over the calibration records of the ten hardware runs; values without parentheses are identical in all records.
		For the fixed-injection-time protocol, all ten hardware runs used the same calibration record.
	}
\end{table}
	
	\subsection{\label{S_section10.1}Noise model}
	We build a noise model using the calibration data in Table~\ref{S_table2} for both protocols.
	The purpose of the noise model is to (i) compare the expected results from the noise model with the experimental results in the main text and to (ii) verify whether the results approach the noiseless mutual information as we reduce the noise.
	
	To simulate the hardware-noise effects, we first prepare the same circuits used in the hardware experiment, which include the variationally prepared TFD state as described in Sec.~\ref{S_section6} and the single-step second-order Suzuki--Trotterization and first-order Lie--Trotterization for the symmetric and fixed-injection-time protocols, respectively.
	At the noise scale $s$, we then build a noise model directly from the device calibration data.
	We assume that, for a given density matrix $\rho$, each gate $g$ acting on a $d_{g}$-dimensional space ($d_{g}=2, 4$ for one- and two-qubit gates) carries a depolarizing noise channel of parameter $\Lambda_{g}$,
	\begin{equation}
		\label{S_eq28}
		\mathcal{E}_{\Lambda_{g}}(\rho) = (1-\Lambda_{g})\,\rho
		+ \Lambda_{g}\,\frac{I}{d_{g}},
	\end{equation}
	whose average gate fidelity is $F = 1 - \Lambda_{g}\,(d_{g}-1)/d_{g}$, i.e. an average gate error $\Lambda_{g}\,(d_{g}-1)/d_{g}$. 
	Taking $\varepsilon_{g}$ to be the calibrated average gate error and requiring this error to equal its scaled value, $\Lambda_{g}(d_{g}-1)/d_{g} = s\,\varepsilon_{g}$, fixes the channel strength 
	\begin{equation}
		\label{S_eq29}
		\Lambda_{g}(s) = s\,\varepsilon_{g}\,\frac{d_{g}}{d_{g}-1}.
	\end{equation}
	For two-qubit gates, the calibrated errors $\varepsilon_{g}$ are obtained by benchmarking each coupler in isolation and therefore do not capture errors arising under simultaneous operation, such as crosstalk.
	We thus multiply all two-qubit gate errors by a common factor
	\begin{equation}
		r=\frac{\mathrm{EPLG}}{\bar{\varepsilon}_{\rm CZ}},
		\qquad
		\mathrm{EPLG}=1-\mathrm{LF}^{1/(n-1)},
	\end{equation}
	where $\mathrm{EPLG}$ is the error per layered gate obtained from the layer fidelity $\mathrm{LF}$ of an $n=10$-qubit chain reported in the calibration data~\cite{McKay:2023nxa}, and $\bar{\varepsilon}_{\rm CZ}$ is the mean calibrated CZ error over the $n-1$ couplers of this chain.
	The factor $r$ thus quantifies the increase of the two-qubit error under simultaneous operation measured on the same couplers, and multiplying each calibrated two-qubit error by $r$ keeps the relative variation among the couplers used in the circuit.
	We obtain $r\simeq1.67$ and $2.06$ for the symmetric and fixed-injection-time protocols, respectively.
	Accordingly, $\varepsilon_{g}$ in Eq.~\eqref{S_eq29} is replaced by $r\varepsilon_{g}$ for two-qubit gates.
	
	Idle periods are obtained from an as-late-as-possible (ALAP) schedule built from the calibrated gate durations, and during each idle interval on qubit $q$ we apply a thermal-relaxation channel parameterized by the scaled relaxation and dephasing times
	\begin{equation}
		\label{S_eq31}
		T_{1}^{q}(s) = \frac{T_{1}^{q}}{s},
		\qquad
		T_{2}^{q}(s) = \frac{\min(T_{2}^{q},\,2T_{1}^{q})}{s},
	\end{equation}
	where $T_{1}^{q}$ and $T_{2}^{q}$ are the energy-relaxation and dephasing times of qubit $q$, respectively.
	The bound $T_{2}^{q}\leq 2T_{1}^{q}$ ensures that the thermal-relaxation channel is physical.
	From Eq.~\eqref{S_eq31}, larger $s$ produces faster relaxation and dephasing over the fixed schedule.
	
	\begin{figure}[t]
		\centering
		\includegraphics[width=0.65\linewidth]{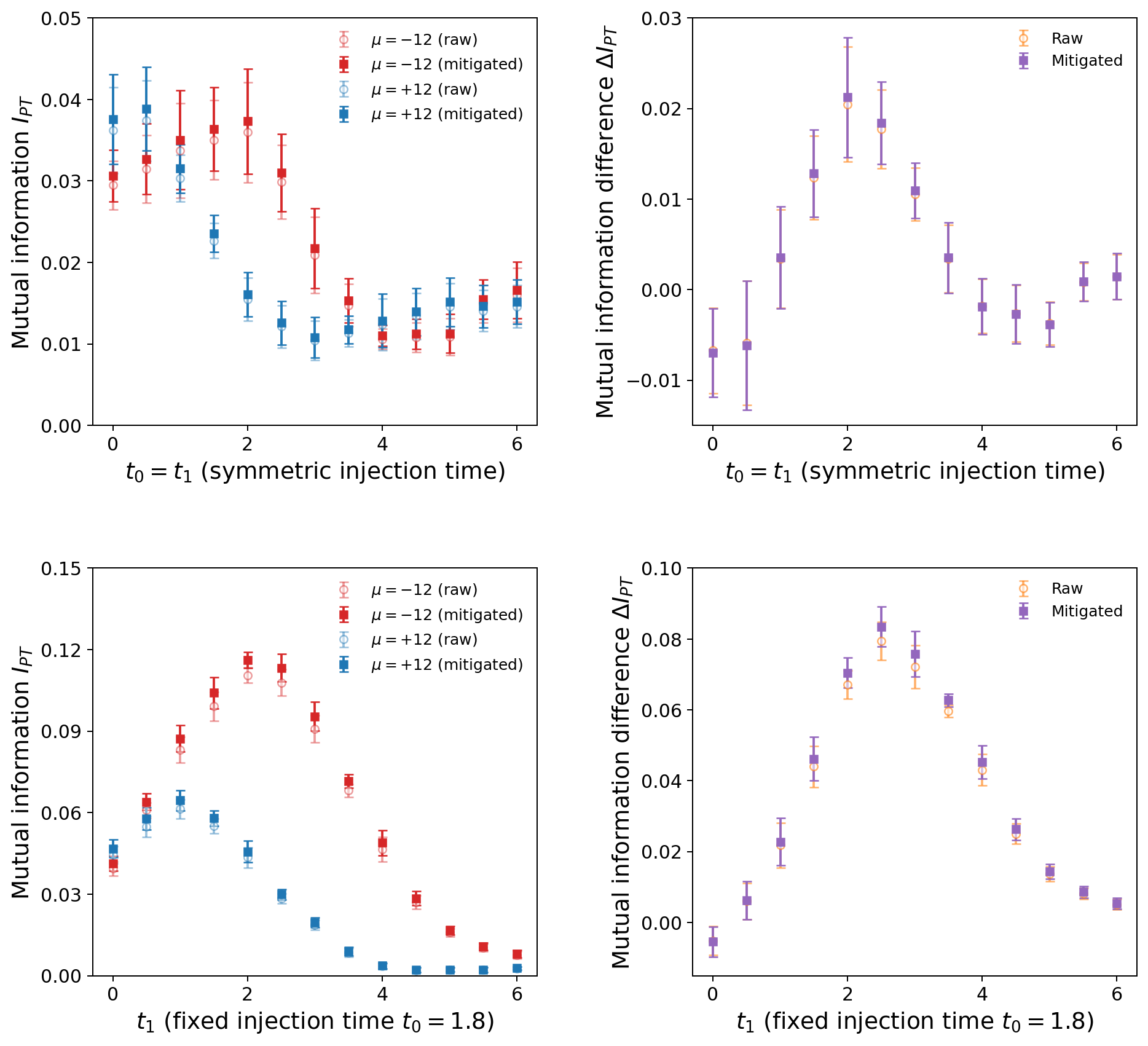}
		\caption{\label{S_figure22}
			Effect of readout-error mitigation on the measured mutual information for the symmetric (top) and fixed-injection-time ($t_{0} = 1.8$, bottom) protocols for $\beta=3$ and $\abs{\mu}=12$.
			Left: mutual information $I_{PT}$ for $\mu=-12$ (red) and $\mu=+12$ (blue), before (open markers) and after (filled markers) readout-error mitigation.
			Right: corresponding mutual-information asymmetry $\Delta I_{PT}$ for the raw (orange) and mitigated (purple) data.
			The points and error bars show the mean and standard deviation over $10$ independent runs, with $10{,}000$ shots per tomography circuit.
		}
	\end{figure}

	Initialization errors are modeled by preparing each qubit in $\ket{1}$ instead of $\ket{0}$ with probability $s\,\varepsilon^{q}_{\rm init}$, where $\varepsilon^{q}_{\rm init}$ is the calibrated initialization error.\footnote{The initialization error of $q_{47}$ is not reported in the calibration data and is replaced by the mean over the other qubits in the noise model.}
	Terminal readout on each measured qubit is modeled as a classical bit-flip of the measured outcome, described by the single-qubit readout assignment matrix
	\begin{equation}
		\label{S_eq32}
		M_{q}(s) =
		\begin{pmatrix}
			1 - s\,p^{q}_{1|0} & s\,p^{q}_{0|1}\\[2pt]
			s\,p^{q}_{1|0} & 1 - s\,p^{q}_{0|1}
		\end{pmatrix},
	\end{equation}
	where $p^{q}_{b|a}$ is the calibrated probability of recording outcome $b$ when the state $\ket{a}$ was prepared, so that $p^{q}_{1|0}$ and $p^{q}_{0|1}$ are the $\ket{0}\!\to\!\ket{1}$ and $\ket{1}\!\to\!\ket{0}$ bit-flip probabilities, respectively.
	The probabilities $p^{q}_{1|0}$ and $p^{q}_{0|1}$ are taken directly from the calibration data, which are 
	\begin{equation}
		\begin{aligned}
			(p^{P}_{1|0}, p^{P}_{0|1}) &= (0.085 \pm 0.059, 0.901 \pm 0.083)\times 10^{-2},\\
			(p^{T}_{1|0}, p^{T}_{0|1}) &= (0.122 \pm 0.118, 0.789 \pm 0.130)\times 10^{-2}
		\end{aligned}
	\end{equation}
	for the symmetric protocol, and 
	\begin{equation}
		\begin{aligned}
			(p^{P}_{1|0}, p^{P}_{0|1}) &= (0.073, 0.879)\times 10^{-2},\\
			(p^{T}_{1|0}, p^{T}_{0|1}) &= (1.074, 0.488)\times 10^{-2}
		\end{aligned}
	\end{equation}
	for the fixed-injection-time protocol.
	The columns of $M_{q}(s)$ are indexed by the prepared state and its rows by the measured outcome, so it maps the ideal outcome probabilities to the measured ones.
	Meanwhile, because the measurement twirling is applied in the hardware experiment, we symmetrize the readout assignment matrix as follows:
	\begin{equation}
		\label{S_eq35}
		\tilde{M}_{q} = \begin{pmatrix}
			1 - s r_{q} & s r_{q}\\
			s r_{q} & 1 - sr_{q}
		\end{pmatrix},
		\qquad
		r_{q} = \dfrac{p_{1|0}^{q} + p_{0|1}^{q}}{2}.
	\end{equation}
	This symmetrized assignment matrix is used both in the noise model and in subsequent readout mitigation.
	
	For each tomography basis, we mitigate readout errors using the tensor-product assignment matrix $\tilde{M}_{PT}=\tilde{M}_{P}\otimes\tilde{M}_{T}$ constructed from the calibrated single-qubit readout probabilities. 
	The measured probability vector is corrected according to
	\begin{equation}
		\mathbf{p}_{\rm corr}=\tilde{M}_{PT}^{-1}\,\mathbf{p}_{\rm meas},
	\end{equation}
	where $\mathbf{p}_{\rm meas}=(p_{00},p_{01},p_{10},p_{11})^{T}$ is the measured probability distribution on the $PT$ registers and $\mathbf{p}_{\rm corr}$ is the corresponding readout-corrected probability vector~\cite{Maciejewski2020mitigationofreadout}.
	The corrected probabilities are then used to reconstruct $\rho_{PT}$ and compute $I_{PT}$.
	The readout-mitigated results are shown in Fig.~\ref{S_figure22}.
	As this figure shows, readout mitigation slightly enhances both $I_{PT}$ and $\Delta I_{PT}$.
	\begin{figure}[t]
		\centering
		\includegraphics[width=\linewidth]{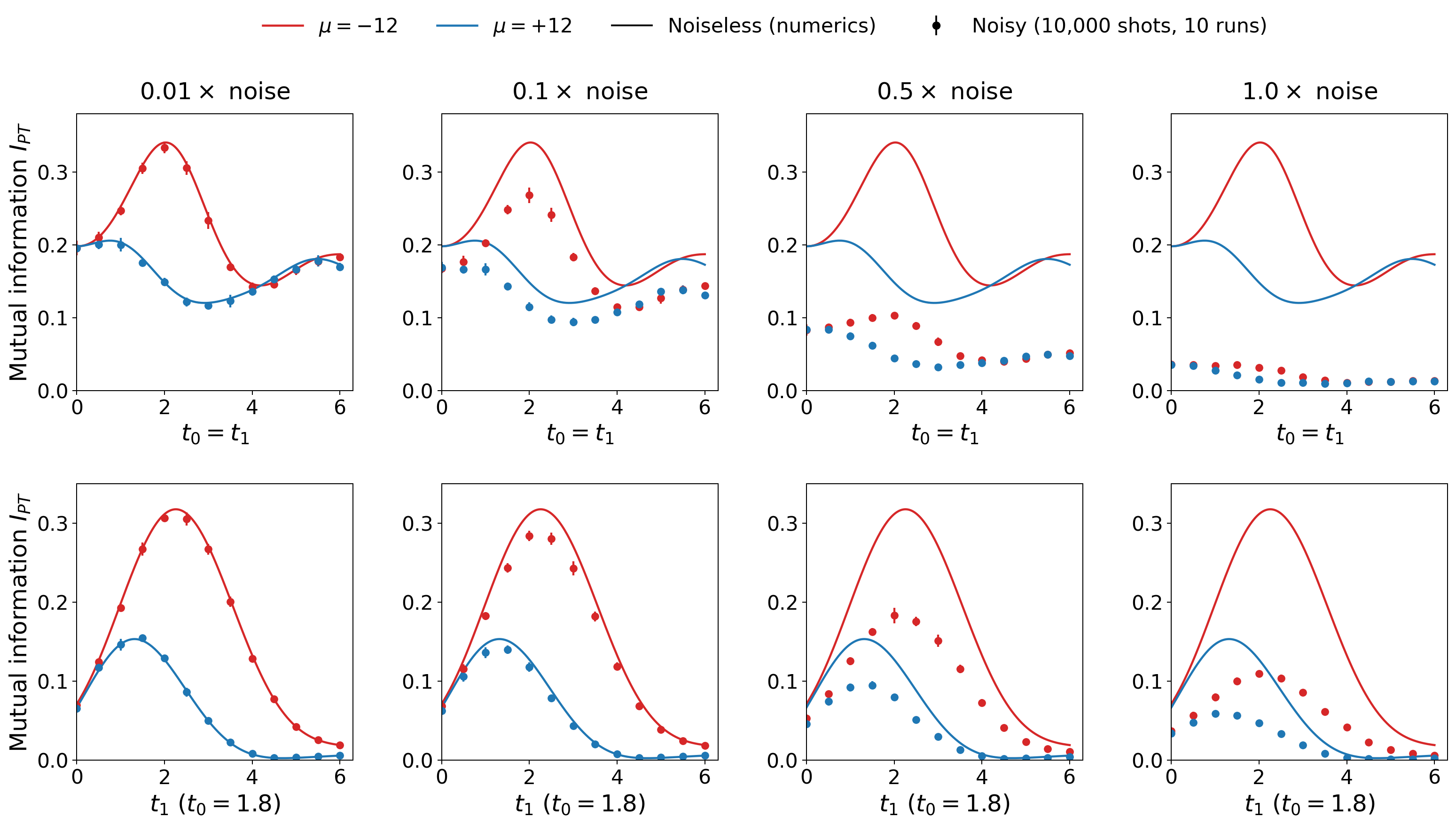}
		\caption{\label{S_figure23}
			Noise-scaling of the mutual information $I_{PT}$ for the symmetric (top) and fixed-injection-time ($t_{0}=1.8$, bottom) protocols at $\beta=3$ and $\abs{\mu}=12$.
			Each panel corresponds to a noise scale $s$ that multiplies all
			calibrated gate, idle, readout, and initialization error rates.
			Solid lines indicate numerical noiseless results from the transpiled circuits with the TFD state from VQA and single-step second-order Suzuki--Trotterization (symmetric) and first-order Lie--Trotterization (fixed-injection-time).
			Points with error bars indicate mean $I_{PT}$ with standard deviations, using $10{,}000$ shots for each tomography circuit over $10$ independent runs.
			Red and blue colors represent $\mu = -12$ and $\mu = +12$, respectively.
		}
	\end{figure}
	
	For each scale $s$, each sign of $\mu$, and each $t_{1}$, we construct the same nine Pauli-basis tomography circuits used in the hardware experiment and evolve the full scheduled circuits under the gate, idle, and initialization noise in Eqs.~\eqref{S_eq29} and \eqref{S_eq31}.
	For each tomography circuit, we draw $10{,}000$ shots including the readout error of Eq.~\eqref{S_eq35}, and reconstruct $\rho_{PT}$ and $I_{PT}$ exactly as for the hardware data in Sec.~\ref{S_section9}.
	The full tomography procedure is repeated $10$ times with independent shot samples, each using the calibration record of the corresponding hardware run in the main text.
	The simulated $I_{PT}$ is reported as the mean and standard deviation (SD) over these $10$ repetitions.
	
	The noise model relies on the following explicit assumptions.
	(i) Gate errors are modeled as independent, calibrated depolarizing channels as in Eq.~\eqref{S_eq29}, with the two-qubit errors rescaled by the layer fidelity; crosstalk is thus included only effectively, and coherent and leakage errors are not included.
	(ii) Circuit timing follows an ALAP schedule inferred from the wire dependencies, with the $R_{ZZ}$ duration set equal to the same-edge CZ duration, and all terminal readouts assumed to start synchronously; no dynamical decoupling (DD) is applied in the model because DD suppresses coherent idle errors, which are not included in the present model.
	(iii) Thermal relaxation acts only during the inferred idle intervals as in Eq.~\eqref{S_eq31}; no additional relaxation is added during active gates beyond the depolarizing noise.
	(iv) The noise scale $s$ multiplies only the gate infidelities, the idle decay rates, and the readout and initialization error probabilities, while all gate and readout durations are held fixed.
	(v) Each repetition uses the calibration record of the corresponding hardware run, so the simulated uncertainty reflects shot noise and the calibration changes between runs, but not drift within a run.
	
	Figure~\ref{S_figure23} shows $I_{PT}$ for $s\in\{0.01,0.1,0.5,1.0\}$.
	For comparison, we numerically reproduce the noiseless reference curves of $I_{PT}$ using the same transpiled circuits used in the hardware experiment, including the circuits for $\ket{\psi_{\mathrm{VQA}}}$ and single-step Lie--Trotterization and Suzuki--Trotterization.
	As $s$ decreases, the simulated points approach the noiseless reference curves, whereas at $s = 1$ the calibrated model reproduces the suppression observed on hardware.
	Importantly, the noiseless reference is obtained from the same variationally prepared TFD state and the same single-step first-order Lie--Trotterization and second-order Suzuki--Trotterization used in the hardware implementation.
	Therefore, the difference between the noiseless reference and the calibrated $s = 1$ simulation isolates the effect of the modeled hardware noise from the TFD-preparation and Trotterization errors already contained in the noiseless circuit.
	
	Figure~6 in the main text directly compares the $s = 1$ calibrated-noise simulation with the experimental results.
	The calibrated model closely reproduces the suppression of $I_{PT}$ and $\Delta I_{PT}$ in both protocols.
	The remaining discrepancies may arise from noise mechanisms not included in the present model, such as coherent errors, residual crosstalk, leakage, and calibration drift.
	These results show that the calibrated depolarizing gate errors (including the effective layered two-qubit errors), thermal relaxation and dephasing during idle periods, and readout errors can account for the dominant suppression of the teleportation signal observed on the device.
	
	\subsection{\label{S_section10.2}Statistical significance}

    To quantify the statistical significance of the experimentally observed mutual-information asymmetry, we analyze the results obtained from the $N_{\rm run}=10$ independent hardware runs using a statistical test.
    Since the data for the two signs of $\mu$ are obtained within the same run, the two measurements are paired within each run to evaluate $\Delta I_{PT}(t_1)$ for each run.
	\begin{figure}[t]
		\centering
		\includegraphics[width=\linewidth]{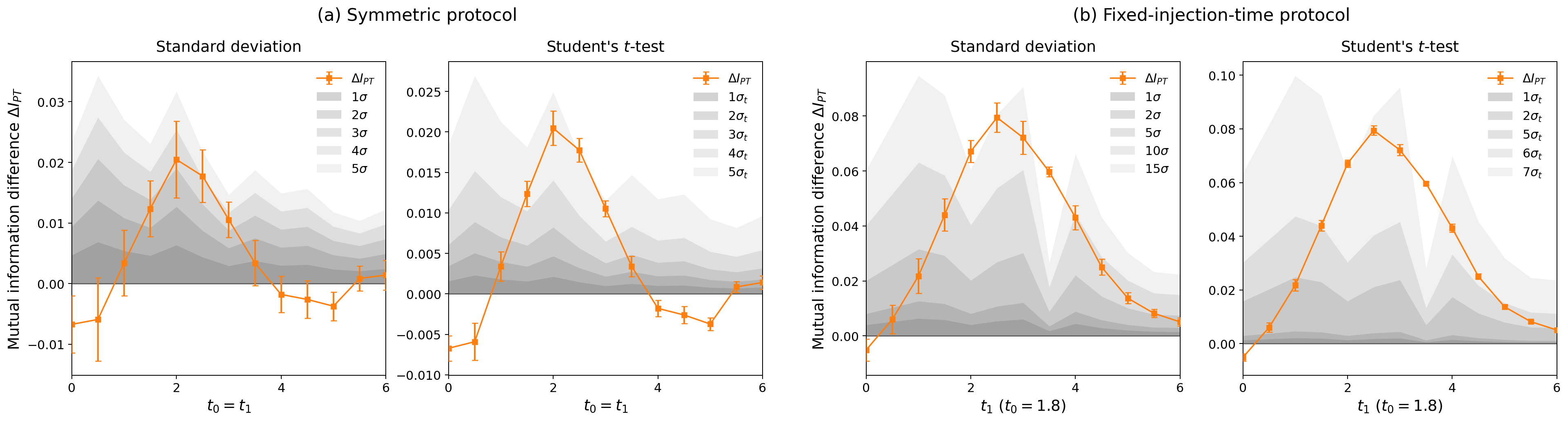}
		\caption{\label{S_figure24}
			Statistical analysis of the experimentally observed mutual-information asymmetry $\Delta I_{PT}$ over $10$ independent runs for the symmetric (a) and fixed-injection-time (b) protocols.
			Each left panel shows the mean $\Delta I_{PT}$ with the standard deviation (SD), together with reference bands corresponding to multiples of the SD.
			Each right panel shows the same experimental data analyzed using a Student's $t$-test, with shaded regions denoting Student's-$t$ thresholds equivalent to the Gaussian two-sided significance levels indicated in the legends.
		}
	\end{figure}
	
	Figure~\ref{S_figure24} shows the mean asymmetry $\overline{\Delta I_{PT}}(t_1)$ together with the SD $\sigma(t_{1})$ and reference positive bands $\left[0, k\sigma(t_1)\right]$ with $k = 1, \ldots, 5$ for the symmetric protocol and the values indicated in the legend for the fixed-injection-time protocol.
	Here, $k$ denotes the number of SDs from zero.
	As the figure shows, the deviations at the peak are above $3\sigma$ and $14\sigma$ for the symmetric and fixed-injection-time protocols, respectively, where the SD reflects the variation among the independent runs.
	
\begin{figure}[t]
		\centering
		\includegraphics[width=0.6\linewidth]{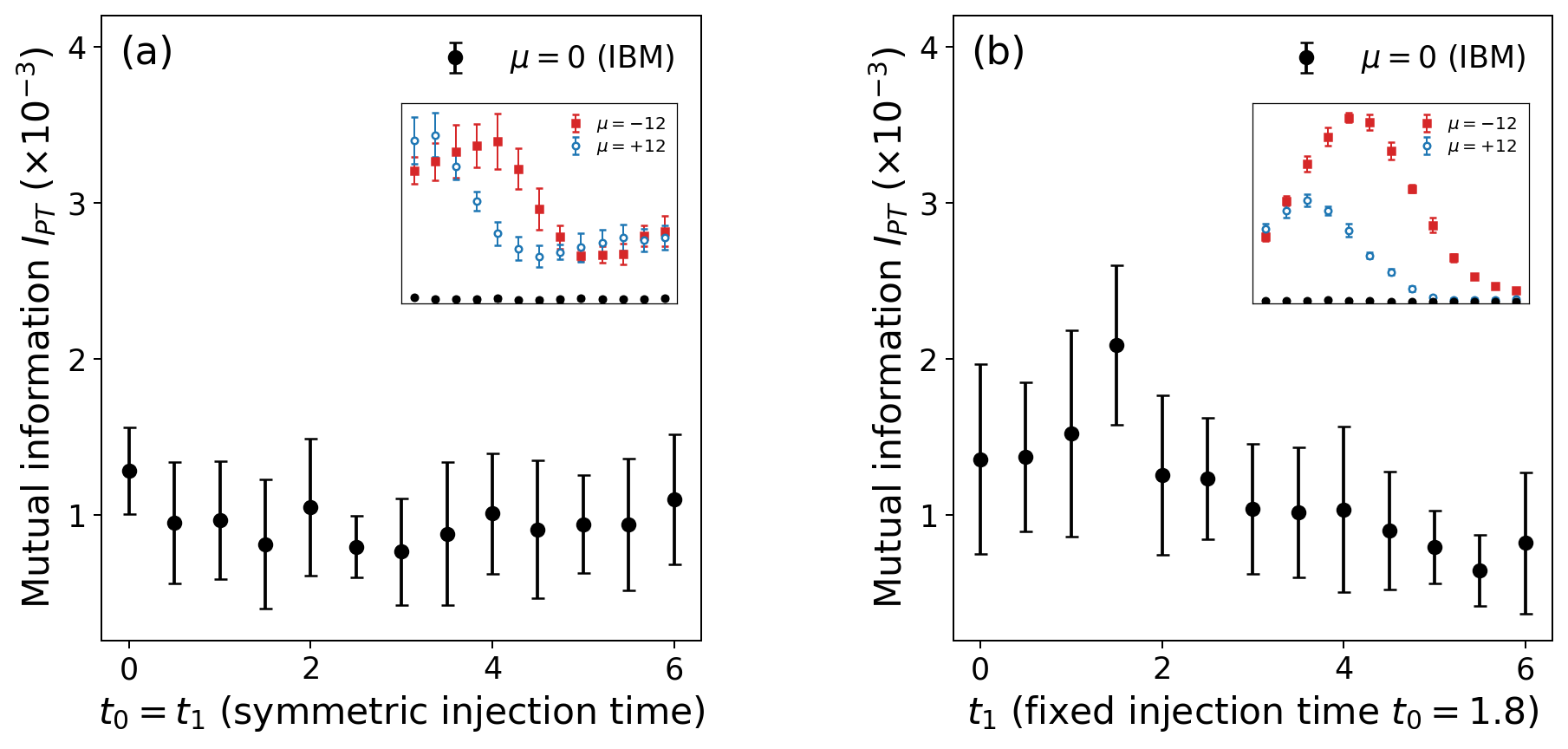}
		\caption{\label{S_figure25}
			Mutual information $I_{PT}$ measured on hardware for $\mu=0$ (black) for (a) the symmetric and (b) the fixed-injection-time ($t_0=1.8$) protocols, averaged over $10$ independent runs with the SD as error bars.
			Insets: the same $\mu=0$ data together with the $\mu=-12$ (red) and $\mu=+12$ (blue) data on a common scale.
		}
	\end{figure}

	To directly quantify the statistical significance and separation of the mean asymmetry from zero, we perform a Student's $t$-test, which is also used in Ref.~\cite{Jafferis2022_2}.
	As a reference, the hardware data for $\mu=0$, for which $I_{PT}$ vanishes in the absence of noise, remain at the level of $10^{-3}$, far below the values for $\mu=\pm12$; see Fig.~\ref{S_figure25}.
    The test assumes that the independent run-to-run values of $\Delta I_{PT}(t_1)$ are approximately normally distributed.
	At each sampled time, the test statistic is
	\begin{equation}
		t(t_1)=\frac{\overline{\Delta I_{PT}}(t_1)}{\sigma(t_1)/\sqrt{N_{\rm run}}},
	\end{equation}
	where $\sigma(t_1)/\sqrt{N_{\rm run}}$ is the standard error of the mean.
	For $N_{\rm run}=10$, the number of degrees of freedom is $\nu=N_{\rm run}-1=9$.
	To express the Student's-$t$ thresholds in terms of the Gaussian $k\sigma$ significance levels, we define the two-sided Gaussian significance level as
	\begin{equation}
		\alpha_k=2[1-\Phi(k)],
	\end{equation}
	where $\Phi(x)$ is the cumulative distribution function (CDF) of the standard normal distribution.
	The corresponding Student's-$t$ critical value is
	\begin{equation}
		c_k=F_{t_\nu}^{-1}(1-\alpha_k/2),
	\end{equation}
	where $F_{t_\nu}(x)$ is the CDF of the Student's $t$ distribution with $\nu$ degrees of freedom and $F_{t_\nu}^{-1}$ denotes its inverse.
	For $\nu=9$, the resulting critical values are $c_1=1.059$, $c_2=2.320$, $c_3=4.094$, $c_4=6.999$, $c_5=12.422$, $c_6=23.766$, and $c_7=50.020$.
	
	In Fig.~\ref{S_figure24}, we also illustrate the $k\sigma$-equivalent Student's-$t$ reference bands $\left[0, \, c_{k}\sigma(t_{1})/\sqrt{N_{\rm run}}\right]$, labeled $k\sigma_{t}$.
	Thus, when $\overline{\Delta I_{PT}}(t_1)$ lies outside this band, its separation from zero exceeds the Student's-$t$ threshold corresponding to the same two-sided significance level as the Gaussian $k\sigma$ level.
	As shown in Fig.~\ref{S_figure24}, the mean asymmetry around the teleportation peak is well separated from zero under the Student's $t$-test.
	In particular, at the largest experimental asymmetry, $t_0=t_1=2.0$ for the symmetric protocol and $t_1=2.5$ for the fixed-injection-time protocol, $\overline{\Delta I_{PT}}$ lies outside the $4\sigma$- and $6\sigma$-equivalent Student's-$t$ reference bands, respectively.
	The positive asymmetry also exceeds the $2\sigma$-equivalent threshold around the peak in both protocols.
	These results demonstrate that the chosen binary sparse $N = 8$ SYK Hamiltonian exhibits a statistically significant nonzero $\Delta I_{PT}$ in the hardware experiment.

	\section{\label{S_section11}Size winding}
	Operator-growth properties in the TW protocol can be estimated by observing \textit{size winding}, a feature of operator-size-dependent phase alignment in the time-evolved injected operator.
	This diagnostic is motivated by the proposed relation between operator size and bulk momentum in the holographic interpretation of the SYK model~\cite{susskind2018thingsfall_2, PhysRevD.98.126016_2, PRXQuantum.4.010320_2, PRXQuantum.4.010321_2}.
	In the TW protocol, size winding provides an operator-level probe of the sign-dependent teleportation signal, showing how the interaction unwinds the coherently aligned phases.
	Although our system size is finite, this analysis remains informative because it probes the phase structure of operator growth directly, rather than relying only on the magnitude of the teleportation signal.
	Here we spell out the mechanism of size winding in the TW protocol, following Refs.~\cite{Qi2019_2, PhysRevX.12.031013_2, PRXQuantum.4.010320_2, PRXQuantum.4.010321_2, Jafferis2022_2}.
	
	\subsection{\label{S_section11.1}Size winding ansatz}
	For a given thermal state $\rho_{\beta}=e^{-\beta H}/\Tr e^{-\beta H}$, we expand the thermalized Majorana operator $\rho_{\beta}^{1/2}\psi^{j}(t)$ with $j \in \{1, \ldots, N\}$ in a basis of Majorana strings~\cite{Qi2019_2}. 
	Here $H$ is the one-sided Hamiltonian, and $\psi^{j}$ denotes either $\psi_{L}^{j}$ or $\psi_{R}^{j}$ on the chosen side.
	We label each string by a binary vector $P=(P_1,\ldots,P_N)$ with $P_{j}\in\{0,1\}$, such that the entry $P_{j}$ specifies whether the Majorana operator $\psi^{j}$ is included in the string. 
	With this notation, the expansion is
	\begin{equation}
		\label{S_eq40}
		\rho_{\beta}^{1/2}\psi^{j}(t) = \sum_{P} c_{P}(t)\psi^{P},
	\end{equation}
	where $\psi^{j}(t) = e^{iHt}\psi^{j} e^{-iHt}$.
	Here $\psi^{P}$ denotes the ordered product of Majorana operators selected by the nonzero entries of $P$, namely
	\begin{equation}
		\label{S_eq41}
		\psi^{P} = 2^{\abs{P}/2}i^{\abs{P}(\abs{P} - 1)/2}\prod_{j:P_j=1}\psi^j, \qquad \Tr\,\!(\psi^{P}{\psi^{Q}}^{\dagger}) = 2^{N}\delta^{PQ},
	\end{equation}
	and $\abs{P}=\sum_{j=1}^{N}P_j$ is the operator size.
	In this convention, $\psi^{P}$ is traceless and Hermitian unless $P$ is the zero vector.
	Here, $\abs{P}$ counts the number of Majorana operators appearing in the string; for example, $\abs{P}=1$ corresponds to a single Majorana operator, while larger $\abs{P}$ corresponds to a many-body Majorana string.
	As time evolves, a chaotic many-body system gives rise to operator size growth due to scrambling.
	
	We now introduce size winding: the phase of $c_{P}(t)$ in Eq.~\eqref{S_eq40} is organized by the size $\abs{P}$.
	We impose a stronger condition, perfect size winding, referring to the phase-alignment ansatz~\cite{PRXQuantum.4.010321_2}
	\begin{equation}
		\label{S_eq42}
		c_{P} = e^{i(\alpha\abs{P}/N+\phi)} r_{P},
		\qquad 
		r_{P}\in\mathbb{R}.
	\end{equation}
	From now on, the time dependence of $c_{P}$, $\alpha$, $\phi$, and $r_{P}$ is implicit.
	This ansatz implies that the phase of each string coefficient is determined solely by the size $\abs{P}$, rather than by the detailed string pattern.
	Equivalently, within each fixed-size sector, all coefficients are phase-aligned, while the common phase varies linearly with the size.
	Thus the operator has a coherent phase winding across size sectors, which is the central structure probed by the time evolution and by the left-right interaction in the protocol.
	
	\subsection{\label{S_section11.2}Winding size distribution}
	For each size sector $\abs{P} = l$, we define the winding size distribution~\cite{PRXQuantum.4.010320_2, PRXQuantum.4.010321_2}
	\begin{equation}
		\label{S_eq43}
		q(l)  = \sum_{P : \abs{P} = l} c_{P}^{2} = \abs{q(l)}e^{i\theta(l)},
	\end{equation}
	which is Eq.~(14) in the main text.
	At a given time, the corresponding size-winding phase is $\theta(l) = \arg q(l)$.
	For convenience, we redefine the phase as $\arg q(l)-\arg q(1)$ at each $l$.
	The usual size distribution is obtained from $\mathcal{P}(l) = \sum_{P \, : \, \abs{P} = l}|c_{P}|^{2}$~\cite{Roberts2018_2} and is insensitive to the complex phases of $c_{P}$. 
	By contrast, the winding size distribution uses $c_P^2$ and therefore retains phase information.
	For near-perfect size winding, this phase exhibits an approximately linear dependence on the operator size $l$.
	
	We also compute the ratio of the winding size distribution to the conventional size distribution,
	\begin{equation}
		R(l) = \dfrac{\abs{q(l)}}{\mathcal{P}(l)},
	\end{equation}
	which is Eq.~(15) in the main text.
	This quantity satisfies $0\le R(l)\le 1$.
	A value $R(l)\simeq1$ means that the phases of $c_{P}^{2}$ are well aligned within the same size sector.
	Therefore, near-perfect size winding is characterized by phase coherence within each size sector, measured by $R(l)\simeq1$, together with an approximately linear dependence of $\arg q(l)$ on $l$ across different size sectors~\cite{PRXQuantum.4.010321_2, Jafferis2022_2}.
	
	\subsection{\label{S_section11.3}Size winding before and after interaction}
	It is known that the SYK model exhibits both an approximately linear dependence of $\arg q(l)$ on $l$ and strong phase alignment~\cite{PRXQuantum.4.010321_2}.
	Thus, near-perfect size winding provides a useful criterion for distinguishing teleportation mediated by gravitational-like dynamics from generic scrambling in the protocol~\cite{PhysRevX.12.031013_2}.
	We therefore analyze the size winding before and after the interaction for our chosen binary sparse Hamiltonian as follows.
	
	We first analyze the size winding before the interaction, for an insertion of a single Majorana operator.
	Using the definition of TFD state, the left and right operators acting on the TFD state can be represented as
	\begin{equation}
		\label{S_eq45}
		\begin{aligned}
			\dfrac{i}{2^{N/2}}\mathcal{O}_{L}(-t)\ket{\mathrm{TFD}} &= \rho_{\beta}^{1/2}\mathcal{O}_{R}(t)\ket{I},\\
			\dfrac{1}{2^{N/2}}\mathcal{O}_{R}(t)\ket{\mathrm{TFD}} &= \mathcal{O}_{R}(t)\rho_{\beta}^{1/2}\ket{I},
		\end{aligned}
	\end{equation}
	where we have used the fact that $i\mathcal{O}_{L}(-t)\ket{I} = \mathcal{O}_{R}(t)\ket{I}$.
	For the single-Majorana insertion considered here, we use $\mathcal{O}_{L}(t) = \psi_{L}^{i}(t)$ and $\mathcal{O}_{R}(t) = \psi_{R}^{i}(t)$ for the $i$th Majorana fermion.
	We define the right-string basis states by
	\begin{equation}
		\label{S_eq46}
		\ket{P} = \psi_R^P\ket{I},
		\qquad
		\innerproduct{P}{P'} = \delta_{PP'}.
	\end{equation}
	Using Eq.~\eqref{S_eq40} and the perfect size-winding ansatz in Eq.~\eqref{S_eq42}, the two states in Eq.~\eqref{S_eq45} can be written in the right-string basis as
	\begin{equation}
		\label{S_eq47}
		\begin{aligned}
			\dfrac{i}{2^{N/2}}\mathcal{O}_{L}(-t)\ket{\text{TFD}} &= \sum_{P}c_{P}\ket{P} = \sum_{P}e^{i(\alpha\abs{P}/N + \phi)}r_{P}\ket{P},\\
			\dfrac{1}{2^{N/2}}\mathcal{O}_{R}(t)\ket{\text{TFD}} &= \sum_{P}c_{P}^{\ast}\ket{P} = \sum_{P}e^{-i(\alpha\abs{P}/N + \phi)}r_{P}\ket{P}.
		\end{aligned}
	\end{equation}
	The two states therefore have opposite winding phases in the same right-string basis.
	Using Eq.~\eqref{S_eq46}, we obtain $c_{P}(t)$ by projecting Eq.~\eqref{S_eq47} onto $\ket{P}$.
	Then $c_{P}(t)$ before the interaction is obtained by
	\begin{equation}
		\label{S_eq48}
		c_{P}(t) = \dfrac{i}{2^{N/2}}\bra{P}\mathcal{O}_{L}(-t)\ket{\mathrm{TFD}}.
	\end{equation}
	Equivalently, using the Frobenius inner product in Eq.~\eqref{S_eq41}, we also have
	\begin{equation}
		c_{P}(t) = 2^{-N}\Tr\,\!(\rho_{\beta}^{1/2}\psi_{L}^{i}(t){\psi_{L}^{P}}^{\dagger}).
	\end{equation}
	Then we obtain the winding size distribution $q(l)$ before the interaction by applying Eq.~\eqref{S_eq43}.
	
	We now analyze the size winding after applying the interaction at $t=0$.
	Similar to Eq.~\eqref{S_eq48}, the post-interaction coefficients $c_{P}^{(\mu)}(t)$ are obtained by projecting the kicked state onto the same right-string basis,
	\begin{equation}
		\label{S_eq50}
		c_{P}^{(\mu)}(t) = \dfrac{i}{2^{N/2}}\bra{P}e^{i\mu V}\mathcal{O}_{L}(-t)\ket{\mathrm{TFD}}.
	\end{equation}
	Then we obtain the post-interaction winding size distribution by applying Eq.~\eqref{S_eq43} with $c_{P}$ replaced by $c_{P}^{(\mu)}$.
	In Eq.~\eqref{S_eq50}, the interaction $e^{i\mu V}$ gives a size-dependent phase to each size sector,
	\begin{equation}
		e^{i\mu V}\ket{P} \sim e^{i\mu\abs{P}/qN}\ket{P},
	\end{equation}
	up to a global phase.
	Therefore, for an appropriate value of $\mu$, the interaction can reverse the winding direction of $\mathcal{O}_{L}(-t)\ket{\mathrm{TFD}}$, changing the sign of the slope of $\arg q(l)$.
	Within the size--momentum interpretation~\cite{susskind2018thingsfall_2, PhysRevD.98.126016_2, PRXQuantum.4.010320_2, PRXQuantum.4.010321_2}, this reversal is interpreted as re-emergence on the other side and serves as a diagnostic of traversability in the TW protocol~\cite{PRXQuantum.4.010321_2, Jafferis2022_2}.
	
	\begin{figure}[t]
		\includegraphics[width=0.8\linewidth]{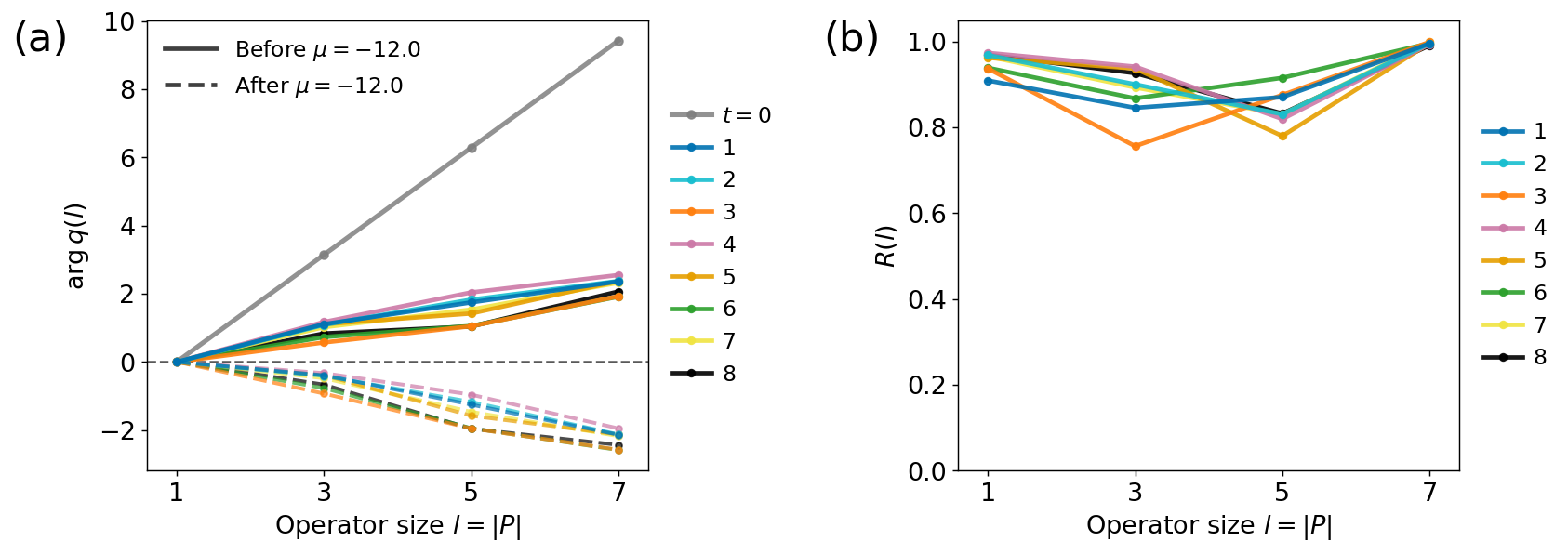}
		\caption{\label{S_figure26}
			Size-winding diagnostics for all single-Majorana insertions in the chosen Hamiltonian in the main text.
			The numerical labels in the legends denote the Majorana index $i$ in $\psi_{L}^{i}$.
			For all insertions, the probe time $t = 2.2$ is fixed to a single common value, given by the mutual-information peak time of the symmetric protocol averaged over all injection pairs.
			(a) Phase of the winding size distribution $\arg q(l)$ before the interaction (solid) and after the interaction with $\mu=-12$ (dashed).
			(b) Ratio of the winding size distribution $R(l)$ before the interaction.
		}
	\end{figure}
	
	Perfect size winding can also be characterized through the two-sided correlator~\cite{PhysRevX.12.031013_2, PRXQuantum.4.010320_2, PRXQuantum.4.010321_2}.
	Schematically, we consider
	\begin{equation}
		C_{\psi}^{(i)}(\mu,t) = \bra{\mathrm{TFD}} \psi_{R}^{i}(t)e^{i\mu V}\psi_{L}^{i}(-t) \ket{\mathrm{TFD}} \sim \sum_{l} q(l)e^{i\varphi_{\mu}(l)} ,
	\end{equation}
	where $\varphi_{\mu}(l)$ is linear in $l$.
	With our sign convention, if $\theta(l)=\arg q(l)$ is also linear in $l$, then $\mu<0$ can cancel the slope of $\theta(l)+\varphi_{\mu}(l)$ and make different size sectors add constructively.
	The opposite sign of $\mu$ increases the winding instead of canceling it, leading to destructive interference between size sectors.
	This phase alignment enhances the overlap between the kicked left insertion and the right-side response, producing a large $|C_{\psi}^{(i)}(\mu,t)|$.
	This is the size-winding interpretation of successful teleportation in the TW protocol~\cite{PRXQuantum.4.010320_2}.
	In this sense, the choice $\mu=-12$ for the chosen Hamiltonian in the main text provides a protocol-relevant interaction strength that reverses the winding direction in the size-sector phase.

	We also checked the size-winding diagnostics for all single-Majorana insertions $\psi_{L}^{i}$ with $i=1, \ldots, 8$ for the chosen Hamiltonian in the main text.
	For all insertions, we use the same probe time $t = 2.2$, fixed to the symmetric-protocol mutual-information peak time for $\mu = -12$ averaged over all injection pairs $(i, j)$ with $1 \leq i < j \leq 8$.
	Although the detailed values of $q(l)$ and $R(l)$ depend on the insertion operator, all eight insertions show the same qualitative behavior at this common probe time, as shown in Fig.~\ref{S_figure26}: the phase of $q(l)$ is approximately linear in $l$, the interaction reverses the winding direction for $\mu < 0$, and $R(l)$ remains close to unity over the relevant size sectors.
	This indicates that the size-winding behavior near the teleportation peak time is not specific to the particular Majorana insertion used in the main text.

\end{document}